\documentclass[11pt,a4paper]{article}
\usepackage[margin=1.0in]{geometry}
\usepackage{aas_macros,amsmath,amssymb,cite,hyperref,mathrsfs,wasysym}
\numberwithin{equation}{section}

\let\originalleft\left
\let\originalright\right
\renewcommand{\left}{\mathopen{}\mathclose\bgroup\originalleft}
\renewcommand{\right}{\aftergroup\egroup\originalright}

\newcommand{\ab}[1]{\left|#1\right|}

\newcommand{\br}[1]{\left[#1\right]}
\newcommand{\cu}[1]{\left\{#1\right\}}
\newcommand{\pa}[1]{\left(#1\right)}
\newcommand{\pd}{\mathop{}\!\partial}
\newcommand{\ed}{\mathop{}\!\mathrm{d}}

\DeclareMathOperator{\HeunC}{HeunC}
\DeclareMathOperator{\im}{Im}
\DeclareMathOperator{\re}{Re}
\newcommand{\mc}[1]{\mathcal{#1}}
\newcommand{\msc}[1]{\mathscr{#1}}
\newcommand{\ol}{\overline}
\newcommand{\tho}{\text{\thorn}}
\newcommand{\wt}{\widetilde}

\begin{document}

\title{\textbf{Gravitational Waves on Kerr Black Holes I:\\
Reconstruction of Linearized Metric Perturbations}}
\author{Roman Berens\footnote{\href{mailto:roman.berens@vanderbilt.edu}{roman.berens@vanderbilt.edu}}, Trevor Gravely\footnote{\href{mailto:trevor.gravely@vanderbilt.edu}{trevor.gravely@vanderbilt.edu}}, and Alexandru Lupsasca\footnote{\href{mailto:alexandru.v.lupsasca@vanderbilt.edu}{alexandru.v.lupsasca@vanderbilt.edu}}}
\date{\textit{Department of Physics \& Astronomy, Vanderbilt University, Nashville TN 37212, USA}}

\maketitle

\begin{abstract}
    The gravitational perturbations of a rotating Kerr black hole are notoriously complicated, even at the linear level.
    In 1973, Teukolsky showed that their physical degrees of freedom are encoded in two gauge-invariant Weyl curvature scalars that obey a separable wave equation.
    Determining these scalars is sufficient for many purposes, such as the computation of energy fluxes.
    However, some applications---such as second-order perturbation theory---require the reconstruction of metric perturbations.
    In principle, this problem was solved long ago, but in practice, the solution has never been worked out explicitly.
    Here, we do so by writing down the metric perturbation (in either ingoing or outgoing radiation gauge) that corresponds to a given mode of either Weyl scalar.
    Our formulas make no reference to the Hertz potential (an intermediate quantity that plays no fundamental role) and involve only the radial and angular Kerr modes, but not their derivatives, which can be altogether eliminated using the Teukolsky--Starobinsky identities.
    We expect these analytic results to prove useful in numerical studies and for extending black hole perturbation theory beyond the linear regime.
\end{abstract}

\newpage
\tableofcontents

\parskip=1em
\newpage

\section{Introduction and summary}
\label{sec:Introduction}

The Kerr metric describes the geometry of spacetime around a rotating astrophysical black hole.
In Boyer-Lindquist coordinates $(t,r,\theta,\phi)$, its line element $ds^2=g_{\mu\nu}\ed x^\mu\ed x^\nu$ takes the form
\begin{subequations}
\label{eq:Kerr}
\begin{gather}
    ds^2=\epsilon_g\pa{-\frac{\Delta}{\Sigma}\pa{\ed t-a\sin^2{\theta}\ed\phi}^2+\frac{\Sigma}{\Delta}\ed r^2+\Sigma\ed\theta^2+\frac{\sin^2{\theta}}{\Sigma}\br{\pa{r^2+a^2}\ed\phi-a\ed t}^2},\\
    \Delta=r^2-2Mr+a^2,\qquad
    \Sigma=r^2+a^2\cos^2{\theta}
    =\zeta\ol{\zeta},\qquad
    \zeta=r-ia\cos{\theta},
\end{gather}
\end{subequations}
where $\epsilon_g=+1$ if the metric has signature $(-,+,+,+)$, or $\epsilon_g=-1$ if its signature is $(+,-,-,-)$.
Throughout this paper, we will keep $\epsilon_g$ arbitrary to accommodate either choice of convention.

The gravitational perturbations of the Kerr geometry \eqref{eq:Kerr} are of great importance to both relativists and gravitational-wave scientists.
Describing the possible perturbations is a famously complicated problem, even at the linear level and in the absence of sources: inserting the ansatz $g_{\mu\nu}+h_{\mu\nu}$ into Einstein's field equations and expanding to linear order in the perturbation $h_{\mu\nu}$ around the Kerr background $g_{\mu\nu}$, one obtains the linearized vacuum Einstein equations (App.~\ref{app:LinearizedGravity})
\begin{align}
    \label{eq:LinearizedEE}
    -\nabla^2h_{\mu\nu}+2\nabla^\rho\nabla_{(\mu}h_{\nu)\rho}-\nabla_\mu\nabla_\nu{h^\rho}_\rho+g_{\mu\nu}\pa{\nabla^2{h^\rho}_\rho-\nabla^\rho\nabla^\sigma h_{\rho\sigma}}=0.
\end{align}
This is a formidable system of ten, coupled, second-order partial differential equations for the ten independent components of the metric perturbation $h_{\mu\nu}=h_{\nu\mu}$.
However, these components do not all represent independent physical degrees of freedom.
The Bianchi identity imposes a set of four constraints, leaving only six free components.
Four more can be fixed using diffeomorphism invariance, which allows for shifts of the perturbation by gauge transformations of the form
\begin{align}
    \label{eq:Diffeomorphisms}
    h_{\mu\nu}\to h_{\mu\nu}+\nabla_\mu\xi_\nu+\nabla_\nu\xi_\mu,
\end{align}
corresponding to linearized diffeomorphisms generated by an arbitrary vector $\xi=\xi^\mu\pd_\mu$.
Hence, a metric perturbation $h_{\mu\nu}$ really contains only two physical (propagating) degrees of freedom.
To tackle the system \eqref{eq:LinearizedEE} directly, this gauge redundancy must be eliminated by fixing a gauge.
Imposing a good choice of gauge conditions can greatly simplify the problem, but finding such a gauge is in itself a nontrivial task.

Regge and Wheeler \cite{Regge1957}, and later Zerilli \cite{Zerilli1970}, successfully carried out this line of attack for the case of a nonrotating black hole and solved the system \eqref{eq:LinearizedEE} in the Schwarzschild background.
The spherical symmetry of the spacetime provides a natural decomposition of any perturbation $h^{\mu\nu}$ into tensor harmonics $h_{\ell m}^{\mu\nu}$ with a definite behavior on the sphere $(\theta,\phi)$.
Time-translation symmetry further enables a decomposition of these harmonics into modes $h_{\omega\ell m}^{\mu\nu}$ behaving as $e^{-i\omega t}$.
This mode ansatz separates out the angular variables, leaving a set of equations for the radial dependence of the metric components.
A suitable choice of gauge (``Regge-Wheeler'' gauge) sets many of these components to zero, reducing the entire system to one radial equation of Schr\"odinger type (with two possible potentials, depending on the parity of the perturbation).

This direct approach crucially relies on spherical symmetry to decouple the system \eqref{eq:LinearizedEE}.
In the Kerr spacetime, this symmetry is broken, and to date no one has found an alternative method for directly decoupling these equations for $h_{\mu\nu}$, leaving them intractable.
In two breakthrough papers \cite{Teukolsky1973,Teukolsky1974}, Teukolsky discovered a new line of attack on the problem.
Rather than treat the components $h_{\mu\nu}$ of the metric perturbation as the fundamental variables, he instead considered the corresponding perturbation $C_{\mu\nu\rho\sigma}^{(1)}$ of the Weyl curvature tensor and identified two scalar projections that encode the two physical degrees of freedom carried by the metric perturbation,%
\begin{subequations}
\label{eq:WeylScalars}
\begin{align}
    \psi_0&=\epsilon_gC_{\mu\nu\rho\sigma}^{(1)}l^\mu m^\nu l^\rho m^\sigma, \\
    \psi_4&=\epsilon_gC_{\mu\nu\rho\sigma}^{(1)}n^\mu\ol{m}^\nu n^\rho\ol{m}^\sigma.
\end{align}
\end{subequations}
Here, $C_{\mu\nu\rho\sigma}^{(1)}$ is the linearized Weyl tensor, which can be expressed in terms of $h_{\mu\nu}$ (see Eq.~\eqref{eq:LinearizedWeyl} below), while $\cu{l,n,m,\ol{m}}$ is a Newman--Penrose tetrad \cite{Newman1962}: a complex null tetrad that consists of a pair of real null vectors $\cu{l,n}$ and a complex null vector $m$ obeying the conditions%
\begin{subequations}
\label{eq:Tetrad}
\begin{gather}
    l^\mu l_\mu=n^\mu n_\mu
    =m^\mu m_\mu
    =\ol{m}^\mu\ol{m}_\mu 
    =l^\mu m_\mu 
    =l^\mu\ol{m}_\mu 
    =n^\mu m_\mu 
    =n^\mu\ol{m}_\mu 
    =0,\\
    -l^\mu n_\mu=m^\mu\ol{m}_\mu
    =\epsilon_g,
\end{gather}
\end{subequations}
which together imply that the metric can be decomposed as
\begin{align}
    \label{eq:MetricDecomposition}
    g_{\mu\nu}=\epsilon_g\pa{-2l_{(\mu}n_{\nu)}+2m_{(\mu}\ol{m}_{\nu)}}.
\end{align}
For the Kerr metric \eqref{eq:Kerr}, a standard choice we will use throughout is the Kinnersley tetrad \cite{Kinnersley1969}
\begin{subequations}
\label{eq:KinnersleyTetrad}
\begin{align}
    l&=l^\mu\pd_\mu
    =\frac{r^2+a^2}{\Delta}\pd_t+\pd_r+\frac{a}{\Delta}\pd_\phi,\\
    n&=n^\mu\pd_\mu
    =\frac{1}{2\Sigma}\br{\pa{r^2+a^2}\pd_t-\Delta\pd_r+a\pd_\phi},\\
    m&=m^\mu\pd_\mu
    =\frac{1}{\sqrt{2}\,\ol{\zeta}}\pa{ia\sin{\theta}\pd_t+\pd_\theta+\frac{i}{\sin{\theta}}\pd_\phi}.
\end{align}
\end{subequations}
In a tour de force, Teukolsky recast the linearized Einstein equations \eqref{eq:LinearizedEE} as equations on the Weyl scalars \eqref{eq:WeylScalars} and showed that the latter miraculously decouple.
In fact, both $\Psi^{(+2)}=\psi_0$ and $\Psi^{(-2)}=\zeta^4\psi_4$ obey a \textit{single} decoupled equation: the Teukolsky master equation
\begin{align}
    \label{eq:TME}
    &\left[\pa{\frac{\pa{r^2+a^2}^2}{\Delta}-a^2\sin^2{\theta}}\pd_t^2+\frac{4aMr}{\Delta}\pd_t\pd_\phi+\pa{\frac{a^2}{\Delta}-\frac{1}{\sin^2{\theta}}}\pd^2_\phi
    \right.\nonumber\\
    &\quad-\Delta^{-s}\pd_r\pa{\Delta^{s+1}\pd_r}-\frac{1}{\sin{\theta}}\pd_\theta\pa{\sin{\theta}\pd_\theta}-2s\pa{\frac{a(r-M)}{\Delta}+\frac{i\cos{\theta}}{\sin^2{\theta}}}\pd_\phi\nonumber\\
    &\quad\left.-2s\pa{\frac{M\pa{r^2-a^2}}{\Delta}-r-ia\cos{\theta}}\pd_t+\pa{\frac{s^2}{\tan^2{\theta}}-s}\right]\Psi^{(s)}=4\pi\Sigma T^{(s)},
\end{align}
where $T^{(s)}$ encodes the source for the spin-$s$ perturbation.\footnote{Besides describing the spin-2 gravitational perturbations when $s=\pm2$, this master equation also governs the perturbations of the Kerr spacetime with arbitrary (half-)integer spin $s$: for instance, when $s=0$, it reduces to the wave equation $\nabla^2\Psi^{(0)}=-4\pi\epsilon_gT^{(0)}$ for a scalar field $\Psi^{(0)}$, and when $s=\pm1$, it describes the two physical modes $\Psi^{(+1)}=F_{\mu\nu}l^\mu m^\nu$ and $\Psi^{(-1)}=\pa{r-ia\cos{\theta}}^2F_{\mu\nu}\ol{m}^\mu n^\nu$ within a spin-1 electromagnetic perturbation $F_{\mu\nu}$.}
In the vacuum (sourceless) case, $T^{(s)}=0$.
In this paper, we will restrict our attention to the linearized Einstein equations \eqref{eq:LinearizedEE} in vacuum, which correspond to the master equation \eqref{eq:TME} with $s=\pm2$ and $T^{(\pm2)}=0$.

In a second miracle, the master equation \eqref{eq:TME} turns out to be separable.
As with the metric perturbations $h_{\mu\nu}$ of the Schwarzschild spacetime, one can use the stationarity and axisymmetry of the Kerr background to decompose the Weyl scalars \eqref{eq:WeylScalars} into modes behaving as $e^{-i\omega t+im\phi}$.
Then, the radial and polar dependence can be further separated by decomposing each $\Psi^{(\pm2)}$ as
\begin{subequations}
\label{eq:ModeDecomposition}
\begin{align}
    \Psi^{(s)}(t,r,\theta,\phi)&=\int\ed\omega\sum_{\ell=\ab{s}}^\infty\sum_{m=-\ell}^{+\ell}c_{\ell m}^{(s)}(\omega)\Psi_{\omega\ell m}^{(s)}(t,r,\theta,\phi),\\
    \label{eq:SingleMode}
    \Psi_{\omega\ell m}^{(s)}(t,r,\theta,\phi)&=e^{-i\omega t+im\phi}R_{\omega\ell m}^{(s)}(r)S_{\omega\ell m}^{(s)}(\theta).
\end{align}
\end{subequations}
Indeed, plugging this ansatz into the master equation \eqref{eq:TME} separates it into two, second-order ordinary differential equations (ODEs) for the radial and angular modes $R_{\omega\ell m}^{(s)}(r)$ and $S_{\omega\ell m}^{(s)}(\theta)$:
\begin{align}
    \label{eq:RadialODE}
    \br{\Delta^{-s}\frac{d}{dr}\pa{\Delta^{s+1}\frac{d}{dr}}+\frac{K^2-2is(r-M)K}{\Delta}+4is\omega r-\lambda_{\omega\ell m}^{(s)}}R_{\omega\ell m}^{(s)}(r)&=0,\\
    \label{eq:AngularODE}
    \br{\frac{1}{\sin{\theta}}\frac{d}{d\theta}\pa{\sin{\theta}\frac{d}{d\theta}}+a^2\omega^2\cos^2{\theta}-\frac{(m+s\cos{\theta})^2}{\sin^2{\theta}}-2sa\omega\cos{\theta}+s+A}S_{\omega\ell m}^{(s)}(\theta)&=0.
\end{align}
Here, $\lambda_{\omega\ell m}^{(s)}$ denotes a separation constant, which is only known numerically, and we introduced
\begin{align}
    \label{eq:Potentials}
    K=\pa{r^2+a^2}\omega-am,\qquad 
    A=\lambda_{\omega\ell m}^{(s)}+2am\omega-(a\omega)^2.
\end{align}
The ODEs \eqref{eq:RadialODE} and \eqref{eq:AngularODE} are of Heun type, and can each be mapped into the confluent Heun equation.
We express their solutions in terms of the confluent Heun function in Section~\ref{sec:Results} below.
This provides a completely analytic solution for the modes appearing in the decomposition \eqref{eq:ModeDecomposition}, and hence for the full Weyl scalars \eqref{eq:WeylScalars} associated with a given metric perturbation $h_{\mu\nu}$.

As an aside, we briefly comment on the origin of the two ``miracles'' uncovered by Teukolsky: 1) the Kerr metric perturbations $h_{\mu\nu}$ are encoded in two gauge-invariant Weyl scalars \eqref{eq:WeylScalars}, each obeying the same decoupled scalar equation \eqref{eq:TME}, and 2) this equation is completely separable.
In fact, Teukolsky showed that the first ``miracle'' occurs in any algebraically special spacetime of Petrov type D \cite{Petrov1954}: in this sense, therefore, the reduction of the spin-$s$ field equations to a pair of decoupled scalar equations on $\Psi^{(s)}$ can be viewed as a consequence of algebraic specialness.\footnote{In general, this decoupling property only holds for fixed $|s|$.
When perturbations of multiple spins are present, their equations may not fully decouple: for example, the Kerr-Newman spacetime is of Petrov type D, but the Teukolsky equations for electromagnetic ($|s|=1$) and gravitational ($|s|=2$) perturbations do not decouple.}
In Schwarzschild, the second ``miracle''---the separability of the spin-$s$ field equations---follows from spherical symmetry.
This symmetry is broken in Kerr, but as Carter discovered \cite{Carter1968}, it is replaced by a ``hidden'' symmetry in the form of an irreducible rank-2 Killing tensor $K_{\mu\nu}$, whose existence implies the separability of the scalar wave equation.
Indeed, the Killing equation for $K_{\mu\nu}$ implies that the differential operator $\mc{K}=\nabla^\mu K_{\mu\nu}\nabla^\nu$ commutes with the Laplacian $\nabla^2=\nabla^\mu g_{\mu\nu}\nabla^\nu$, and both commute with the action of the Killing vectors $\pd_t$ and $\pd_\phi$ generating the isometries of Kerr (stationarity and axisymmetry, respectively); since these four differential operators all commute, they are simultaneously diagonalizable, and their joint eigenvectors are precisely the separated eigenmodes $\Psi_{\omega\ell m}^{(0)}=e^{-i\omega t+im\phi}R_{\omega\ell m}^{(0)}(r)S_{\omega\ell m}^{(0)}(\theta)$;\footnote{The eigenvalues corresponding to $i\pd_t$, $i\pd_\phi$, $\nabla^2$, and $\mc{K}$ are respectively the energy $\omega$, azimuthal angular momentum $m$, invariant mass $\mu^2$, and separation constant $\lambda_{\omega\ell m}^{(0)}$ of the mode solution to $\pa{\nabla^2-\mu^2}\Psi_{\omega\ell m}^{(0)}=0$.
Likewise, the geodesic equation (which is the eikonal limit of the wave equation) is completely integrable due to the existence of four conserved quantities associated with these operators (e.g., $\mc{K}_{\mu\nu}p^\mu p^\nu$ is the Carter constant).} see, e.g.,\cite{Frolov2017} for a more detailed review.
More generally, the separability of the spin-$s$ field equations can be derived from the existence of commuting differential operators, all of which are ``hidden'' in the sense that they are associated with symmetries of higher rank, rather than being geometrically realized as isometries.\footnote{The symmetry operators that separate the spin-$s$ equations in Kerr are derived by Carter and McLenaghan \cite{Carter1979} for $|s|=\frac{1}{2}$ and by Torres del Castillo \cite{delCastillo1988} for $|s|=1$.
Aksteiner and B\"ackdahl \cite{Aksteiner2019} tackled the case $|s|=2$.}
While satisfying, symmetry-based arguments do not fully explain these ``miracles'': for one, the existence of ``hidden'' symmetries is sufficient but \textit{not} necessary for separability.\footnote{The C-metric describing a pair of accelerating black holes \cite{Griffiths2006} provides a counter-example: it is algebraically special of Petrov type D and therefore admits a conformal Killing tensor, but it does not possess an exact one \cite{Vollmer2017}.
We are not aware of any method to derive separability from conformal symmetries and their differential operators.}
Dudley and Finley \cite{Dudley1977} proved that all spacetimes in the Pleba\'nski-Demia\'nski class \cite{Plebanski1976}, which includes all vacuum type D solutions, share these two properties (decoupling and separability).

The determination of the Weyl scalars \eqref{eq:WeylScalars} is sufficient for answering many questions about the linearized gravitational perturbations of Kerr.
As a prime example, the energy or angular momentum that a metric perturbation carries into the horizon or out to infinity can be extracted from either Weyl scalar \cite{Teukolsky1974}.
More precisely, given a metric perturbation $h_{\mu\nu}$ with a single mode \eqref{eq:SingleMode} of $\Psi^{(-2)}=\zeta^4\psi_4$, the outgoing energy flux through a 2-sphere at infinity is \cite{Teukolsky1973}
\begin{align}
    \label{eq:OutgoingFlux}
    \frac{d^2E_{\rm out}}{dt\,d\Omega}=\lim_{r\to\infty}\frac{r^2}{4\pi\omega^2}\ab{\psi_4}^2,
\end{align}
which is time-independent (holds on any spacelike time slice) as long as the frequency $\omega$ is real.

\noindent Likewise, given a metric perturbation $h_{\mu\nu}$ corresponding to a single, real-frequency mode \eqref{eq:SingleMode} of $\Psi^{(+2)}=\psi_0$, the ingoing energy flux through a 2-sphere at infinity is \cite{Teukolsky1973}
\begin{align}
    \label{eq:IngoingFlux}
    \frac{d^2E_{\rm in}}{dt\,d\Omega}=\lim_{r\to\infty}\frac{r^2}{64\pi\omega^2}\ab{\psi_0}^2,
\end{align}
while the energy flux through the horizon $r_+=M+\sqrt{M^2-a^2}$ (whose sign is set by $k$) is \cite{Teukolsky1974}
\begin{align}
    \frac{d^2E_{\rm hole}}{dt\,d\Omega}=\frac{1}{2\pi}\frac{\omega}{32k\pa{k^2+4\epsilon^2}}\frac{\Delta^4}{\pa{2Mr_+}^3}\ab{\psi_0}^2\bigg|_{r=r_+},
    \label{eq:flux_hole}
\end{align}
where $k$ is the deviation from the superradiant bound and $\Omega_+$ the angular velocity of the horizon,
\begin{align}
    \label{eq:Parameters}
    k=\omega-m\Omega_+,\qquad
    \Omega_+=\frac{a}{2Mr_+},\qquad
    \epsilon=\frac{\sqrt{M^2-a^2}}{4Mr_+}. 
\end{align}
Analogous formulas for the flux of angular momentum $J$ follow from the substitution $dJ=\frac{m}{\omega}dE$.
Conservation of total energy and angular momentum imply that \cite{Teukolsky1974}
\begin{align}
    \frac{dE_{\rm hole}}{dt}=\frac{dE_{\rm in}}{dt}-\frac{dE_{\rm out}}{dt},\qquad
    \frac{dJ_{\rm hole}}{dt}=\frac{dJ_{\rm in}}{dt}-\frac{dJ_{\rm out}}{dt}.
\end{align}
It is possible to express each of the above fluxes (in, out, or hole) in terms of either Weyl scalar.\footnote{At infinity, $R_{\omega\ell m}^{(+2)}(r)\stackrel{r\to\infty}{\approx}\frac{Y_{\rm in}}{r}e^{-i\omega r^*}+\frac{Y_{\rm out}}{r^5}e^{+i\omega r^*}$, so taking the limit \eqref{eq:IngoingFlux} directly extracts the ingoing flux from the leading term of $\psi_0$, while the outgoing flux is encoded in its subleading term; the opposite holds for $\psi_4$.}
Indeed, $\psi_0$ and $\psi_4$ are \textit{not} independent of each other: a common misconception is that these two scalars correspond to the two degrees of freedom carried by the metric perturbation $h_{\mu\nu}$, but this fails to account for their complex nature.
In fact, a real perturbation $h_{\mu\nu}$, which carries two \textit{real} degrees of freedom, is fully encoded in a single \textit{complex} Weyl scalar (which carries one complex degree of freedom, or two real ones).\footnote{Even a real metric perturbation $h_{\mu\nu}$ gives rise to complex Weyl scalars \eqref{eq:WeylScalars} since the tetrad \eqref{eq:KinnersleyTetrad} is complex.}
Indeed, Wald \cite{Wald1973} showed that $\psi_0$ and $\psi_4$ individually contain all the information in $h_{\mu\nu}$, up to global ($\ell=0,1$) perturbations of the black hole mass $M$ and angular momentum $J=aM$.
As the above formulas demonstrate, for certain purposes, it may be more convenient to use $\psi_0$ rather than $\psi_4$ (or vice versa), but it is always possible to reconstruct either one from the other, and hence to use only one of them.
However, as we will soon see, the precise relation between them is subtle and was not given by Teukolsky.\footnote{Teukolsky \cite{Teukolsky1974} seems to suggest that a metric perturbation with $\psi_0$ given by his Eq.~(3.29) has a corresponding $\psi_4$ given by his Eq.~(3.30), but this is not the case: the true relation is significantly more complicated---see Sec.~\ref{subsec:WeylRelations}.}

Although knowledge of the Weyl scalars $\psi_0$ and/or $\psi_4$ is adequate for many purposes (such as the computation of energy and angular momentum fluxes), the full metric perturbation $h_{\mu\nu}$ is often required to determine other aspects of the perturbed spacetime geometry (such as the induced metric on the horizon), or to push perturbation theory beyond first order \cite{Loutrel2021,Ripley2021}.
It is therefore useful to reconstruct the specific metric perturbation $h_{\mu\nu}$ associated with a given Weyl scalar.
Various methods for carrying out this metric reconstruction have been developed, but the explicit form of the metric perturbation $h_{\mu\nu}$ that corresponds to a single mode of $\psi_0$ or $\psi_4$ has---to our knowledge---never been published. 
The main purpose of this paper is to close this gap in the literature.
Before describing our results, we briefly summarize preexisting work.

The problem of metric reconstruction was first tackled by Chrzanowski \cite{Chrzanowski1975} and Cohen and Kegeles \cite{Cohen1975}, who found explicit formulas for solutions $h_{\mu\nu}$ of the linearized Einstein equations \eqref{eq:LinearizedEE} in terms of solutions $\Psi^{(\pm2)}$ of the Teukolsky master equation \eqref{eq:TME}.
Soon after, Wald \cite{Wald1978} used a notion of adjointness (a $\dag$ operation) for tensorial differential operators to give a short and elegant derivation of these results (see App.~C of \cite{Dias2009} for a concise summary of his approach).
Similar equations for constructing spacetime fields associated with spin-$s$ perturbations from $\Psi^{(\pm s)}$ were also derived by Cohen and Kegeles \cite{Kegeles1979} and many other later authors.

The main upshot of these works is that there exist two complex, symmetric, differential operators ${\mc{S}_0^\dag}_{\mu\nu}$ and ${\mc{S}_4^\dag}_{\mu\nu}$---that we write out in Eqs.~\eqref{eq:AdjointOperator0}--\eqref{eq:AdjointOperator4} below---with the following properties.
Given a solution $\Psi^{(-2)}$ to the Teukolsky equation \eqref{eq:TME}, the real, symmetric 2-tensor
\begin{align}
    \label{eq:MetricReconstructionIRG}
    h_{\mu\nu}^{\rm IRG}=2\epsilon_g\re\pa{{\mc{S}_0}_{\mu\nu}^\dag\Psi_{\rm H}}
    =2\epsilon_g\re\pa{{\mc{S}_0}_{\mu\nu}^\dag\Psi^{(-2)}}
\end{align}
solves the linearized Einstein equations \eqref{eq:LinearizedEE} in the ``ingoing radiation gauge'' where (App.~\ref{subsec:LinearizedEE})
\begin{align}
    \label{eq:IRG}
    \text{IRG}:\qquad
    l^\mu h_{\mu\nu}^{\rm IRG}=0,\qquad
    g^{\mu\nu}h_{\mu\nu}^{\rm IRG}=0.
\end{align}
Likewise, given a solution $\Psi^{(+2)}$ to the Teukolsky equation \eqref{eq:TME}, the real, symmetric 2-tensor
\begin{align}
    \label{eq:MetricReconstructionORG}
    h_{\mu\nu}^{\rm ORG}=2\epsilon_g\re\pa{{\mc{S}_4}_{\mu\nu}^\dag\Psi_{\rm H}'}
    =2\epsilon_g\re\pa{{\mc{S}_4}_{\mu\nu}^\dag\zeta^4\Psi^{(+2)}}
\end{align}
solves the linearized Einstein equations \eqref{eq:LinearizedEE} in the ``outgoing radiation gauge'' where
\begin{align}
    \label{eq:ORG}
    \text{ORG}:\qquad
    n^\mu h_{\mu\nu}^{\rm ORG}=0,\qquad
    g^{\mu\nu}h_{\mu\nu}^{\rm ORG}=0.
\end{align}
While remarkable, these formulas fall short of a complete solution to the ``metric reconstruction'' problem because of the unfortunate feature that the ``reconstructed'' metrics \eqref{eq:MetricReconstructionIRG} and \eqref{eq:MetricReconstructionORG} do not have as their Weyl scalars the $\Psi^{(\pm2)}$ from which they were built.
That is, if one starts with a given $\Psi^{(-2)}=\zeta^4\psi_4$ (or $\Psi^{(+2)}=\psi_0$) and uses the above formulas to construct an associated metric perturbation $h_{\mu\nu}^{\rm IRG}$ (or $h_{\mu\nu}^{\rm ORG}$) with  corresponding linearized Weyl tensor $C_{\mu\nu\rho\sigma}^{(1)}$, and then computes the projections \eqref{eq:WeylScalars}, then the resulting Weyl scalar $\tilde{\psi}_4$ (or $\tilde{\psi}_0$) will not reproduce the scalar $\psi_4=\zeta^{-4}\Psi^{(-2)}$ (or $\psi_0=\Psi^{(+2)}$) with which one originally started.

To properly reconstruct the metric perturbation $h_{\mu\nu}$ associated with a given Weyl scalar $\psi_4$ or $\psi_0$, one must therefore plug a \textit{different} solution $\Psi^{(-2)}$ or $\Psi^{(+2)}$ into Eqs.~\eqref{eq:MetricReconstructionIRG} or \eqref{eq:MetricReconstructionORG}: one that is specifically engineered for the projections \eqref{eq:WeylScalars} to recover the desired Weyl scalars.
Such auxiliary solutions to the Teukolsky master equation \eqref{eq:TME} are known as \textit{Hertz potentials}.

In keeping with tradition, we will use $\Psi_{\rm H}$ to denote the solution $\Psi^{(-2)}$ to the master equation \eqref{eq:TME} with $s=-2$ that provides a Hertz potential for $h_{\mu\nu}^{\rm IRG}$, and will use $\Psi_{\rm H}'=\zeta^4\Psi^{(+2)}$ to denote the (rescaled) solution $\Psi^{(+2)}$ to the master equation \eqref{eq:TME} with $s=+2$ that provides a Hertz potential for $h_{\mu\nu}^{\rm ORG}$.
The Weyl scalars $\psi_0$ and $\psi_4$ are related to the IRG Hertz potential $\Psi_{\rm H}$ by
\begin{subequations}
\label{eq:IngoingPotential}
\begin{align}
    \label{eq:IngoingPotential0}
    \psi_0&=\frac{1}{4}l^4\ol{\Psi}_{\rm H},\\
    \label{eq:IngoingPotential4}
    \zeta^4\psi_4&=\frac{1}{4}\pa{\frac{1}{4}\mc{L}_{-1}\mc{L}_0\mc{L}_1\mc{L}_2\ol{\Psi}_{\rm H}-3M\pd_t\Psi_{\rm H}},
\end{align}
\end{subequations}
and similarly, they are also related to the ORG Hertz potential $\Psi_{\rm H}'$ by
\begin{subequations}
\label{eq:OutgoingPotential}
\begin{align}
    \label{eq:OutgoingPotential0}
    \psi_0&=\frac{1}{4}\pa{\frac{1}{4}\ol{\mc{L}}_{-1}\ol{\mc{L}}_0\ol{\mc{L}}_1\ol{\mc{L}}_2\frac{\ol{\Psi}_{\rm H}'}{\ol{\zeta}^4}+3M\pd_t\frac{\Psi_{\rm H}'}{\zeta^4}},\\
    \label{eq:OutgoingPotential4}
    \zeta^4\psi_4&=\frac{1}{4}\Delta^2\pa{\frac{\Sigma}{\Delta}n}^4\Delta^2\frac{\ol{\Psi}_{\rm H}'}{\ol{\zeta}^4},
\end{align}
\end{subequations}
where the differential operators $l$, $n$, and $m$ were defined in Eq.~\eqref{eq:KinnersleyTetrad}, and we also introduced
\begin{align}
    \label{eq:CurlyLn}
    \mc{L}_n&=\sqrt{2}\zeta\ol{m}+n\cot{\theta}
    =\pd_\theta-ia\sin{\theta}\pd_t-\frac{i}{\sin{\theta}}\pd_\phi+n\cot{\theta}.
\end{align}
We re-derive these relations in Sec.~\ref{subsec:Kerr} below.
By manipulating them, one may completely eliminate the Hertz potentials $\Psi_{\rm H}$ and $\Psi_{\rm H}'$ to obtain two coupled, fourth-order, differential relations between the Weyl scalars $\psi_0$ and $\psi_4$ only: the famous Teukolsky--Starobinsky identities%
\begin{subequations}
\label{eq:TeukolskyStarobinskyIdentities}
\begin{align}
    l^4\zeta^4\psi_4&=\frac{1}{4}\mc{L}_{-1}\mc{L}_0\mc{L}_1\mc{L}_2\psi_0-3M\pd_t\ol{\psi}_0,\\
    \Delta^2\pa{\frac{\Sigma}{\Delta}n}^4\Delta^2\psi_0&=\frac{1}{4}\ol{\mc{L}}_{-1}\ol{\mc{L}}_0\ol{\mc{L}}_1\ol{\mc{L}}_2\zeta^4\psi_4 +3M\pd_t\ol{\zeta}^4\ol{\psi}_4.
\end{align}
\end{subequations}
Here, they appear in their ``first form'' \cite{PriceThesis}; we give their ``second form'' in Eqs.~\eqref{eq:TSI2} below.

In summary, given a $\psi_0$ (or $\psi_4$), one can solve Eq.~\eqref{eq:IngoingPotential0} (or \eqref{eq:IngoingPotential4}, resp.) for its IRG Hertz potential $\Psi_{\rm H}$, from which one can then recover the corresponding $\psi_4$ (or $\psi_0$) via Eq.~\eqref{eq:IngoingPotential4} (or \eqref{eq:IngoingPotential0}, resp.), and also reconstruct the physical metric perturbation $h_{\mu\nu}^{\rm IRG}$ in IRG via Eq.~\eqref{eq:MetricReconstructionIRG}.

Alternatively, given the same $\psi_0$ (or $\psi_4$), one can instead solve Eq.~\eqref{eq:OutgoingPotential0} (or \eqref{eq:OutgoingPotential4}, resp.) for its ORG Hertz potential $\Psi_{\rm H}'$, from which one can then recover the corresponding $\psi_4$ (or $\psi_0$) via Eq.~\eqref{eq:OutgoingPotential4} (or \eqref{eq:OutgoingPotential0}, resp.), and also reconstruct the physical metric perturbation $h_{\mu\nu}^{\rm ORG}$ in the ORG via Eq.~\eqref{eq:MetricReconstructionORG}.
These steps completely describe the procedure for metric reconstruction (in either ingoing or outgoing radiation gauge) from either one of the Weyl scalars ($\psi_0$ or $\psi_4$).
To our knowledge, however, no one has ever chained all of these steps together---until this paper.

\subsection{Mode inversion and comparison with previous work}

To carry out the metric reconstruction procedure explicitly, one must invert the relations \eqref{eq:IngoingPotential} and \eqref{eq:OutgoingPotential} to obtain the Hertz potentials $\Psi_{\rm H}$ and $\Psi_{\rm H}'$ in terms of the Weyl scalars $\psi_0$ and $\psi_4$.
Although this inversion is difficult---or perhaps even impossible---to perform in full generality, it is straightforward (if messy) to invert these equations at the mode level, that is, for the case where $\Psi^{(+2)}=\psi_0$ or $\Psi^{(-2)}=\zeta^4\psi_4$ consists of a single mode \eqref{eq:SingleMode}.
For real frequencies $\omega$, the results of this mode inversion have appeared previously in the literature, though scattered across multiple papers \cite{Ori2003,vandeMeent2015,Pound2022,Nichols2012}, as we review below.
Here, we tackle the general case of arbitrary (and possibly complex) frequency $\omega$, and completely invert all of the relations \eqref{eq:IngoingPotential} and \eqref{eq:OutgoingPotential}.

At this stage, it may be helpful to point out the redundancies inherent in this mode inversion.
At a very basic level, since every metric perturbation $h_{\mu\nu}$ is fully encoded in either Weyl scalar, in principle, it is really only necessary to use half of the relations \eqref{eq:IngoingPotential} and \eqref{eq:OutgoingPotential}: more precisely, if one works with $\psi_0$ exclusively, then only Eqs.~\eqref{eq:IngoingPotential0} and \eqref{eq:OutgoingPotential0} are useful; conversely, if one works with $\psi_4$ exclusively, then only Eqs.~\eqref{eq:IngoingPotential4} and \eqref{eq:OutgoingPotential4} are useful.
Moreover, if one is content to fix a gauge (either IRG or ORG), then one need only invert one of the four relations.

For instance, in his pioneering work on metric reconstruction, Ori \cite{Ori2003} chose to work with $\psi_0$ exclusively, and in IRG only.
As a result, he only needed to invert Eq.~\eqref{eq:IngoingPotential0}---which is his Eq.~(11)---to extract the Hertz potential $\Psi_{\rm H}$ (in his notation, $\Psi_{\rm IRG}$) corresponding to a single mode of $\psi_0$; see his Eqs.~(12) and (14).
By contrast, in their later study of metric reconstruction, van de Meent and Shah \cite{vandeMeent2015} chose to work with $\psi_4$ exclusively, and in ORG only.
As such, they only needed to invert Eq.~\eqref{eq:OutgoingPotential4}---which is their Eq.~(71)---to derive the Hertz potential $\Psi_{\rm H}'$ (in their notation, $\Psi_{\rm ORG}$) associated with a single mode of $\psi_4$; see their Eqs.~(94)--(97).

In their excellent review, Pound and Wardell \cite{Pound2022} inverted precisely the other two relations.
Their Eq.~(106) inverts our Eq.~\eqref{eq:OutgoingPotential0} for $\Psi_{\rm H}'$ (in their notation, $\psi^{\rm ORG}$) in terms of $\Psi^{(+2)}=\psi_0$, which they denote $\psi_2$---see their Eq.~(81).
Likewise, their Eq.~(107) inverts our Eq.~\eqref{eq:IngoingPotential4} for $\Psi_{\rm H}$ (in their notation, $\psi^{\rm IRG}$) in terms of $\Psi^{(-2)}=\zeta^4\psi_4$, which they label $\psi_{-2}$---see their Eq.~(82).
Together, the three papers \cite{Ori2003,vandeMeent2015,Pound2022} completely invert all four of the relations \eqref{eq:IngoingPotential} and \eqref{eq:OutgoingPotential}, but only for the case of real frequency $\omega$, which can be a serious limitation for many applications.
An earlier work \cite{Nichols2012} also inverted one of these relations for complex $\omega$ (see App.~C therein).

Here, we invert all four of these relations for complex frequency $\omega$.
As a technical aside, there is another redundancy inherent in mode inversion that can simplify the problem.
The key observation is that, even if one wishes to use both $\psi_0$ and $\psi_4$, and work in both IRG and ORG, it is still enough to invert only two of the relations \eqref{eq:IngoingPotential} and \eqref{eq:OutgoingPotential}: that is, it suffices to invert Eqs.~\eqref{eq:IngoingPotential4} and \eqref{eq:OutgoingPotential0} only, as Pound and Wardell did, or else Eqs.~\eqref{eq:IngoingPotential0} and \eqref{eq:OutgoingPotential4} only.

In the former case, given a mode of $\psi_0$, one can obtain the ORG Hertz potential $\Psi_{\rm H}'$ by inverting Eq.~\eqref{eq:OutgoingPotential0}, and then plug it into Eq.~\eqref{eq:OutgoingPotential4} to directly obtain (without inversion) the corresponding $\psi_4$, from which the IRG Hertz potential $\Psi_{\rm H}$ may then be recovered by inverting Eq.~\eqref{eq:IngoingPotential4}; if instead, one is given a mode of $\psi_4$, then one can obtain the IRG Hertz potential $\Psi_{\rm H}$ by inverting Eq.~\eqref{eq:IngoingPotential4}, and then plug it into Eq.~\eqref{eq:IngoingPotential0} to directly obtain (without inversion) the corresponding $\psi_0$, from which the ORG Hertz potential $\Psi_{\rm H}'$ may then be recovered by inverting Eq.~\eqref{eq:OutgoingPotential0}.
Either way, two inversions are sufficient to obtain both Hertz potentials, and hence recover the metric perturbation $h_{\mu\nu}^{\rm IRG/ORG}$ in both gauges via Eqs.~\eqref{eq:MetricReconstructionIRG} and \eqref{eq:MetricReconstructionORG}.

Similarly, one can also complete the full metric reconstruction using only the inverses of Eqs.~\eqref{eq:IngoingPotential0} and \eqref{eq:OutgoingPotential4}.
In fact, there is a slight technical advantage to using the latter two inverses (rather than those used by Pound and Wardell): the reason is that inverting Eq.~\eqref{eq:IngoingPotential0} with one mode of $\psi_0$ results in a single mode of $\Psi_{\rm H}$, and likewise, inverting Eq.~\eqref{eq:OutgoingPotential4} with one mode of $\psi_4$ yields a single mode of $\Psi_{\rm H}'$; on the other hand, inverting Eq.~\eqref{eq:IngoingPotential4} with one mode of $\psi_4$ (or Eq.~\eqref{eq:OutgoingPotential0} with one mode of $\psi_0$) produces two modes of $\Psi_{\rm H}$ (or $\Psi_{\rm H}'$, resp.).
For this reason, we favor the more economical pair of equations, which yield simpler formulas.

In summary, the full solution of the linearized vacuum Einstein equations \eqref{eq:LinearizedEE} in the Kerr background can be broken down into three steps:
\begin{enumerate}
    \item First, one must solve the Teukolsky master equation \eqref{eq:TME} for either one of the Weyl scalars $\psi_0=\Psi^{(+2)}$ or $\psi_4=\zeta^{-4}\Psi^{(-2)}$.
    The desired solution (the one obeying appropriate boundary conditions for the physical problem at hand) is decomposable into modes \eqref{eq:SingleMode}.
    \item Second, one must carry out---mode by mode---the inversion procedure summarized above to obtain the other Weyl scalar and both of the associated Hertz potentials $\Psi_{\rm H}$ and $\Psi_{\rm H}'$.
    \item Third, one must plug these potentials into Eqs.~\eqref{eq:MetricReconstructionIRG} or \eqref{eq:MetricReconstructionORG} to reconstruct the metric perturbation $h_{\mu\nu}$ in ingoing radiation gauge \eqref{eq:IRG} or outgoing radiation gauge \eqref{eq:ORG}.
\end{enumerate}
The first step was completely worked out by Teukolsky \cite{Teukolsky1973} (including in the non-vacuum case) in the manner described below Eq.~\eqref{eq:ModeDecomposition}.
Here, our treatment offers only one minor technical advance: following Borissov and Fiziev \cite{Borissov2009}, we take the spin-$s$ mode solutions \eqref{eq:SingleMode} of the Teukolsky master equation \eqref{eq:TME} and reduce both the radial ODE \eqref{eq:RadialODE} and the angular ODE \eqref{eq:AngularODE} to the confluent Heun equation in its canonical form (which we review in App.~\ref{app:ConfluentHeun}).
This enables us to express both the radial modes $R_{\omega\ell m}^{(s)}(r)$ and the angular modes $S_{\omega\ell m}^{(s)}(\theta)$ in terms of the confluent Heun function $\HeunC(z)$, which was recently implemented in \textsc{Mathematica}.
As a result, one can symbolically manipulate the Weyl scalar modes $\Psi^{(s)}$ and the reconstructed metric components $h_{\mu\nu}$, without loading the ``Black Hole Perturbation Toolkit'' \textsc{BHPToolKit} \cite{BHPToolkit}.

The second step was completely worked out for real frequency $\omega$ \cite{Ori2003,vandeMeent2015,Pound2022}, and in one case also for complex $\omega$ \cite{Nichols2012}.
Here, we extend mode inversion with complex frequency $\omega$ to all cases.
We also derive direct relations between modes of $\psi_0$ and $\psi_4$, which are to our knowledge novel.

Finally, the third step was worked out by Chrzanowski, Cohen and Kegeles \cite{Chrzanowski1975,Cohen1975}.
Here, our main contribution is simply to chain all three steps together in order to explicitly write down the components $h_{\mu\nu}$ of the metric perturbation associated with a given Weyl scalar mode.

In our view, this final result adds value to the three steps taken in isolation (as ``the whole is greater than the sum of its parts''), and we hope our formulas will be of wide interest even beyond the field of black hole perturbation theory.
At the same time, in putting everything together, we have taken great care to clearly lay out the logical structure of metric reconstruction, in the hope that our treatment will also serve as a valuable entry point to this beautiful problem.

\subsection{Structure of the paper}

In this paper, we present---for the first time---the explicit form of the metric perturbation $h_{\mu\nu}$ that is associated with a single mode of a Weyl scalar $\psi_0$ or $\psi_4$.
For completeness, we give $h_{\mu\nu}$ in both IRG \eqref{eq:IRG} and ORG \eqref{eq:ORG}, and in three different coordinate systems: Boyer-Lindquist coordinates (Sec.~\ref{sec:Results}), as well as ingoing and outgoing Kerr coordinates (App.~\ref{app:IngoingOutgoing}).
We hope these formulas will prove useful in many applications and serve as a starting point for future studies.

The results we obtain are derived using the machinery of black hole perturbation theory, but having arrived at a final formula for $h_{\mu\nu}$, its validity can be checked directly and independently.
For this reason, we begin by presenting all the relevant equations in Sec.~\ref{sec:Results}, without dwelling on their origin.
After some preliminary definitions in Sec.~\ref{subsec:Definitions}, we describe the angular modes $S_{\omega\ell m}^{(s)}(\theta)$ in Sec.~\ref{subsec:AngularModes} and the radial modes $R_{\omega\ell m}^{(s)}(r)$ in Sec.~\ref{subsec:RadialModes}.
We express these modes in terms of the confluent Heun function and list their most salient properties, before introducing some of their associated constants in Sec.~\ref{subsec:TeukolskyStarobinskyConstants}.
The latter play an essential role in writing down the radial and angular Teukolsky--Starobinsky identities, which we summarize in Sec.~\ref{subsec:TeukolskyStarobinsky}.
Then, in Sec.~\ref{subsec:WeylRelations}, we present an explicit relation between the modes of the Weyl scalars $\psi_0$ and $\psi_4$ that are associated with a given metric perturbation $h_{\mu\nu}$.
This formula, which is to our knowledge new in the literature, lets one directly reconstruct either Weyl scalar from the other, explicitly demonstrating their interdependence.
Next, we give the reconstructed components of the metric perturbation $h_{\mu\nu}$ associated with a mode of $\psi_0$ or $\psi_4$---first in IRG in Sec.~\ref{subsec:MetricReconstructionIRG}, and then in the ORG in Sec.~\ref{subsec:MetricReconstructionORG}---before describing how our results may be directly checked in Sec.~\ref{subsec:ConsistencyChecks}.
Finally, we conclude in Sec.~\ref{subsec:TeukolskyStarobinskyForms} with a summary of the Teukolsky--Starobinsky identities in all their forms, as well as their logical interrelations.
This part of the paper can be read on its own.

The rest of the paper is devoted to a derivation of these results.
The key steps are condensed into a concise Sec.~\ref{sec:Derivation}, which assumes prior familiarity with the techniques developed by Newman and Penrose \cite{Newman1962}, and by Geroch, Held, and Penrose \cite{Geroch1973}; their formalisms are summarized in an excellent review by Whiting and Price \cite{Whiting2005}.
We closely follow the development in Price's comprehensive thesis \cite{PriceThesis} and quote its principal results in Sec.~\ref{subsec:Review}.
Price worked strictly in the ``mostly minus'' convention for the metric signature (with $\epsilon_g=-1$), so we take care to generalize his treatment to arbitrary sign of $\epsilon_g$.
Then, in Sec.~\ref{subsec:Kerr}, we specialize these general formulas, which apply to Petrov type D spacetime, to the case of the Kerr metric \eqref{eq:Kerr}.
Finally, this allows us to carry out mode inversion in Sec.~\ref{subsec:ModeInversion} and metric reconstruction in Sec.~\ref{subsec:KerrMetricReconstruction}.

We relegate some technical results to appendices.
In App.~\ref{app:LinearizedGravity}, we review the linearization of the Einstein equations and discuss some of the subtleties associated with radiation gauge (which imposes five conditions rather than the expected four).
In App.~\ref{app:IngoingOutgoing}, we coordinate transform the components of $h_{\mu\nu}^{\rm IRG/ORG}$ to ingoing and outgoing Kerr coordinates, which have the distinct advantage of remaining regular across the horizon (unlike their Boyer-Lindquist counterparts).

The components of $h_{\mu\nu}$ resulting from the metric reconstruction procedure do not depend on the Hertz potentials $\Psi_{\rm H}$ or $\Psi_{\rm H}'$, which serve merely as intermediate quantities and can be altogether eliminated after the mode inversion step.
Hence, they do not appear at all in Sec.~\ref{sec:Results}.
The components $h_{\mu\nu}$ do depend on the first and second derivatives of the radial and angular mode functions $R_{\omega\ell m}^{(s)}(r)$ and $S_{\omega\ell m}^{(s)}(\theta)$, since the operators ${\mc{S}_0^\dag}_{\mu\nu}$ and ${\mc{S}_4^\dag}_{\mu\nu}$ appearing in \eqref{eq:MetricReconstructionIRG} and \eqref{eq:MetricReconstructionORG} involve up to two derivatives.
Naturally, the second derivatives may be easily eliminated using the ODEs \eqref{eq:RadialODE} and \eqref{eq:AngularODE}.
Less obviously, it is also possible to eliminate their first derivatives using a separated form of the Teukolsky--Starobinsky identities \eqref{eq:TeukolskyStarobinskyIdentities}, as we show in App.~\ref{app:FirstDerivatives}.
This results in components $h_{\mu\nu}$ that involve only the mode functions $R_{\omega\ell m}^{(s)}(r)$ and $S_{\omega\ell m}^{(s)}(\theta)$, but not their derivatives, whose computation may introduce numerical error.
This form of the metric components, which may improve numerical stability, is given in App.~\ref{app:NoDerivatives}.

Lastly, we summarize the theory of the confluent Heun function, and carefully derive the quantization conditions for its eigenvalues, in App.~\ref{app:ConfluentHeun}.
We also discuss the orthogonality of the confluent Heun functions and derive an analytic formula for their normalization, following the work of Becker \cite{Becker1997} on the full Heun equation.
Finally, we apply these results to the angular ODE \eqref{eq:AngularODE} in App.~\ref{app:AngularHeun} and to the radial ODE \eqref{eq:RadialODE} in App.~\ref{app:RadialHeun}.
Beyond this paper, we also provide \textsc{Mathematica} notebooks containing our formulas and their checks on \href{https://github.com/Metric-Reconstruction/metric-reconstruction}{this Github}.

\subsection{Present limitations and future directions}

While we strove to maintain full generality in our treatment, it does suffer from some limitations.
First, the most glaring limitation is that we only considered the linearized Einstein equations \eqref{eq:LinearizedEE} in \textit{vacuum}.
Metric reconstruction in the presence of sources is just as interesting (if not more) and has immediate practical implications for the problem of self-force, which bears strong relevance to current and planned gravitational-wave observations.
Progress along those lines has been mostly focused on the Schwarzschild case \cite{Keidl2010,Shah2011a}, with a few notable exceptions \cite{Shah2011b,vandeMeent2018,Green2020,Toomani2022} beyond the work of Ori \cite{Ori2003} and van de Meent and Shah \cite{vandeMeent2015}.

The non-vacuum case poses at least two new difficulties.
First, the Weyl scalars and their associated Hertz potentials no longer obey the same equation: while in the vacuum case, they all solved the homogeneous form \eqref{eq:TME} of the Teukolsky master equation, in the presence of sources, they must now solve its inhomogeneous version, but with \textit{different} sources.\footnote{Moreover, the reconstruction of a Weyl scalar from the other now involves the source terms, unlike in Sec.~\ref{subsec:WeylRelations}.}
Though challenging to deal with, this complication is not insuperable.
The second difficulty, however, is less easily overcome: the radiation gauges \eqref{eq:IRG} and \eqref{eq:ORG} cease to work in the presence of sources.
More precisely, any metric perturbation satisfying the gauge conditions \eqref{eq:IRG} or \eqref{eq:ORG} will be singular in some neighborhood of the source \cite{Barack2001,Ori2003}.
This phenomenon occurs because, when they are combined with the non-vacuum Einstein equations, these gauge conditions place constraints on $T_{\mu\nu}$ that cannot in general be met: even in the simplest case of a stationary point-like source in flat spacetime, the metric perturbation must develop a string-like singularity extending from the source particle \cite{Barack2001}.
Pound, Merlin, and Barack \cite{Pound2014} classified the different types of radiation gauge singularities for a point mass in a general spacetime.
They also developed a method for stitching together two gauges that are regular in two different regions to construct a gauge with no string singularities, but at the cost of introducing a jump discontinuity across a surface intersecting the worldline of the particle.
van de Meent and Shah \cite{vandeMeent2015} used this procedure to compute the gravitational waves emitted by a particle on a non-circular equatorial orbit around a Kerr black hole, and van de Meent \cite{vandeMeent2018} later extended the analysis to generic Kerr orbits.
To our knowledge, an analogous procedure for non-point-like sources has yet to be developed.

We briefly note that Green, Hollands, and Zimmerman developed a formalism for perturbing the Kerr spacetime to higher order \cite{Green2020}, which they applied (together wtih Toomani, Spiers, and Pound) to a self-force calculation \cite{Toomani2022}.
This formalism adds to the reconstructed metric an additional ``corrector'' tensor that is designed to eliminate any restrictions on $T_{\mu\nu}$, and which results in a regular gauge that evades some problems of the ``no-string'' gauge described above.

We also point out that the metric reconstruction procedure carries a fundamental ambiguity: perturbations for which both $\psi_0$ and $\psi_4$ vanish cannot be captured.
Wald \cite{Wald1973} showed that in the vacuum case, such perturbations are (gauge-equivalent to) perturbations of the mass and angular momentum of the black hole (often called ``zero modes'').
These two ``missing'' pieces must be determined separately, a task known as metric completion.
This problem subsists in the case of a point particle source; see, e.g., \cite{Merlin2016,vandeMeent2017} for further discussion.

Second, here we systematically investigated perturbations of the subextreme Kerr geometry, given by Eq.~\eqref{eq:Kerr} with $0<|a|<M$.
The case of an extreme Kerr black hole (with $|a|=M$) necessitates a separate analysis.
Mathematically, the main difference lies in the radial modes $R_{\omega\ell m}^{(s)}(r)$, which degenerate to doubly confluent Heun functions and require dedicated treatment.
Physically, one may also expect additional complications to arise due to the emergence of the ``extremal throat'': a region of divergent proper volume that develops at the coordinate radius of the horizon as a black hole approaches extremality.
The spacetime metric in the throat, first discovered by Bardeen and Horowitz \cite{Bardeen1999}, is known as the NHEK (near-horizon extreme Kerr) geometry.
It is a vacuum Einstein solution in its own right and displays an enhanced conformal symmetry.
This observation forms the basis for a conjectured holographic duality for Kerr black holes \cite{Guica2009} and enables the solution of gravitational-wave emission problems that would otherwise be analytically intractable \cite{Porfyriadis2014,Hadar2014,Hadar2015}; see, e.g., Sec.~III of \cite{Kapec2020} for a recent review.
Understanding how gravitational perturbations of NHEK \cite{Dias2009,Amsel2009} connect to their extreme Kerr counterparts is a rich problem \cite{Starobinski1974,Hadar2021,Castro2021} (which for Reissner-Nordstr\"om has been solved by Porfyriadis \cite{Porfyriadis2018,Porfyriadis2019}).

Finally, our analysis applies to all Weyl scalar modes with ``generic'' complex frequencies $\omega$, but excludes the algebraically special ones \cite{Chandrasekhar1984}, for which the real perturbation $h_{\mu\nu}$ has only one nonzero Weyl scalar.
Mathematically, these modes arise at special frequencies that are zeroes of the radial Teukolsky--Starobinsky constant \eqref{eq:RadialConstant}, whose vanishing qualitatively changes the nature of the radial Teukolsky--Starobinsky identities, and hence of the mode inversion problem.
Physically, the flux formulas \eqref{eq:OutgoingFlux} and \eqref{eq:IngoingFlux} show that these modes correspond to waves that are either purely ingoing (if $\psi_4$ vanishes) or purely outgoing (if $\psi_0$ vanishes) at spatial infinity.
For this reason, modes with these algebraically special frequencies can sometimes be totally transmitted through the gravitational potential barrier of the black hole \cite{Andersson1994}, a phenomenon of considerable intrinsic interest; for further discussion, see, e.g., recent work by Cook and Lu \cite{Cook2023}.
These modes require special treatment and will be the subject of future work.
Since there exist no real algebraically special frequencies for $|s|\le2$ \cite{TeixeiradaCosta2021}, our handling of complex $\omega$ will be key.

\section{Statement of results}
\label{sec:Results}

In this section, we explicitly write down the metric components $h_{\mu\nu}$ for perturbations of the Kerr spacetime \eqref{eq:Kerr} that obey the linearized vacuum Einstein equations \eqref{eq:LinearizedEE} with a particular associated Weyl scalar $\psi_0$ or $\psi_4$.
The presentation is telegraphic: we only provide the essential formulas needed to express $h_{\mu\nu}$, and defer their derivation to later sections.
These derivations require the full machinery of the Newman--Penrose and Geroch--Held--Penrose formalisms, which we eschew in this section.
We also make no mention of the Hertz potentials, which are purely intermediate quantities that are no longer necessary once the metric has been reconstructed.

After some preliminary definitions in Sec.~\ref{subsec:Definitions}, we describe the angular modes $S_{\omega\ell m}^{(s)}(\theta)$ of the Kerr geometry in Sec.~\ref{subsec:AngularModes} and then its radial modes $R_{\omega\ell m}^{(s)}(r)$ in Sec.~\ref{subsec:RadialModes}.
Next, in Sec.~\ref{subsec:TeukolskyStarobinskyConstants}, we introduce the corresponding angular constants $D_{\omega\ell m}$, $D_{\omega\ell m}'$, and $\mc{D}_{\omega\ell m}$, together with radial constants $\msc{C}_{\omega\ell m}$, $\msc{C}_{\omega\ell m}'$, and $\mc{C}_{\omega\ell m}$, which appear in the angular and radial Teukolsky--Starobinsky identities that are presented in Sec.~\ref{subsec:TeukolskyStarobinsky} along with their unseparated analogue.
These identities can be used to check the physical relations between the Weyl scalars $\psi_0$ and $\psi_4$ given in Sec.~\ref{subsec:WeylRelations}.
At last, we provide explicit metric reconstruction formulas from the modes of either Weyl scalar in Sec.~\ref{subsec:MetricReconstructionIRG} for ingoing radiation gauge and Sec.~\ref{subsec:MetricReconstructionORG} for outgoing radiation gauge.
Finally, we discuss some consistency checks in Sec.~\ref{subsec:ConsistencyChecks}, before concluding in Sec.~\ref{subsec:TeukolskyStarobinskyForms} with a discussion of the various forms of the Teukolsky--Starobinsky identities and their logical interrelations. 

\subsection{Preliminary definitions}
\label{subsec:Definitions}

The Kerr metric \eqref{eq:Kerr} has an outer event horizon $r=r_+$ and an inner event horizon $r=r_-$ at the two roots of $\Delta=(r-r_+)(r-r_-)$; these horizons rotate with angular velocity $\Omega_\pm$ where
\begin{align}
    r_\pm=M\pm\sqrt{M^2-a^2},\qquad
    \Omega_\pm=\frac{a}{2Mr_\pm}
    =\frac{a}{r_\pm^2+a^2}.
\end{align}
The Boyer-Lindquist metric \eqref{eq:Kerr} is singular at the horizons.
Regular coordinates that ensure the metric remains smooth across the horizons are introduced in App.~\ref{app:IngoingOutgoing} using a tortoise coordinate
\begin{align}
    \label{eq:TortoiseCoordinate}
    r_*=r+c_+\ln\pa{\frac{r-r_+}{2M}}-c_-\ln\pa{\frac{r-r_-}{2M}},\qquad
    c_\pm=\frac{2Mr_\pm}{r_+-r_-},
\end{align}
as well as another coordinate
\begin{align}
    \label{eq:SharpCoordinate}
    r_\sharp=\frac{a}{r_+-r_-}\ln\pa{\frac{r-r_+}{r-r_-}},
\end{align}
which are defined such that
\begin{align}
    \frac{dr_*}{dr}=\frac{r^2+a^2}{\Delta},\qquad
    \frac{dr_\sharp}{dr}=\frac{a}{\Delta}.
\end{align}
For any choice of tetrad vectors $a,b\in\cu{l,n,m,\ol{m}}$, we will write $h_{ab}$ to denote the projection
\begin{align}
    \label{eq:TetradProjection}
    h_{ab}\equiv a^\mu b^\nu h_{\mu\nu}.
\end{align}
Since we only consider real metric perturbations $h_{\mu\nu}$, these projections are such that $\ol{h}_{ab}=h_{\ol{a}\ol{b}}$.

\subsection{Angular modes}
\label{subsec:AngularModes}

For every choice of (continuous) frequency $\omega\in\mathbb{C}$ and (quantized) azimuthal angular momentum $m\in\mathbb{Z}$, there is an infinite---but discrete---set of separation constants $\lambda^{(s)}(a\omega,m)$ for which the angular ODE \eqref{eq:AngularODE} admits a solution that is regular at both the northern pole $\theta=0$ and at the southern pole $\theta=\pi$.
Each of these solutions has some number $n\ge0$ of zeros over the range $0<\theta<\pi$, which defines an index $\ell\equiv n+\max\pa{|s|,|m|}$.
Clearly, every solution has a well-defined label $\ell$, and the converse statement (namely, that every integer $\ell\ge\max\pa{|s|,|m|}$ picks a unique solution) is also true: this is Theorem 4.1 of \cite{Breuer1977}, generalized by Eq.~(4.1) of \cite{Casals2005}.

Thus, we can index the discrete set of regular solutions by $(\omega,\ell,m)$, with $\omega\in\mathbb{C}$, $\ell\ge|s|$, and $-\ell\le m\le\ell$.\footnote{Since the number of zeros $n$ must be nonnegative, $\ell\ge\max\pa{|s|,|m|}$, which implies both $\ell\ge|s|$ and $\ell\ge|m|$.}
These solutions are the eigenmodes $S_{\omega\ell m}^{(s)}(\theta)$ of the angular ODE \eqref{eq:AngularODE}, with associated eigenvalues $\lambda_{\omega\ell m}^{(s)}$.
These ``spin-weighted spheroidal eigenfunctions'' take the form
\begin{align}
    \label{eq:AngularModes}
    \hat{S}_{\omega\ell m}^{(s)}(\theta)=\pa{1-\cos{\theta}}^{\mu_1}\pa{1+\cos{\theta}}^{\mu_2}e^{a\omega\pa{1+\cos{\theta}}}\,H\pa{\frac{1+\cos{\theta}}{2}},
\end{align}
where the hat indicates that we are referring to the modes with this specific normalization, and
\begin{align}
\label{eq:AngularHeun}
    H(z)=\HeunC\pa{-p+\beta,2\beta,2\mu_2+1,2\mu_1+1,4a\omega;z}
\end{align}
denotes the confluent Heun function as implemented in \textsc{Mathematica} (see App.~\ref{app:ConfluentHeun} for details), which is normalized such that $H(0)=1$.
Here, we also introduced the parameters
\begin{subequations}
\begin{gather}
    \mu_1=\frac{|s+m|}{2},\qquad
    \mu_2=\frac{|s-m|}{2},\qquad
    \beta=2a\omega\pa{\mu_1+\mu_2+s+1},\\
    p=-\lambda_{\omega\ell m}^{(s)}-s(s+1)+2a\omega(\mu_1-\mu_2-m)+(\mu_1+\mu_2)^2+\mu_1+\mu_2.
\end{gather}
\end{subequations}

The derivation of these modes is presented below in App.~\ref{app:AngularHeun}, which also describes a method for determining the eigenvalues $\lambda_{\omega\ell m}^{(s)}$ from the confluent Heun function.
Unfortunately, the $\lambda_{\omega\ell m}^{(s)}$ do not admit a convenient analytical representation, and must always be solved for numerically.
The ``Black Hole Perturbation Toolkit''  \textsc{BHPToolKit} \cite{BHPToolkit} automatically returns them via the command \texttt{SpinWeightedSpheroidalEigenvalue}$[s,\ell,m,a\omega]$.

For fixed $\omega$, $m$, and $s$, the hatted modes \eqref{eq:AngularModes} are orthogonal (but not quite orthonormal):
\begin{align}
    \label{eq:Orthogonality}
    \int_0^\pi\hat{S}_{\omega\ell m}^{(s)}(\theta)\hat{S}_{\omega\ell'm}^{(s)}(\theta)\sin{\theta}\ed\theta=\delta_{\ell\ell'}I_{\omega\ell m}^{(s)},
\end{align}
where the constants $I_{\omega\ell m}^{(s)}$ can be computed numerically  but also admit a closed-form expression in terms of the confluent Heun function and its derivatives, given in App.~\ref{app:AngularNormalization} below.

If the frequency $\omega$ is real, then these modes are complete over $\theta\in[0,\pi]$.\footnote{Stewart \cite{Stewart1975} proves strong completeness for $\omega\in\mathbb{R}$ and at least weak completeness for $\omega$ in a complex disk.}
Thus, by Eq.~\eqref{eq:Orthogonality},
\begin{align}
    \sum_{\ell=\max\pa{|s|,|m|}}^\infty\frac{\hat{S}_{\omega\ell m}^{(s)}(\theta)\hat{S}_{\omega\ell m}^{(s)}(\theta')}{I_{\omega\ell m}^{(s)}}=\delta\pa{\cos{\theta}-\cos{\theta'}}.
\end{align}
Our hatted modes \eqref{eq:AngularModes} define the traditional ``spin-weighted spheroidal harmonics'' \cite{Breuer1977} via
\begin{align}
    Z_{\omega\ell m}^{(s)}(\theta,\phi)=\frac{1}{\sqrt{2\pi I_{\omega\ell m}^{(s)}}}\hat{S}_{\omega\ell m}^{(s)}(\theta)e^{im\phi}.
\end{align}
For real frequency $\omega$, these harmonics form a complete, orthonormal set over the 2-sphere:
\begin{subequations}
\begin{gather}
    \label{eq:Orthnormality}
    \int_0^\pi\int_0^{2\pi}Z_{\omega\ell m}^{(s)}(\theta,\phi)\ol{Z}_{\omega\ell'm'}^{(s)}(\theta,\phi)\sin{\theta}\ed\theta\ed\phi=\delta_{\ell\ell'}\delta_{mm'},\\
    \sum_{\ell=|s|}^\infty\sum_{m=-\ell}^{+\ell}Z_{\omega\ell m}^{(s)}(\theta,\phi)\ol{Z}_{\omega\ell m}^{(s)}(\theta',\phi')=\delta\pa{\cos{\theta}-\cos{\theta'}}\delta\pa{\phi-\phi'}.
\end{gather}
\end{subequations}
As $a\to0$ and spherical symmetry is restored, the spin-weighted spheroidal harmonics $Z_{\omega\ell m}^{(s)}(\theta,\phi)$ reduce (up to signs) to their standard spherical counterparts $Y_{\ell m}^{(s)}(\theta,\phi)=P_{\ell m}^{(s)}(\cos{\theta})e^{im\phi}$, where
\begin{align}
    P_{\ell m}^{(s)}(\cos{\theta})=N_{\ell m}^{(s)}\sin^{2\ell}\pa{\frac{\theta}{2}}\sum_{k=0}^{\ell-s}(-1)^{k}\binom{\ell-s}{k}\binom{\ell+s}{k+s-m}\cot^{2k+s-m}\pa{\frac{\theta}{2}},
\end{align}
are ``spin-weighted associated Legendre polynomials'' with normalization\footnote{This choice follows from the orthonormality condition \eqref{eq:Orthnormality}, that is, $\int_{S^2}Y^{(s)}_{\ell m}(\Omega)\ol{Y}^{(s)}_{\ell'm'}(\Omega)\ed\Omega=\delta_{\ell\ell'}\delta_{mm'}$.}
\begin{align}
    N_{\ell m}^{(s)}=(-1)^{\ell+m-s}\sqrt{\frac{2\ell+1}{4\pi}\frac{(\ell+m)!}{(\ell+s)!}\frac{(\ell-m)!}{(\ell-s)!}}.
\end{align}
In the limit $a\to0$, the eigenvalues $\lambda_{\omega\ell m}^{(s)}$ lose their frequency-dependence and degenerate to
\begin{align}
    \lambda_{0\ell m}^{(s)}=\ell(\ell+1)-s(s+1).
\end{align}
Analytic expansions of $\lambda_{\omega\ell m}^{(s)}$ to high order in small $c=a\omega$, as well as large real or imaginary $c$, have been derived by Berti, Cardoso and Casals \cite{Berti2006}.

Finally, the symmetries of the angular ODE \eqref{eq:AngularODE} imply that our hatted angular eigenmodes \eqref{eq:AngularModes} and their eigenvalues obey the following identities:\footnote{The proportionality factor in the last identity is set by the specific normalization of our hatted modes \eqref{eq:AngularModes}.}
\begin{align}
    \label{eq:AngularSymmetries}
    \lambda_{\omega\ell m}^{(-s)}=\lambda_{\omega\ell m}^{(s)}+2s,\qquad
    \ol{\lambda}_{\omega\ell m}^{(s)}=\lambda_{-\ol{\omega},\ell,-m}^{(s)},\qquad
    \ol{\hat{S}}_{\omega\ell m}^{(s)}=\hat{S}_{-\ol{\omega},\ell,-m}^{(-s)}.
\end{align}

\subsection{Radial modes}
\label{subsec:RadialModes}

For every choice of (continuous) frequency $\omega\in\mathbb{C}$ and integer harmonics $(\ell,m)$ with $\ell\ge|s|$ and $-\ell\le m\le\ell$, the second-order radial ODE \eqref{eq:RadialODE} admits a two-dimensional space of solutions.
A convenient choice of basis for this space is provided by the ``in'' and ``out'' solutions, which correspond to modes that are purely ingoing or purely outgoing at the horizon, respectively:\footnote{\label{fn:UpDown}Other common modes are the ``up'' and ``down'' solutions, which are purely ingoing or outgoing at infinity (see, e.g., Fig.~1 in \cite{Pound2022}).
They correspond to solutions of the confluent Heun equation \eqref{eq:ConfluentHeun} around its irregular singular point $z=\infty$, which we unfortunately do not know how to represent in terms of the function $\HeunC(z)$.}
\begin{subequations}
\label{eq:RadialModes}
\begin{align}
    \hat{R}_{\omega\ell m}^{(s)\,{\rm in}}(r)&=\pa{r_+-r_-}^{-\xi_2-s}\pa{r-r_+}^{-\xi_1-s}\pa{r-r_-}^{\xi_2}e^{i\omega(r-r_+)}H^{\rm in}\pa{-\frac{r-r_+}{r_+-r_-}},\\
    \hat{R}_{\omega\ell m}^{(s)\,{\rm out}}(r)&=\pa{r_+-r_-}^{-\xi_2}\pa{r-r_+}^{\xi_1}\pa{r-r_-}^{\xi_2} e^{i\omega(r-r_+)}H^{\rm out}\pa{-\frac{r-r_+}{r_+-r_-}},
\end{align}
\end{subequations}
where the hat indicates that we are referring to the modes with this specific normalization, and%
\begin{subequations}
\begin{align}
    H^{\rm in}(z)&=\HeunC\pa{q+(\epsilon-\delta)(1-\gamma),\alpha+\epsilon(1-\gamma),2-\gamma,\delta,\epsilon;z},\\
    H^{\rm out}(z)&=\HeunC\pa{q,\alpha,\gamma,\delta,\epsilon;z}
\end{align}
\end{subequations}
with parameters (recall that the $c_\pm$ were defined in Eq.~\eqref{eq:TortoiseCoordinate}, while Eq.~\eqref{eq:Parameters} introduced $k$)
\begin{subequations}
\begin{gather}
    \xi_1=ic_+\pa{\omega-m\Omega_+}=ic_+k,\qquad
    \xi_2=-ic_-\pa{\omega-m\Omega_-},\\
    \gamma=2\xi_1+s+1,\qquad 
    \delta=2\xi_2+s+1,\qquad
    \epsilon=-2i\omega\pa{r_+-r_-},\\
    \alpha=-2i\omega(2s+1)\pa{r_+-r_-},\qquad 
    q=-2i\omega r_+(2s+1)+\lambda_{\omega\ell m}^{(s)}.
\end{gather}
\end{subequations}
The derivation of these modes is presented below in App.~\ref{app:RadialHeun}.
Their behavior near the horizon is best described using the tortoise coordinate \eqref{eq:TortoiseCoordinate}:
\begin{subequations}
\label{eq:HorizonBehavior}
\begin{align}
    \hat{R}_{\omega\ell m}^{(s)\,{\rm in}}(r)&\stackrel{r\to r_+}{\approx}\pa{r_+-r_-}^{-s}\pa{r-r_+}^{-ic_+k-s}
    \approx\br{e^{ikr_+}\frac{\pa{r_+-r_-}^{-ic_-k}}{\pa{r_++r_-}^{2ikM}}}\Delta^{-s}e^{-ikr_*},\\
    \hat{R}_{\omega\ell m}^{(s)\,{\rm out}}(r) &\stackrel{r\to r_+}{\approx}\pa{r-r_+}^{ic_+k}
    \approx\br{e^{-ikr_+}\frac{\pa{r_+-r_-}^{ic_-k}}{\pa{r_++r_-}^{-2ikM}}}e^{ik r_*}.
\end{align}
\end{subequations}
In practice, physical boundary conditions must be purely ingoing at the horizon.
This excludes the ``out'' mode, leaving only the ``in'' mode as physical.
Nonetheless, the ``out'' mode remains mathematically interesting, for instance in the study of totally transmitted modes (TTMs) \cite{Cook2014}.

Finally, the symmetries of the radial ODE \eqref{eq:RadialODE} imply that our hatted radial eigenmodes \eqref{eq:RadialModes} obey the following identities (here, the proportionality factors depend on normalization):
\begin{align}
    \label{eq:RadialSymmetries}
     \ol{\hat{R}}_{\omega\ell m}^{(s)\,{\rm in/out}}=\hat{R}_{-\ol{\omega},\ell,-m}^{(s)\,{\rm in/out}} 
    =\Delta^{-s}\hat{R}_{\ol{\omega}\ell m}^{(-s)\,{\rm out/in}}.
\end{align}

\subsection{Teukolsky--Starobinsky constants}
\label{subsec:TeukolskyStarobinskyConstants}

The Teukolsky--Starobinsky constants are derived in Sec.~\ref{subsec:TeukolskyStarobinsky} below; here, we merely quote them.
For some authors, the term refers to the quantities\footnote{\label{fn:TS}We have expressed these constants in terms of $\lambda_{\omega\ell m}^{(+2)}$, but could have also used $\lambda_{\omega\ell m}^{(-2)}=\lambda_{\omega\ell m}^{(+2)}+4$; see Eq.~\eqref{eq:AngularSymmetries}.}
\begin{align}
    \label{eq:TeukolskyStarobinskyD}
    \mc{D}_{\omega\ell m}&=\pa{\lambda_{\omega\ell m}^{(+2)}+4}^2\pa{\lambda_{\omega\ell m}^{(+2)}+6}^2+8a\omega\pa{m-a\omega}\pa{\lambda_{\omega\ell m}^{(+2)}+4}\pa{5\lambda_{\omega\ell m}^{(+2)}+26}\notag\\
    &\phantom{=}\ +48(a\omega)^2\br{2\lambda_{\omega\ell m}^{(+2)}+8+3\pa{m-a\omega}^2}.
\end{align}
For other authors, the term instead refers to the frequency-shifted version
\begin{align}
    \label{eq:RadialConstant}
    \mc{C}_{\omega\ell m}=\mc{D}_{\omega\ell m}+\pa{12\omega M}^2.
\end{align}
Both versions factorize into products of other quantities with important properties described in Sec.~\ref{subsec:TeukolskyStarobinsky}.
In particular, $\mc{D}_{\omega\ell m}$ is the product of ``angular Teukolsky--Starobinsky constants''
\begin{align}
    \label{eq:Factorization}
    \mc{D}_{\omega\ell m}=\hat{D}_{\omega\ell m}\hat{D}_{\omega\ell m}',
\end{align}
which take the explicit form\footref{fn:TS}
\begin{align}
    \label{eq:ConstantsD}
    \hat{D}_{\omega\ell m}=
    \begin{cases}
        \frac{(m-2)!}{(m+2)!}\mc{D}_{\omega\ell m}
        &m\geq2,\\
        -\frac{1}{6}p_\omega^{(3)}\pa{\lambda_{\omega\ell m}^{(+2)}}
        &m=1,\\
        p_{-\omega}^{(2)}\pa{\lambda_{\omega\ell m}^{(+2)}}
        &m=0,\\
        -6p_{-\omega}^{(1)}\pa{\lambda_{\omega\ell m}^{(+2)}}
        &m=-1,\\
        \frac{(m+1)!}{(m-3)!}
        &m\leq-2,
    \end{cases}\qquad
    \hat{D}_{\omega\ell m}'=
    \begin{cases}
        \frac{(m+2)!}{(m-2)!}
        &m\geq2,\\
        -6p_\omega^{(1)}\pa{\lambda_{\omega\ell m}^{(+2)}}
        &m=1,\\
        p_\omega^{(2)}\pa{\lambda_{\omega\ell m}^{(+2)}}
        &m=0,\\
        -\frac{1}{6}p_{-\omega}^{(3)}\pa{\lambda_{\omega\ell m}^{(+2)}}
        &m=-1,\\
        \frac{(m-3)!}{(m+1)!}\mc{D}_{\omega\ell m}
        &m\leq-2,
    \end{cases}
\end{align}
where $\frac{(m-3)!}{(m+1)!}=\frac{1}{(m+1)m(m-1)(m-2)}$, etc., and in terms of $c\equiv a\omega$, we introduced the polynomials\footnote{The factorization \eqref{eq:Factorization} is nontrivial for $|m|\le1$ and relies on the constants \eqref{eq:TeukolskyStarobinskyD} and polynomials \eqref{eq:Polynomials} obeying, for $x=\lambda_{\omega\ell m}^{(+2)}$, the identities $\mc{D}_{\omega\ell,1}=p_\omega^{(1)}(x)p_\omega^{(3)}(x)$, $\mc{D}_{\omega\ell,0}=p_\omega^{(2)}(x)p_{-\omega}^{(2)}(x)$, and $\mc{D}_{\omega\ell,-1}=p_{-\omega}^{(1)}(x)p_{-\omega}^{(3)}(x)$.}%
\begin{subequations}
\label{eq:Polynomials}
\begin{align}
    p_\omega^{(1)}(x)&=x+6c+4,\\
    p_\omega^{(2)}(x)&=x^2+2\pa{4c+5}x+4\pa{3c^2+8c+6},\\
    p_\omega^{(3)}(x)&=x^3-2\pa{3c-8}x^2-4\pa{c^2+8c-21}x+8\pa{3c^3-8c^2-c+18}.
\end{align}
\end{subequations}

Likewise, $\mc{C}_{\omega\ell m}$ is the product of ``radial Teukolsky--Starobinsky constants''
\begin{align}
    \mc{C}_{\omega\ell m}=\hat{\msc{C}}_{\omega\ell m}^{\rm in}\hat{\msc{C}}_{\omega\ell m}^{{\rm in}\,\prime}
    =\hat{\msc{C}}_{\omega\ell m}^{\rm out}\hat{\msc{C}}_{\omega\ell m}^{{\rm out}\,\prime},
\end{align}
which are separately defined for the ``in'' and ``out'' solutions as
\begin{align}
    \label{eq:ConstantsC}
    \hat{\msc{C}}_{\omega\ell m}^{\rm in}=\Gamma,\qquad
    \hat{\msc{C}}_{\omega\ell m}^{{\rm in}\,\prime}=\frac{\mc{C}_{\omega\ell m}}{\Gamma},\qquad
    \hat{\msc{C}}_{\omega\ell m}^{\rm out}=\frac{\mc{C}_{\omega\ell m}}{\wt{\Gamma}},\qquad
    \hat{\msc{C}}_{\omega\ell m}^{{\rm out}\,\prime}=\wt{\Gamma},
\end{align}
with
\begin{subequations}
\label{eq:GammaSigmaW}
\begin{gather}
    \Gamma=(w+2i\sigma)(w+i\sigma)w(w-i\sigma),\qquad
    \wt{\Gamma}=(w-2i\sigma)(w-i\sigma)w(w+i\sigma),\\
    \sigma=r_+-r_-,\qquad
    w=4Mkr_+.
\end{gather}
\end{subequations}
The precise form of our angular and radial Teukolsky--Starobinsky constants depends on the specific normalization of our angular modes \eqref{eq:AngularModes} and radial modes \eqref{eq:RadialModes}, but this normalization choice drops out of the products that define the Teukolsky--Starobinsky constants $\mc{D}_{\omega\ell m}$ or $\mc{C}_{\omega\ell m}$.

We will henceforth suppress the labels ``in'' or ``out'' on  $\hat{\msc{C}}_{\omega\ell m}$, $\hat{\msc{C}}_{\omega\ell m}'$, and $\hat{R}_{\omega\ell m}^{(s)}(r)$, as our equations will hold for either choice (but not for a linear combination thereof---caveat lector).

Finally, by careful inspection and use of the symmetries of $\lambda_{\omega\ell m}^{(s)}$ given in Eq.~\eqref{eq:AngularSymmetries}, one has%
\begin{subequations}
\label{eq:TeukolskyStarobinskySymmetries}
\begin{gather}
    \ol{\hat{D}}_{\ol{\omega}\ell m}=\hat{D}_{\omega\ell m},\qquad
    \ol{\hat{D}_{}'}_{\!\!\ol{\omega}\ell m}=\hat{D}_{\omega\ell m}',\qquad
    \hat{D}_{-\omega,\ell,-m}=\hat{D}_{\omega\ell m}',\\
    \ol{\hat{\msc{C}}}_{\omega\ell m}=\hat{\msc{C}}_{-\ol{\omega},\ell,-m},\qquad
    \ol{\hat{\msc{C}}_{}'}_{\!\!\!\omega\ell m}=\hat{\msc{C}}_{-\ol{\omega},\ell,-m}'.
\end{gather}
\end{subequations}
The first two identities simply reflect the fact that $\hat{D}_{\omega\ell m}$ and $\hat{D}_{\omega\ell m}'$ are real if the frequency $\omega$ is real.
On the other hand, $\hat{\msc{C}}_{\omega\ell m}$ and $\hat{\msc{C}}_{\omega\ell m}'$ are generally complex, even for real $\omega$.

\subsection{Radial and angular Teukolsky--Starobinsky identities}
\label{subsec:TeukolskyStarobinsky}

Using the radial ``potential'' $K$ defined in Eq.~\eqref{eq:Potentials}, we now define radial operators
\begin{align}
    \label{eq:Dn}
    \msc{D}_n=\pd_r-\frac{iK}{\Delta}+2n\frac{r-M}{\Delta},\qquad
    \msc{D}_n^\dag=\pd_r+\frac{iK}{\Delta}+2n\frac{r-M}{\Delta}.
\end{align}
Likewise, using an angular ``potential'' $Q$, we also define angular operators
\begin{align}
    \label{eq:Ln}
    \msc{L}_n=\pd_\theta+Q+n\cot{\theta},\qquad
    \msc{L}_n^\dag=\pd_\theta-Q+n\cot{\theta},\qquad
    Q=-a\omega\sin{\theta}+\frac{m}{\sin{\theta}}.
\end{align}
These operators are the ``mode versions'' of $l$, $\frac{\Sigma}{\Delta}n$, $\mc{L}_n$, and $\ol{\mc{L}}_n$ of Eqs.~\eqref{eq:KinnersleyTetrad} and \eqref{eq:CurlyLn}, meaning%
\begin{subequations}
\label{eq:ModeVersions}
\begin{align}
    l\br{f(r,\theta)e^{-i\omega t+im\phi}}&=\msc{D}_0f(r,\theta)e^{-i\omega t+ im\phi},\\
    \frac{\Sigma}{\Delta}n\br{f(r,\theta)e^{-i\omega t+im\phi}}&=-\frac{1}{2}\msc{D}_0^\dag f(r,\theta)e^{-i\omega t+im\phi},\\
    \mc{L}_n\br{f(r,\theta)e^{-i\omega t+ im\phi}}&=\msc{L}_nf(r,\theta)e^{-i\omega t+im\phi},\\
    \ol{\mc{L}}_n\br{f(r,\theta)e^{-i\omega t+im\phi}}&=\msc{L}_n^\dag f(r,\theta)e^{-i\omega t+im\phi}.
\end{align}
\end{subequations}
In terms of these differential operators, the radial and angular ODEs \eqref{eq:AngularODE} and \eqref{eq:RadialODE} are just
\begin{subequations}
\begin{align}
    \pa{\Delta\msc{D}_1\msc{D}_2^\dag+6i\omega r-\lambda_{\omega\ell m}^{(+2)}-4}R_{\omega\ell m}^{(+2)}&=0,\\
    \pa{\Delta\msc{D}_{-1}^\dag\msc{D}_0-6i\omega r-\lambda_{\omega\ell m}^{(-2)}}R_{\omega\ell m}^{(-2)}&=0,\\
    \pa{\msc{L}_{-1}^\dag\msc{L}_2-6a\omega\cos{\theta}+\lambda_{\omega\ell m}^{(+2)}+4}S_{\omega\ell m}^{(+2)}&=0,\\
    \pa{\msc{L}_{-1}\msc{L}_2^\dag+6a\omega\cos{\theta}+\lambda_{\omega\ell m}^{(-2)}}S_{\omega\ell m}^{(-2)}&=0.
\end{align}
\end{subequations}

A direct computation reveals that the application of a particular fourth-order differential operator to any solution $S_{\omega\ell m}^{(\pm2)}$ of the angular ODE \eqref{eq:AngularODE} with $s=\pm2$ yields another solution $S_{\omega\ell m}^{(\mp2)}$ of the same ODE but with opposite spin $s=\mp2$:
\begin{subequations}
\begin{align}
    \msc{L}_{-1}\msc{L}_0\msc{L}_1\msc{L}_2S_{\omega\ell m}^{(+2)}&\propto S_{\omega\ell m}^{(-2)},\\
    \msc{L}_{-1}^\dag\msc{L}_0^\dag\msc{L}_1^\dag\msc{L}_2^\dag S_{\omega\ell m}^{(-2)}&\propto S_{\omega\ell m}^{(+2)}.
\end{align}
\end{subequations}
Likewise, the action of a certain fourth-order differential operator on any solution $R_{\omega\ell m}^{(\pm2)}$ of the radial ODE \eqref{eq:RadialODE} with $s=\pm2$ yields a solution $R_{\omega\ell m}^{(\mp2)}$ of the ODE with opposite spin $s=\mp2$:
\begin{subequations}
\begin{align}
    \msc{D}_0^4R_{\omega\ell m}^{(-2)}&\propto R_{\omega\ell m}^{(+2)},\\
    \Delta^2\pa{\msc{D}_0^\dag}^4\Delta^2R_{\omega\ell m}^{(+2)}&\propto R_{\omega\ell m}^{(-2)}.
\end{align}
\end{subequations}
These statements follow from the assumption that $R_{\omega\ell m}^{(\pm2)}$ and $S_{\omega\ell m}^{(\pm2)}$ obey the radial and angular ODEs \eqref{eq:RadialODE} and \eqref{eq:AngularODE} and hold regardless of the linear combination of modes being considered or their normalization. However, if one considers specific modes with a particular normalization, then the proportionality factors become fixed.
For the angular modes $\hat{S}_{\omega\ell m}^{(\pm2)}$ defined in Eq.~\eqref{eq:AngularModes},%
\begin{subequations}
\label{eq:ATSI1}
\begin{align}
    \label{eq:ATSI1a}
    \msc{L}_{-1}\msc{L}_0\msc{L}_1\msc{L}_2\hat{S}_{\omega\ell m}^{(+2)}&=\hat{D}_{\omega\ell m}\hat{S}_{\omega\ell m}^{(-2)},\\
    \label{eq:ATSI1b}
    \msc{L}_{-1}^\dag\msc{L}_0^\dag\msc{L}_1^\dag\msc{L}_2^\dag\hat{S}_{\omega\ell m}^{(-2)}&=\hat{D}_{\omega\ell m}'\hat{S}_{\omega\ell m}^{(+2)},
\end{align}
\end{subequations}
where the proportionality constants $\hat{D}_{\omega\ell m}$ and $\hat{D}'_{\omega\ell m}$ are given in Eq.~\eqref{eq:ConstantsD}.
As for the radial modes $\hat{R}_{\omega\ell m}^{(\pm2)}$ defined in Eq.~\eqref{eq:RadialModes}, it luckily turns out that ``in'' and ``out'' modes do not mix:\footnote{This property also holds for the ``up'' and ``down'' modes: they do not mix under the action of these operators.}%
\begin{subequations}
\label{eq:RTSI1}
\begin{align}
    \label{eq:RTSI1a}
    \msc{D}_0^4\hat{R}_{\omega\ell m}^{(-2)\,{\rm in/out}}&=\hat{\msc{C}}_{\omega\ell m}^{\rm in/out}\hat{R}_{\omega\ell m}^{(+2)\,{\rm in/out}},\\
    \label{eq:RTSI1b}
    \Delta^2\pa{\msc{D}_0^\dag}^4\Delta^2\hat{R}_{\omega\ell m}^{(+2)\,{\rm in/out}}&=\hat{\msc{C}}_{\omega\ell m}^{{\rm in/out}\,\prime}
    \hat{R}_{\omega\ell m}^{(-2)\,{\rm in/out}},
\end{align}
\end{subequations}
where the proportionality constants $\hat{\msc{C}}_{\omega\ell m}$ and $\hat{\msc{C}}_{\omega\ell m}'$ are given in Eq.~\eqref{eq:ConstantsC}.
We will refer to Eqs.~\eqref{eq:ATSI1} and \eqref{eq:RTSI1} as the angular and radial Teukolsky--Starobinsky identities in first form.
We review their relation to the Teukolsky--Starobinsky identities in first form \eqref{eq:TeukolskyStarobinskyIdentities} in Sec.~\ref{subsec:ConsistencyChecks}.
We give a slick derivation of the constants in Eq.~\eqref{eq:RTSI1} in App.~\ref{app:RadialHeun} (for ``up/down'' modes too).

Plugging Eqs.~\eqref{eq:ATSI1a} and \eqref{eq:ATSI1b} into each other yields an eighth-order differential relation for each mode $\hat{S}_{\omega\ell m}^{(\pm2)}$.
These are the angular Teukolsky--Starobinsky identities in second form:
\begin{subequations}
\label{eq:ATSI2}
\begin{align}
    \msc{L}_{-1}^\dag\msc{L}_0^\dag\msc{L}_1^\dag\msc{L}_2^\dag\msc{L}_{-1}\msc{L}_0\msc{L}_1\msc{L}_2\hat{S}_{\omega\ell m}^{(+2)}&=\mc{D}_{\omega\ell m}\hat{S}_{\omega\ell m}^{(+2)},\\
    \msc{L}_{-1}\msc{L}_0\msc{L}_1\msc{L}_2\msc{L}_{-1}^\dag\msc{L}_0^\dag\msc{L}_1^\dag\msc{L}_2^\dag\hat{S}_{\omega\ell m}^{(-2)}&=\mc{D}_{\omega\ell m}\hat{S}_{\omega\ell m}^{(-2)}.
\end{align}
\end{subequations}
Likewise, plugging Eqs.~\eqref{eq:RTSI1a} and \eqref{eq:RTSI1b} into each other results in an eighth-order differential relation for each $\hat{R}_{\omega\ell m}^{(\pm2)}$.
These are the radial Teukolsky--Starobinsky identities in second form:
\begin{subequations}
\label{eq:RTSI2}
\begin{align}
    \msc{D}_0^4\Delta^2\pa{\msc{D}_0^\dag}^4\Delta^2 \hat{R}_{\omega\ell m}^{(+2)\,{\rm in/out}}
    &=\mc{C}_{\omega\ell m}\hat{R}^{(+2)\,{\rm in/out}}_{\omega\ell m},\\
    \Delta^2\pa{\msc{D}_0^\dag}^4\Delta^2\msc{D}_0^4 \hat{R}_{\omega\ell m}^{(-2)\,{\rm in/out}} 
    &=\mc{C}_{\omega\ell m}\hat{R}_{\omega\ell m}^{(-2)\,{\rm in/out}}.
\end{align}
\end{subequations}
There is a subtle but important distinction between the first-form identities \eqref{eq:ATSI1} and \eqref{eq:RTSI1} and their second-form analogues \eqref{eq:ATSI2} and \eqref{eq:RTSI2}: whereas the former only hold for our particular choice of angular and radial modes $\hat{S}_{\omega\ell m}^{(\pm2)}$ and $\hat{R}_{\omega\ell m}^{(\pm2)}$, the latter apply to any solution of their respective ODEs (that is, for any linear combination of hatted modes).
Thus, we could take the hats off the modes in the second-form identities \eqref{eq:ATSI2} and \eqref{eq:RTSI2}, and they would continue to hold.
By contrast, if we changed the modes in the first-form identities \eqref{eq:ATSI1} and \eqref{eq:RTSI1}, then we would have to correspondingly change the constants appearing therein.
For this reason, the hatted constants have a different logical status than their unhatted analogues: while the constants $\hat{D}_{\omega\ell m}$, $\hat{D}_{\omega\ell m}'$, $\hat{\msc{C}}_{\omega\ell m}$, and $\hat{\msc{C}}_{\omega\ell m}'$ depend on our choice of hatted modes $\hat{S}_{\omega\ell m}^{(\pm2)}$ and $\hat{R}_{\omega\ell m}^{(\pm2)}$ (and are hence also hatted), their products $\mc{D}_{\omega\ell m}=\hat{D}_{\omega\ell m}\hat{D}_{\omega\ell m}'$ and $\mc{C}_{\omega\ell m}=\hat{\msc{C}}_{\omega\ell m}\hat{\msc{C}}_{\omega\ell m}'$ are independent of this choice of modes (which is why they are unhatted).

In this sense, the unhatted constants are properties of their respective equations, while their hatted factors belong to specific mode solutions of the equations.
This points to another logical difference between these identities, namely that the first forms imply the second forms, whereas the converse is not true.
Indeed, multiplying the fourth-order operators in \eqref{eq:ATSI1} and \eqref{eq:RTSI1} to obtain eighth-order ones in \eqref{eq:ATSI2} and \eqref{eq:RTSI2} leads to information loss: one is left only with the unhatted products $\mc{D}_{\omega\ell m}=\hat{D}_{\omega\ell m}\hat{D}_{\omega\ell m}'$ and $\mc{C}_{\omega\ell m}=\hat{\msc{C}}_{\omega\ell m}\hat{\msc{C}}_{\omega\ell m}'$, from which their hatted factors cannot be recovered, since the specific modes appearing in the first-form identities are forgotten.

\subsection{Relation between modes of the Weyl scalars \texorpdfstring{$\psi_0$}{ψ0} and \texorpdfstring{$\psi_4$}{ψ4}}
\label{subsec:WeylRelations}

After mode inversion and elimination of the Hertz potentials in Sec.~\ref{subsec:ModeInversion} below, we find the (two modes of) $\psi_4$ associated with a given (single mode of) $\psi_0$, and vice versa.

Given a single mode of $\psi_0$ of the form
\begin{align}
    \label{eq:Mode0}
    \psi_0=e^{-i\omega t+im\phi}\hat{R}_{\omega\ell m}^{(+2)}\hat{S}_{\omega\ell m}^{(+2)},
\end{align}
the corresponding $\psi_4$ is given by
\begin{align}
    \label{eq:CorrespondingMode4}
    \zeta^4\psi_4&=\frac{\hat{D}_{\omega\ell m}}{4\hat{\msc{C}}_{\omega\ell m}}e^{-i\omega t+im\phi}\hat{R}_{\omega\ell m}^{(-2)}\hat{S}_{\omega\ell m}^{(-2)}-\frac{3i\ol{\omega}M}{\ol{\hat{\msc{C}}}_{\omega\ell m}}e^{i\ol{\omega}t-im\phi}\hat{R}_{-\ol{\omega},\ell,-m}^{(-2)}\hat{S}_{-\ol{\omega},\ell,-m}^{(-2)}.
\end{align}
Conversely, given a single mode of $\zeta^4\psi_4$ of the form
\begin{align}
    \label{eq:Mode4}
    \zeta^4\psi_4=e^{-i\omega t+im\phi}\hat{R}_{\omega\ell m}^{(-2)}\hat{S}_{\omega\ell m}^{(-2)},
\end{align}
the corresponding $\psi_0$ is
\begin{align}
    \label{eq:CorrespondingMode0}
    \psi_0&=\frac{4\hat{D}_{\omega\ell m}'}{\hat{\msc{C}}_{\omega\ell m}'}e^{-i\omega t+im\phi}\hat{R}_{\omega\ell m}^{(+2)}\hat{S}_{\omega\ell m}^{(+2)}+\frac{48i\ol{\omega}M}{\ol{\hat{\msc{C}}_{}'}_{\!\!\!\omega\ell m}}e^{i\ol{\omega}t-im\phi}\hat{R}_{-\ol{\omega},\ell,-m}^{(+2)}\hat{S}_{-\ol{\omega},\ell,-m}^{(+2)}.
\end{align}
As a consistency check, one can verify that each of these pairs of $(\psi_0,\psi_4)$---that is, the one given by Eqs.~\eqref{eq:Mode0}--\eqref{eq:CorrespondingMode4}, or the one given by Eqs.~\eqref{eq:Mode4}--\eqref{eq:CorrespondingMode0}---obeys the Teukolsky--Starobinsky identities \eqref{eq:TeukolskyStarobinskyIdentities}.
This requires the use of the symmetry properties \eqref{eq:AngularSymmetries} and \eqref{eq:RadialSymmetries}, together with the radial and angular Teukolsky--Starobinsky identities \eqref{eq:ATSI1} and \eqref{eq:RTSI1}.

\subsection{Metric reconstruction in ingoing radiation gauge}
\label{subsec:MetricReconstructionIRG}

Here, we explicitly reconstruct the metric perturbation $h_{\mu\nu}$ in ingoing radiation gauge \eqref{eq:IRG} that is associated with a single mode \eqref{eq:Mode0} of $\psi_0$ or with a single mode \eqref{eq:Mode4} of $\zeta^4\psi_4$.

To treat both cases simultaneously, we express the components of $h_{\mu\nu}^{\rm IRG}$ in terms of constants $A^{\rm IRG}$ and $B^{\rm IRG}$.
These must take different values according to whether one reconstructs the metric perturbation from a single mode \eqref{eq:Mode0} of $\psi_0$, in which case one must set
\begin{align}
\label{eq:IRGMode0Solution}
    A^{\rm IRG}=0,\qquad
    B^{\rm IRG}=\frac{4}{\overline{\hat{\msc{C}}}_{\omega\ell m}},
\end{align}
or from a single mode \eqref{eq:Mode4} of $\zeta^4\psi_4$, in which case one must set
\begin{align}
\label{eq:IRGMode4Solution}
    A^{\rm IRG}=-\frac{192i\omega M}{\mc{C}_{\omega\ell m}},\qquad
    B^{\rm IRG}=\frac{16\ol{\hat{D}_{}'}_{\!\!\omega\ell m}}{\ol{\mc{C}}_{\omega\ell m}}.
\end{align}
The Boyer-Lindquist components of the real metric perturbation $h_{\mu\nu}^{\rm IRG}$ are then given by\footnote{\label{fn:Trivial}These expressions are in some sense trivial, as they are directly obtained by inverting Eq.~\eqref{eq:TetradProjection}.
The nontrivial part of metric reconstruction comes in Eqs.~\eqref{eq:HIRG} and \eqref{eq:HORG}, whose derivation is sketched in Sec.~\ref{subsec:KerrMetricReconstruction}.}
\begin{subequations}
\label{eq:MetricComponentsIRG}
\begin{align}
    h_{tt}^{\rm IRG}&=-a^2\sin^2{\theta}\,\mc{M}_+-2a\sin{\theta}\,\mc{N}_-+h_{nn},\\
    h_{rr}^{\rm IRG}&=\frac{\Sigma^2}{\Delta^2}h_{nn},\\
    h_{\theta\theta}^{\rm IRG}&=\Sigma^2\mc{M}_+,\\
    h_{\phi\phi}^{\rm IRG}&=-\sin^2{\theta}\br{\pa{r^2+a^2}^2\mc{M}_++2a\pa{r^2+a^2}\sin{\theta}\,\mc{N}_--a^2\sin^2{\theta}\,h_{nn}},\\
    h_{tr}^{\rm IRG}&=-\frac{\Sigma}{\Delta}\br{h_{nn}-a\sin{\theta}\,\mc{N}_-},\\
    h_{t\theta}^{\rm IRG}&=\Sigma\br{\mc{N}_+-a\sin{\theta}\,\mc{M}_-},\\
    h_{t\phi}^{\rm IRG}&=a\sin^2{\theta}\br{\pa{r^2+a^2}\mc{M}_++\pa{\frac{\Sigma}{a\sin{\theta}}+2a\sin{\theta}}\mc{N}_--h_{nn}},\\
    h_{r\theta}^{\rm IRG}&=-\frac{\Sigma^2}{\Delta}\mc{N}_+,\\
    h_{r\phi}^{\rm IRG}&=-\frac{\Sigma\sin{\theta}}{\Delta}\br{\pa{r^2+a^2}\mc{N}_--a\sin{\theta}\,h_{nn}},\\
    h_{\theta\phi}^{\rm IRG}&=\Sigma\sin{\theta}\br{\pa{r^2+a^2}\mc{M}_--a\sin{\theta}\,\mc{N}_+},
\end{align}
\end{subequations}
where we introduced the manifestly real projections $\mc{M}_+\equiv\re\pa{\zeta^{-2}h_{mm}}$, $\mc{M}_-\equiv\im\pa{\zeta^{-2}h_{mm}}$, $\mc{N}_+\equiv\sqrt{2}\re\pa{\zeta^{-1}h_{nm}}$, and $\mc{N}_-\equiv\sqrt{2}\im\pa{\zeta^{-1}h_{nm}}$: more explicitly, recalling Eq.~\eqref{eq:TetradProjection},
\begin{subequations}
\label{eq:RealProjections}
\begin{gather}
    \mc{M}_+=\frac{1}{2}\pa{\frac{h_{mm}}{\zeta^2}+\frac{h_{\ol{m}\ol{m}}}{\ol{\zeta}^2}},\qquad
    \mc{M}_-=\frac{1}{2i}\pa{\frac{h_{mm}}{\zeta^2}-\frac{h_{\ol{m}\ol{m}}}{\ol{\zeta}^2}},\\
    \mc{N}_+=\frac{1}{\sqrt{2}}\pa{\frac{h_{nm}}{\zeta}+\frac{h_{n\ol{m}}}{\ol{\zeta}}},\qquad
    \mc{N}_-=\frac{1}{\sqrt{2}i}\pa{\frac{h_{nm}}{\zeta}-\frac{h_{n\ol{m}}}{\ol{\zeta}}}.
\end{gather}
\end{subequations}
The projections $h_{nn}$, $h_{nm}$, $h_{n\ol{m}}$, $h_{mm}$, and $h_{\ol{m}\ol{m}}$ (the only ones needed in IRG) decompose as
\begin{align}
    \label{eq:TetradComponentsIRG}
    h_{ab}=e^{-i\omega t+im\phi}h_{ab}^{(+)}(r,\theta)+e^{i\ol{\omega}t-im\phi}h_{ab}^{(-)}(r,\theta),
\end{align}
where each $h_{ab}^{(\pm)}(r,\theta)$ is expressible in terms of a single function $H_{\omega\ell m}^{ab}$.

In fact, only three such functions are needed, since
\begin{subequations}
\label{eq:TetradProjectionsIRG}
\begin{align}
    h_{nn}^{(+)}&=A^{\rm IRG}H_{\omega\ell m}^{nn}+\ol{B}^{\rm IRG}\ol{H}_{-\ol{\omega},\ell,-m}^{nn},\\
    h_{nn}^{(-)}&=\ol{h}_{nn}^{(+)},\\
    h_{nm}^{(+)}&=\ol{B}^{\rm IRG}\ol{H}_{-\ol{\omega},\ell,-m}^{nm},\\
    h_{nm}^{(-)}&=\ol{A}^{\rm IRG}\ol{H}_{\omega\ell m}^{nm},\\
    h_{n\ol{m}}^{(\pm)}&=\ol{h}_{nm}^{(\mp)},\\
    h_{mm}^{(+)}&=\ol{B}^{\rm IRG}\ol{H}_{-\ol{\omega},\ell,-m}^{mm},\\
    h_{mm}^{(-)}&=\ol{A}^{\rm IRG}\ol{H}_{\omega\ell m}^{mm},\\
    h_{\ol{m}\ol{m}}^{(\pm)}&=\ol{h}_{mm}^{(\mp)}.
\end{align}
\end{subequations}
Specifying the three functions $H_{\omega\ell m}^{nn}$, $H_{\omega\ell m}^{nm}$, and $H_{\omega\ell m}^{mm}$ finally determines the metric components:%
\begin{subequations}
\label{eq:HIRG}
\begin{align}
    H_{\omega\ell m}^{nn}&\equiv-\frac{\epsilon_g}{4\ol{\zeta}^2}\pa{\msc{L}_1^\dag-\frac{2ia\sin{\theta}}{\zeta}}\msc{L}_2^\dag \hat{R}_{\omega\ell m}^{(-2)}\hat{S}_{\omega\ell m}^{(-2)},\\
    H_{\omega\ell m}^{nm}&\equiv-\frac{\epsilon_g}{2\sqrt{2}\ol{\zeta}}\pa{\msc{D}_0\msc{L}^\dag_2+\frac{a^2\sin{2\theta}}{\Sigma}\msc{D}_0-\frac{2r}{\Sigma}\msc{L}^\dag_2}\hat{R}_{\omega\ell m}^{(-2)}\hat{S}_{\omega\ell m}^{(-2)},\\
    H_{\omega\ell m}^{mm}&\equiv-\frac{\epsilon_g}{2}\pa{\msc{D}_0-\frac{2}{\zeta}}\msc{D}_0\hat{R}_{\omega\ell m}^{(-2)}\hat{S}_{\omega\ell m}^{(-2)}.
\end{align}
\end{subequations}
We give fully explicit forms of these functions---with and without mode derivatives---in App.~\ref{app:NoDerivatives}.

\subsection{Metric reconstruction in outgoing radiation gauge}
\label{subsec:MetricReconstructionORG}

Here, we explicitly reconstruct the metric perturbation $h_{\mu\nu}$ in outgoing radiation gauge \eqref{eq:ORG} that is associated with a single mode \eqref{eq:Mode0} of $\psi_0$ or with a single mode \eqref{eq:Mode4} of $\zeta^4\psi_4$.

To treat both cases simultaneously, we express the components of $h_{\mu\nu}^{\rm ORG}$ in terms of constants $A^{\rm ORG}$ and $B^{\rm ORG}$.
These must take different values according to whether one reconstructs the metric perturbation from a single mode \eqref{eq:Mode0} of $\psi_0$, in which case one must set
\begin{align}
\label{eq:ORGMode0Solution}
    A^{\rm ORG}=\frac{192i\omega M}{\mc{C}_{\omega\ell m}},\qquad
    B^{\rm ORG}=\frac{16\ol{\hat{D}}_{\omega\ell m}}{\ol{\mc{C}}_{\omega\ell m}},
\end{align}
or from a single mode \eqref{eq:Mode4} of $\zeta^4\psi_4$, in which case one must set
\begin{align}
\label{eq:ORGMode4Solution}
    A^{\rm ORG}=0,\qquad
    B^{\rm ORG}=\frac{64}{\ol{\hat{\msc{C}}_{}'}_{\!\!\!\omega\ell m}}.
\end{align}
The Boyer-Lindquist components of the real metric perturbation $h_{\mu\nu}^{\rm ORG}$ are then given by\footref{fn:Trivial}
\begin{subequations}
\label{eq:MetricComponentsORG}
\begin{align}
    h_{tt}^{\rm ORG}&=-a^2\sin^2{\theta}\,\mc{M}_+-\frac{a\Delta\sin{\theta}}{\Sigma}\mc{L}_-+\frac{\Delta^2}{4\Sigma^2}h_{ll},\\
    h_{rr}^{\rm ORG}&=\frac{h_{ll}}{4},\\
    h_{\theta\theta}^{\rm ORG}&=\Sigma^2\mc{M}_+,\\
    h_{\phi\phi}^{\rm ORG}&=-\sin^2{\theta}\br{\pa{r^2+a^2}^2\mc{M}_++\frac{a\Delta\pa{r^2+a^2}\sin{\theta}}{\Sigma}\mc{L}_--\frac{a^2\Delta^2\sin^2{\theta}}{4\Sigma^2}h_{ll}},\\
    h_{tr}^{\rm ORG}&=\frac{\Delta}{4\Sigma}h_{ll}-\frac{a\sin{\theta}}{2}\mc{L}_-,\\
    h_{t\theta}^{\rm ORG}&=\frac{\Delta}{2}\mc{L}_+-a\Sigma\sin{\theta}\,\mc{M}_-,\\
    h_{t\phi}^{\rm ORG}&=a\sin^2{\theta}\br{\pa{r^2+a^2}\mc{M}_++\frac{\Delta}{2\Sigma}\pa{\frac{\Sigma}{a\sin{\theta}}+2a\sin{\theta}}\mc{L}_--\frac{\Delta^2}{4\Sigma^2}h_{ll}},\\
    h_{r\theta}^{\rm ORG}&=\frac{\Sigma}{2}\mc{L}_+,\\
    h_{r\phi}^{\rm ORG}&=\frac{\sin{\theta}}{2}\br{\pa{r^2+a^2}\mc{L}_--\frac{a\Delta\sin{\theta}}{2\Sigma}h_{ll}},\\
    h_{\theta\phi}^{\rm ORG}&=\Sigma\sin{\theta}\br{\pa{r^2+a^2}\mc{M}_--\frac{a\Delta\sin{\theta}}{2\Sigma}\mc{L}_+},
\end{align}
\end{subequations}
where the manifestly real projections $\mc{M}_\pm$ were defined in Eq.~\eqref{eq:RealProjections}, and we also introduced
\begin{gather}
    \label{eq:RealProjectionsBis}
    \mc{L}_+\equiv\sqrt{2}\re\frac{h_{lm}}{\zeta}
    =\frac{1}{\sqrt{2}}\pa{\frac{h_{lm}}{\zeta}+\frac{h_{l\ol{m}}}{\ol{\zeta}}},\qquad
    \mc{L}_-\equiv\sqrt{2}\im\frac{h_{lm}}{\zeta}
    =\frac{1}{\sqrt{2}i}\pa{\frac{h_{lm}}{\zeta}-\frac{h_{l\ol{m}}}{\ol{\zeta}}}.
\end{gather}
The projections $h_{ll}$, $h_{lm}$, $h_{l\ol{m}}$, $h_{mm}$, and $h_{\ol{m}\ol{m}}$ (the only ones needed in ORG) decompose as
\begin{align}
    \label{eq:TetradComponentsORG}
    h_{ab}=e^{-i\omega t+im\phi}h_{ab}^{(+)}(r,\theta)+e^{i\ol{\omega}t-im\phi}h_{ab}^{(-)}(r,\theta),
\end{align}
where once again each $h_{ab}^{(\pm)}(r,\theta)$ is expressible in terms of a single function $H_{\omega\ell m}^{ab}$:
\begin{subequations}
\label{eq:TetradProjectionsORG}
\begin{align}
    h_{ll}^{(+)}&=A^{\rm ORG}H_{\omega\ell m}^{ll}+\ol{B}^{\rm ORG}\ol{H}_{-\ol{\omega},\ell,-m}^{ll},\\
    h_{ll}^{(-)}&=\ol{h}_{ll}^{(+)},\\
    h_{lm}^{(+)}&=A^{\rm ORG}H_{\omega\ell m}^{lm},\\
    h_{lm}^{(-)}&=B^{\rm ORG}H_{-\ol{\omega},\ell,-m}^{lm},\\
    h_{l\ol{m}}^{(\pm)}&=\ol{h}_{lm}^{(\mp)},\\
    h_{mm}^{(+)}&=A^{\rm ORG}H_{\omega\ell m}^{mm},\\
    h_{mm}^{(-)}&=B^{\rm ORG}H_{-\ol{\omega},\ell,-m}^{mm},\\
    h_{\ol{m}\ol{m}}^{(\pm)}&=\ol{h}_{mm}^{(\mp)}.
\end{align}
\end{subequations}
Specifying the three functions $H_{\omega\ell m}^{nn}$, $H_{\omega\ell m}^{nm}$, and $H_{\omega\ell m}^{mm}$ finally determines the metric components:%
\begin{subequations}
\label{eq:HORG}
\begin{align}
    H_{\omega\ell m}^{ll}&\equiv-\frac{\epsilon_g}{4}\zeta^2\pa{\msc{L}_1-\frac{2ia\sin{\theta}}{\zeta}} \msc{L}_2 \hat{R}_{\omega\ell m}^{(+2)}\hat{S}_{\omega\ell m}^{(+2)},\\
    H_{\omega\ell m}^{lm}
    &\equiv\frac{\epsilon_g}{4\sqrt{2}}\frac{\zeta^2}{\ol{\zeta}\Delta}\pa{\msc{D}_0^\dag\msc{L}_2+\frac{a^2\sin{2\theta}}{\Sigma}\msc{D}_0^\dag-\frac{2r}{\Sigma}\msc{L}_2}\Delta^2\hat{R}_{\omega\ell m}^{(+2)}\hat{S}_{\omega\ell m}^{(+2)},\\
    H_{\omega\ell m}^{mm}
    &\equiv-\frac{\epsilon_g}{8}\frac{\zeta^2}{\ol{\zeta}^2}\pa{\msc{D}_0^\dag-\frac{2}{\zeta}} \msc{D}_0^\dag \Delta^2 \hat{R}_{\omega\ell m}^{(+2)}\hat{S}_{\omega\ell m}^{(+2)}.
\end{align}
\end{subequations}
We give fully explicit forms of these functions---with and without mode derivatives---in App.~\ref{app:NoDerivatives}.

\subsection{Consistency checks}
\label{subsec:ConsistencyChecks}

We have reconstructed the metric perturbation $h_{\mu\nu}$ corresponding to the modes of a given Weyl scalar $\psi_0$ or $\psi_4$, in either ingoing radiation gauge (in Sec.~\ref{subsec:MetricReconstructionIRG}) or outgoing radiation gauge (in Sec.~\ref{subsec:MetricReconstructionORG}).
The derivation of the metric components relies on the use of the Newman--Penrose and Geroch--Held--Penrose formalisms, which we review in Sec.~\ref{subsec:Review} below, before applying them to the Kerr metric \eqref{eq:Kerr} in Sec.~\ref{subsec:Kerr}.
However, given the results of this section, it is possible to check them directly without using these formalisms, as we now briefly explain.

First, one can directly check that the metric perturbations $h_{\mu\nu}^{\rm IRG}$ and $h_{\mu\nu}^{\rm ORG}$ given in Eqs.~\eqref{eq:MetricComponentsIRG} and \eqref{eq:MetricComponentsORG} really do solve the vacuum linearized Einstein equations: one can plug them into Eq.~\eqref{eq:LinearizedEE}---or equivalently, the simpler version \eqref{eq:TracelessEE} for traceless perturbations---and verify that the equations are satisfied.
Then, to ``close the loop,'' one can also check that the reconstructed metric perturbations $h_{\mu\nu}^{\rm IRG/ORG}$ really do reproduce the desired Weyl scalars: one can plug them into Eq.~\eqref{eq:WeylScalars}---using the formula \eqref{eq:LinearizedWeyl} for the linearized Weyl tensor $C_{\mu\nu\rho\sigma}^{(1)}$---and verify that this recovers the expected modes of $\psi_0$ and $\psi_4$, and that moreover these scalars are related by the equations of Sec.~\ref{subsec:WeylRelations}.
As such, the results presented in this section can stand alone.

\subsection{Different forms of the Teukolsky--Starobinsky identities}
\label{subsec:TeukolskyStarobinskyForms}

Finally, for completeness, we present other forms of the Teukolsky--Starobinsky identities \eqref{eq:TeukolskyStarobinskyIdentities}, and then briefly discuss their (rather subtle) logical interrelations.

For the sake of brevity, we will give these various identities abbreviated names.
For instance, we will refer to the first form of the Teukolsky--Starobinsky identities \eqref{eq:TeukolskyStarobinskyIdentities} as the TSI$_1$.
These can be derived from Eqs.~\eqref{eq:IngoingPotential} and \eqref{eq:OutgoingPotential} by eliminating the Hertz potentials $\Psi_{\rm H}$ and $\Psi_{\rm H}'$.
Eliminating the Weyl scalars $\psi_0$ and $\psi_4$ instead yields the TSI$_1$ for the Hertz potentials,
\begin{subequations}
\begin{align}
    l^4\Psi_{\rm H}&=\frac{1}{4}\mc{L}_{-1}\mc{L}_0\mc{L}_1\mc{L}_2\frac{\Psi_{\rm H}'}{\zeta^4}+3M\pd_t\frac{\ol{\Psi}_{\rm H}'}{\ol{\zeta}^4},\\
    \Delta^2\pa{\frac{\Sigma}{\Delta}n}^4\Delta^2\frac{\Psi_{\rm H}'}{\zeta^4}&=\frac{1}{4}\ol{\mc{L}}_{-1}\ol{\mc{L}}_0\ol{\mc{L}}_1\ol{\mc{L}}_2\Psi_{\rm H}-3M\pd_t\ol{\Psi}_{\rm H}.
\end{align}
\end{subequations}
One can also plug the TSI$_1$ into each other to obtain eighth-order differential relations involving only one Weyl scalar.
These are the Teukolsky--Starobinsky identities in second form: the TSI$_2$,%
\begin{subequations}
\label{eq:TSI2}
\begin{align}
    \Delta^2\pa{\frac{\Sigma}{\Delta}n}^4\Delta^2l^4\zeta^4\psi_4&=\frac{1}{16}\mc{L}_{-1}\mc{L}_0\mc{L}_1\mc{L}_2\ol{\mc{L}}_{-1}\ol{\mc{L}}_0\ol{\mc{L}}_1\ol{\mc{L}}_2\zeta^4\psi_4-9M^2\pd_t^2\zeta^4\psi_4,\\
    l^4\Delta^2\pa{\frac{\Sigma}{\Delta}n}^4\Delta^2\psi_0&=\frac{1}{16}\ol{\mc{L}}_{-1}\ol{\mc{L}}_0\ol{\mc{L}}_1\ol{\mc{L}}_2\mc{L}_{-1}\mc{L}_0\mc{L}_1\mc{L}_2\psi_0-9M^2\pd_t^2\psi_0,
\end{align}
\end{subequations}
Analogous relations for the Hertz potentials are obtained by replacing $(\zeta^4\psi_0,\zeta^4\psi_4)\to(\Psi_{\rm H}',\Psi_{\rm H})$.

We now wish to expound upon the subtle logical interrelations between the TSI$_1$, TSI$_2$, and their angular and radial counterparts from Sec.~\ref{subsec:TeukolskyStarobinsky}, namely: the angular Teukolsky--Starobinsky identities in first (ATSI$_1$) form \eqref{eq:ATSI1} and second (ATSI$_2$) form \eqref{eq:ATSI2}, and the radial Teukolsky--Starobinsky identities in first (RTSI$_1$) form \eqref{eq:RTSI1} and second (RTSI$_2$) form \eqref{eq:RTSI2}.
In a nutshell:
\begin{enumerate}
    \item The TSI$_1$/ATSI$_1$/RTSI$_1$ imply the TSI$_2$/ATSI$_2$/RTSI$_2$, resp., but the converse is not true.
    \item $\text{ATSI}_2+\text{RTSI}_2=\text{TSI}_2$, but the joint ATSI$_1$ and RTSI$_1$ are not equivalent to the TSI$_1$.
\end{enumerate}
Let us explain the first claim.
First, we have just seen that the TSI$_1$, in which $\psi_0$ and $\psi_4$ appear coupled, imply the TSI$_2$, in which these scalars are decoupled.
The TSI$_1$ encode how the Weyl scalars $\psi_0$ and $\psi_4$ associated with a single perturbation are related: in other words, the TSI$_1$ are not satisfied by any pair of $(\psi_0,\psi_4)$ that solve the Teukolsky master equation \eqref{eq:TME}, but only by those specific pairs that arise from the same metric perturbation $h_{\mu\nu}$.
By contrast, the TSI$_2$ have lost this information: in decoupling the two Weyl scalars, one is left with equations that are satisfied by any pair of $(\psi_0,\psi_4)$ that solve the Teukolsky master equation \eqref{eq:TME}, regardless of whether or not they arise from the same metric perturbation $h_{\mu\nu}$.
Mathematically, recovering this information would amount to refactoring the eighth-order TSI$_2$ into the fourth-order TSI$_1$; one can think of the extra derivatives in the TSI$_2$ as having erased this information.

Similarly, we saw in Sec.~\ref{subsec:TeukolskyStarobinsky} that the ATSI$_1$ imply the ATSI$_2$, and likewise that the RTSI$_1$ imply the RTSI$_2$.
However, as with the TSI, going from first to second form involves the taking of additional derivatives that erase information.
In this case, it is the specific choice of hatted modes $\hat{S}_{\omega\ell m}^{(\pm2)}(\theta)$ and $\hat{R}_{\omega\ell m}^{(\pm2)}(r)$ that is forgotten: multiplying the hatted constants $(\hat{D}_{\omega\ell m},\hat{D}_{\omega\ell m}')$ and $(\hat{\msc{C}}_{\omega\ell m},\hat{\msc{C}}_{\omega\ell m}')$ to obtain their unhatted products $\mc{D}_{\omega\ell m}$ and $\mc{C}_{\omega\ell m}$ is an operation that cannot be undone, since the refactorization is mode-dependent.
Said differently, the ATSI$_1$ and RTSI$_1$ only hold for specific modes, and for this reason cannot be recovered from the ATSI$_2$ and RTSI$_2$, which are satisfied by any solution to their respective ODEs \eqref{eq:RadialODE} and \eqref{eq:AngularODE}.

Now, we turn to the second claim.
The ATSI$_1$ and RTSI$_1$ are obeyed by mode solutions to the ODEs \eqref{eq:RadialODE} and \eqref{eq:AngularODE}, which are obtained by separating the decoupled Teukolsky master equation \eqref{eq:TME} for a \textit{single} scalar $\Psi^{(s)}$.
As such, they contain no coupling information to relate the Weyl scalars $\psi_0$ and $\psi_4$ associated with a given metric perturbation $h_{\mu\nu}$, which is precisely the essence of the TSI$_1$.
Hence, the TSI$_1$ knows more than the ATSI$_1$ and RTSI$_1$ put together, and cannot be derived from them.
As expected, plugging the mode ansatz \eqref{eq:SingleMode} into the TSI$_1$, one can separate them into the ATSI$_1$ and RTSI$_1$, but trying to go backwards results in an incomplete form of the TSI$_1$ with three undetermined parameters (we omit the details here).

As an aside, we note that the TSI$_1$ were first derived by Torres del Castillo \cite{delCastillo1994} using the ATSI$_1$ and RTSI$_1$ (rather than in the more modern way, which uses the Geroch-Held-Penrose formalism \cite{PriceThesis}).
Crucially, however, he had to plug them into Eqs.~\eqref{eq:IngoingPotential} [his Eqs.~(25)], which do know about the coupling between $\psi_0$ and $\psi_4$.
The first modern derivation of the TSI$_1$ was given by Silva-Ortigoza \cite{Silva-Ortigoza1997}, who emphasized the key role they play in coupling $\psi_0$ and $\psi_4$.

Finally, the ATSI$_2$ and RTSI$_2$ are together equivalent to the TSI$_2$, which should not be too surprising, since neither carries coupling information on $\psi_0$ and $\psi_4$.
Indeed, inserting the single-mode ansatz \eqref{eq:SingleMode} for both $\psi_0$ and $\psi_4$ into the TSI$_2$ results in
\begin{subequations}
\begin{align}
    \frac{\Delta^2\pa{\msc{D}_0^\dag}^4\Delta^2\msc{D}_0^4\Delta^2R_{\omega\ell m}^{(-2)}}{R_{\omega\ell m}^{(-2)}}
    &=\frac{\msc{L}_{-1}\msc{L}_0\msc{L}_1\msc{L}_2\msc{L}_{-1}^\dag\msc{L}_0^\dag\msc{L}_1^\dag\msc{L}_2^\dag S_{\omega\ell m}^{(-2)}}{S_{\omega\ell m}^{(-2)}}+(12\omega M)^2,\\
    \frac{\msc{D}_0^4\Delta^2\pa{\msc{D}_0^\dag}^4\Delta^2R_{\omega\ell m}^{(+2)}}{R_{\omega\ell m}^{(+2)}}&=\frac{\msc{L}_{-1}^\dag\msc{L}_0^\dag\msc{L}_1^\dag\msc{L}_2^\dag\msc{L}_{-1}\msc{L}_0\msc{L}_1\msc{L}_2S_{\omega\ell m}^{(+2)}}{S_{\omega\ell m}^{(+2)}}+(12\omega M)^2.
\end{align}
\end{subequations}
Since the last terms on each line are constant, the fractions must be constants as well, which is the content of the ATSI$_2$ and RTSI$_2$.
Hence, the TSI$_2$ imply the ATSI$_2$ and RTSI$_2$.
To go in the other direction, one can invoke the ATSI$_2$ and RTSI$_2$ to write the above identities---this crucially requires the use of the relation $\mc{C}_{\omega\ell m}=\mc{D}_{\omega\ell m}+(12\omega M)^2$, given in Eq.~\eqref{eq:RadialConstant}, which can be checked directly from the explicit forms of $\mc{C}_{\omega\ell m}$ and $\mc{D}_{\omega\ell m}$.
Thus, one can establish that the TSI$_2$ holds at the mode level, and hence (by linearity) also holds in full generality.

\section{Derivation of results}
\label{sec:Derivation}

In this section, we derive the identities quoted in Sec.~\ref{sec:Results}.
Throughout, we assume familiarity with the formalisms developed by Newman and Penrose \cite{Newman1962}, and by Geroch, Held, and Penrose \cite{Geroch1973}.
Whiting and Price \cite{Whiting2005} provide a succinct summary of this technology---we closely follow the treatment presented in Price's encyclopedic thesis \cite{PriceThesis}, whose main results we review in Sec.~\ref{subsec:Review}.
In particular, we write down the metric reconstruction operators ${\mc{S}_0^\dag}_{\mu\nu}$ and ${\mc{S}_4^\dag}_{\mu\nu}$, together with the relations between the Weyl scalars $(\psi_0,\psi_4)$ and the Hertz potentials $(\Psi_{\rm H},\Psi_{\rm H}')$, in a class of algebraically special spacetimes with arbitrary sign $\epsilon_g$.
In Sec.~\ref{subsec:Kerr}, we then apply these general formulas to the special case of a Kerr black hole to recover the formulas of the previous sections.
Finally, in Sec.~\ref{subsec:ModeInversion}, we derive the mode inversion formula that leads to the relations in Sec.~\ref{subsec:WeylRelations}, and in Sec.~\ref{subsec:KerrMetricReconstruction} we sketch how to obtain the metric reconstruction formulas in Secs.~\ref{subsec:MetricReconstructionIRG} and \ref{subsec:MetricReconstructionORG}.

\subsection{Review of black hole perturbation theory}
\label{subsec:Review}

Here, we briefly review key results from Price's thesis \cite{PriceThesis}, extending them when necessary to accommodate an arbitrary choice of sign $\epsilon_g$ (since Price uses the ``mostly minus'' convention $\epsilon_g=-1$, we need to modify several of his equations).
We can now explain why we included the sign $\epsilon_g$ in our normalization of the tetrad \eqref{eq:Tetrad} and in our definition of the Weyl scalars \eqref{eq:WeylScalars}: this was done to ensure that the majority of equations in Price's thesis remain unchanged for either sign convention.
We also introduced some factors of two to match other sources \cite{Pound2022}.
 
We have checked that all of the equations in Chapters 2, 3, 5, and in App.~A of Price's thesis \cite{PriceThesis} continue to hold for arbitrary sign $\epsilon_g$ provided that one makes the following modifications:\footnote{We have also checked all of the equations in his App.~B, but only for the Kerr background specifically.}
\begin{itemize}
    \item In his Eq.~(2-1), the $1$ on the right-hand side becomes $-\epsilon_g$, consistent with our Eq.~\eqref{eq:Tetrad}.
    (There is also a small typo, as his second inner product should be between $m$ and $\ol{m}$.)
    Likewise, his expression for the metric as an outer product of the tetrad vectors in Eq.~(2-3) must gain a factor of $-\epsilon_g$ on its right-hand side.
    \item The Newman--Penrose spin coefficients in his Eqs.~(2-6) and (2-7) all gain a factor of $-\epsilon_g$, so we have, for example,
    \begin{align}
        \label{eq:SpinCoefficients}
        \rho\equiv-\epsilon_g\ol{m}^\mu m^\nu\nabla_\mu l_\nu,\qquad
        \text{etc.}
    \end{align}
    \item The five Weyl scalars in his Eq.~(2-11) gain a factor of $-\epsilon_g$, so we have, for example,
    \begin{align}
        \Psi_0\equiv\epsilon_gC_{\mu\nu\rho\sigma}l^\mu m^\nu l^\rho m^\sigma,\qquad
        \text{etc.}
    \end{align}
    (There is also a typo in his $\psi_2$, as we explain below.)
    As with the tetrad, we add these factors to the definitions of the spin coefficients and Weyl scalars to ensure that no changes are needed in the equations involving only these quantities, such as those in Price's App.~A.
    \item Price does not label the operators we call $\mc{S}_0$ and $\mc{S}_4$, but he gives them implicitly in his Eqs.~(3-23) and (3-24).
    Our operators differ from his by a factor of $\frac{1}{2}$.\footnote{We never write them explicitly in this work, as we only need their adjoints in Eqs.~\eqref{eq:GHP0} and \eqref{eq:GHP4} below.}
    This rescaling cancels out of $h_{\mu\nu}$ due to the additional factor of $2$ in our Eqs.~\eqref{eq:MetricReconstructionIRG} and \eqref{eq:MetricReconstructionORG}.
    We made this choice to match the conventions of Pound and Wardell \cite{Pound2022}, who use the ``mostly plus'' convention $\epsilon_g=1$, and also to ensure that the source term in every equation comes with a factor of $8\pi$, as in Einstein's field equations.
    Thus, with our definitions, the factors of $4\pi$ on the right-hand sides of Price's Eqs.~(3-25), (3-26), and (3-29) become factors of $8\pi$.
    \item Due to the factor of $-\epsilon_g$ in our definition \eqref{eq:WeylScalars} of the Weyl scalars, Price's Eqs.~(3-27) and (3-28) for $\psi_0$ and $\psi_4$ in terms of the metric perturbation $h_{\mu\nu}$ must gain a factor of $-\epsilon_g$.
    \item The normalization of the Hertz potentials is not fixed \textit{a priori} but is set by their relation to the metric components $h_{\mu\nu}$.
    In our Eqs.~\eqref{eq:MetricReconstructionIRG} and \eqref{eq:MetricReconstructionORG}, we included a factor of $2$ (to match Pound and Wardell's formulas, as aforementioned) as well as a factor of $\epsilon_g$, which we inserted so that the relations \eqref{eq:IngoingPotential} and \eqref{eq:OutgoingPotential} between our Weyl scalars and our Hertz potentials carry no factors of $\epsilon_g$.
    Price chose a different normalization, so his Eq.~(3-35) must have its right-hand side multiplied by $-\frac{1}{2}\epsilon_g$ in order to recover ours.
    \item As a consequence of this different normalization for the Hertz potentials, all the factors of $\frac{1}{2}$ in his Eqs.~(5-1), (5-2), (5-3), (5-12), and (5-13) become factors of $\frac{1}{4}$.
\end{itemize}
We also correct a few minor typos in Price's thesis \cite{PriceThesis}:
\begin{itemize}
    \item In his Eq.~(2-11), the second term in his expression for $\psi_2$ (which we will denote $\Psi_2$) should have a minus sign (or, alternatively, $m$ and $\ol{m}$ should be swapped).
    By the symmetries of the Weyl tensor (see, e.g., Chapter of 8 of \cite{Chandrasekhar1983}), this corrected two-term expression matches the single-term one given by Chandrasekhar \cite{Chandrasekhar1983}, or Pound and Wardell \cite{Pound2022}:
    \begin{align}
        \Psi_2\equiv\epsilon_gC_{\mu\nu\rho\sigma}l^\mu m^\nu \ol{m}^\rho n^\sigma,
    \end{align}
    where we have also multiplied by $-\epsilon_g$, as explained above.
    \item In his Eq.~(2.12), every definition is off by a sign.
    (This has no effect on any calculations involving a vacuum spacetime.)
    \item On the right-hand sides of his Eqs.~(5-14) and (5-15), the second terms have wrong signs.
    \item  Finally, in his Eq.~(A-7), the $\tau'$ in the first term on the right-hand side should be $\ol{\tau}'$.
\end{itemize}

We can now quote the results that we need, in a form that applies to any non-accelerating Petrov type D spacetime, and using the formalism of Geroch, Held, and Penrose (GHP) \cite{Geroch1973}.
The operators ${\mc{S}_0^\dag}_{\mu\nu}$ and ${\mc{S}_4^\dag}_{\mu\nu}$, which we have invoked in Eqs.~\eqref{eq:MetricReconstructionIRG} and \eqref{eq:MetricReconstructionORG}, are
\begin{subequations}
\begin{align}
    \label{eq:GHP0}
    {\mc{S}_0^\dag}_{\mu\nu}\Psi&=-\frac{1}{2}l_\mu l_\nu\pa{\eth-\tau}\pa{\eth+3\tau}\Psi-\frac{1}{2}m_\mu m_\nu\pa{\tho-\rho}\pa{\tho+3\rho}\Psi\notag\\
    &\phantom{=}\ +\frac{1}{2}l_{(\mu}m_{\nu)}\br{\pa{\tho-\rho+\ol{\rho}}\pa{\eth+3\tau}+\pa{\eth-\tau+\ol{\tau}'}\pa{\tho+3\rho}}\Psi,\\
    \label{eq:GHP4}
    {\mc{S}_4^\dag}_{\mu\nu}\Psi&=-\frac{1}{2}n_\mu n_\nu\pa{\eth'-\tau'}\pa{\eth'+3\tau'}\Psi-\frac{1}{2}\ol{m}_\mu\ol{m}_\nu\pa{\tho'-\rho'}\pa{\tho'+3\rho'}\Psi\notag\\
    &\phantom{=}\ +\frac{1}{2}n_{(\mu}\ol{m}_{\nu)}\br{\pa{\tho'-\rho'+\ol{\rho}'}\pa{\eth'+3\tau'}+\pa{\eth'-\tau'+\ol{\tau}}\pa{\tho'+3\rho'}}\Psi.
\end{align}
\end{subequations}
The fourth-order relations between the Weyl scalars and the IRG Hertz potential are
\begin{subequations}
\label{eq:WeylFromHertzIRG}
\begin{align}
    \psi_0&=\frac{1}{4}\tho^4\ol{\Psi}_{\rm H},\\
    \psi_4&=\frac{1}{4}\pa{{\eth'}^4\ol{\Psi}_{\rm H}-3\Psi_2^{4/3}\mc{V}\Psi_{\rm H}},
\end{align}
\end{subequations}
while the fourth-order relations between the Weyl scalars and the ORG Hertz potential are
\begin{subequations}
\label{eq:WeylFromHertzORG}
\begin{align}
    \psi_0&=\frac{1}{4}\pa{\eth^4\ol{\Psi}_{\rm H}'+3\Psi_2^{4/3}\mc{V}\Psi_{\rm H}'},\\
    \psi_4&=\frac{1}{4}{\tho'}^4\ol{\Psi}_{\rm H}'.
\end{align}
\end{subequations}
Finally, the first form of the Teukolsky--Starobinsky identities is
\begin{subequations}
\label{eq:TeukolskyStarobinskyGHP}
\begin{align}
    \tho^4\Psi_2^{-4/3}\psi_4 
    &= {\eth'}^4\Psi_2^{-4/3}\psi_0-3\mc{V} \ol{\psi}_0,\\
    {\tho'}^4\Psi_2^{-4/3}\psi_0 
    & =\eth^4\Psi_2^{-4/3}\psi_4+3\mc{V} \ol{\psi}_4.
\end{align}
\end{subequations}

\subsection{Application to the Kerr black hole}
\label{subsec:Kerr}

The formulas in the previous section are applicable to any spacetime of Petrov type D that is also non-accelerating (or equivalently, that admits a rank-2 Killing tensor).
Now, we specialize them to the particular case of the Kerr metric \eqref{eq:Kerr}, which falls within that class.

With the factors of $\epsilon_g$ in our definitions \eqref{eq:SpinCoefficients}, the spin coefficients take their usual form:
\begin{subequations}
\begin{gather} 
    \rho=-\frac{1}{\zeta},\qquad 
    \rho'=\frac{\Delta}{2\zeta\Sigma},\qquad
    \tau=-\frac{ia\sin{\theta}}{\sqrt{2}\Sigma},\qquad
    \tau'=-\frac{ia\sin{\theta}}{\sqrt{2}\zeta^2},\\
    \beta=\frac{\cot{\theta}}{2\sqrt{2}\,\ol{\zeta}},\qquad
    \beta'=\frac{\cot{\theta}-\frac{2ia\sin{\theta}}{\zeta}}{2\sqrt{2}\zeta},\qquad
    \epsilon=0,\qquad
    \epsilon'=\frac{\frac{\Delta}{\zeta}-\pa{r-M}}{2\Sigma},
\end{gather}
\end{subequations}
with all others coefficients zero.
The single, nonzero Weyl scalar of the Kerr background is
\begin{align} 
    \Psi_2=-M\zeta^{-3}.
\end{align}
The operators \eqref{eq:GHP0} and \eqref{eq:GHP4} are only well-defined when acting on quantities $\Psi$ with definite GHP weights.
If ${\mc{S}_0^\dag}_{\mu\nu}$ acts on the IRG Hertz potential $\Psi_{\rm H}$ (or any quantity of weight $\cu{-4,0}$), or if ${\mc{S}_4^\dag}_{\mu\nu}$ acts on the ORG Hertz potential $\Psi_{\rm H}'$ (or any quantity of weight $\cu{4,0}$), then we have
\begin{subequations}
\label{eq:KerrAdjointOperators}
\begin{align}
    \label{eq:AdjointOperator0}
    {\mc{S}_0^\dag}_{\mu\nu}\Psi_{\rm H}&=\Bigg[-\frac{1}{4\ol{\zeta}^2}l_\mu l_\nu\pa{\ol{\mc{L}}_1-\frac{2ia\sin{\theta}}{\zeta}}\ol{\mc{L}}_2-\frac{1}{2}m_\mu m_\nu\pa{l-\frac{2}{\zeta}}l\notag\\
    &\phantom{=}\quad\,
    +\frac{1}{\sqrt{2}\,\ol{\zeta}}l_{(\mu}m_{\nu)}\pa{l\ol{\mc{L}}_2+\frac{a^2\sin{2\theta}}{\Sigma}l-\frac{2r}{\Sigma}\ol{\mc{L}}_2}\Bigg]\Psi_{\rm H},\\
    \label{eq:AdjointOperator4}
    {\mc{S}_4^\dag}_{\mu\nu}\Psi_{\rm H}'&=\zeta^2\Bigg[-\frac{1}{4}n_\mu n_\nu\pa{\mc{L}_1-\frac{2ia\sin{\theta}}{\zeta}}\mc{L}_2-\frac{1}{2\ol{\zeta}^2}\ol{m}_\mu\ol{m}_\nu\pa{\frac{\Sigma}{\Delta}n+\frac{1}{\zeta}}\frac{\Sigma}{\Delta}n\Delta^2\notag\\
    &\phantom{=}\qquad\,+\frac{1}{\sqrt{2}\,\ol{\zeta}\Delta}n_{(\mu}\ol{m}_{\nu)}\pa{\frac{\Sigma}{\Delta}n\mc{L}_2+\frac{a^2\sin{2\theta}}{\Delta}n+\frac{r}{\Sigma}\mc{L}_2}\Delta^2\bigg]\frac{\Psi_{\rm H}'}{\zeta^4}.
\end{align}
\end{subequations}
In the Kerr background, the GHP forms of the relations in Sec.~\ref{subsec:Review} reduce to their forms given in Sec.~\ref{sec:Introduction}: that is, Eqs.~\eqref{eq:WeylFromHertzIRG}, \eqref{eq:WeylFromHertzORG}, and \eqref{eq:TeukolskyStarobinskyGHP}, reduce to Eqs.~\eqref{eq:IngoingPotential}, \eqref{eq:OutgoingPotential}, and \eqref{eq:TeukolskyStarobinskyIdentities}, respectively.

\subsection{Mode inversion}
\label{subsec:ModeInversion}

In this section, we derive the formula given in Sec.~\ref{subsec:WeylRelations} that directly relates modes of the Weyl scalars $\psi_0$ and $\psi_4$ associated with the same metric perturbation $h_{\mu\nu}$.
To do so, we must first carry out the mode inversion procedure: that is, we must invert Eqs.~\eqref{eq:IngoingPotential} and \eqref{eq:OutgoingPotential} to obtain the Hertz potentials $\Psi_{\rm H}$ and $\Psi_{\rm H}'$ that correspond to these modes.

This requires us to consider a linear combination of two modes in the Hertz potentials:
\begin{subequations}
\label{eq:HertzModes}
\begin{align}
    \label{eq:HertzModeIRG}
    \Psi_{\rm H}&=A^{\rm IRG}e^{-i\omega t+im\phi}\hat{R}_{\omega\ell m}^{(-2)}(r)\hat{S}_{\omega\ell m}^{(-2)}(\theta)+B^{\rm IRG}e^{i\ol{\omega}t-im\phi}\hat{R}_{-\ol{\omega},\ell,-m}^{(-2)}(r)\hat{S}^{(-2)}_{-\ol{\omega},\ell,-m}(\theta),\\
    \label{eq:HertzModeORG}
    \zeta^{-4}\Psi_{\rm H}'&=A^{\rm ORG}e^{-i\omega t+im\phi}\hat{R}_{\omega\ell m}^{(+2)}(r)\hat{S}_{\omega\ell m}^{(+2)}(\theta)+B^{\rm ORG}e^{i\ol{\omega}t-im\phi}\hat{R}_{-\ol{\omega},\ell,-m}^{(+2)}(r)\hat{S}_{-\ol{\omega},\ell,-m}^{(+2)}(\theta).
\end{align}
\end{subequations}
Here, $A^{\rm IRG/ORG}$ and $B^{\rm IRG/ORG}$ are yet-to-be-determined constants.
Inserting these expressions into Eqs.~\eqref{eq:IngoingPotential0} and \eqref{eq:OutgoingPotential0}, and then applying the angular and radial Teukolsky--Starobinsky identities \eqref{eq:ATSI1b} and \eqref{eq:RTSI1a} yields 
\begin{subequations}
\begin{align}
    \label{eq:Weyl0ModeIRG}
    \psi_0&=\frac{1}{4}\pa{\hat{\msc{C}}_{\omega\ell m}\ol{B}^{\rm IRG}e^{-i\omega t+im\phi}\hat{R}_{\omega\ell m}^{(+2)}\hat{S}_{\omega\ell m}^{(+2)}+\ol{\hat{\msc{C}}}_{\omega\ell m}\ol{A}^{\rm IRG}e^{i\ol{\omega}t-im\phi}\hat{R}_{-\ol{\omega},\ell,-m}^{(+2)}\hat{S}_{-\ol{\omega},\ell,-m}^{(+2)}},\\
    \label{eq:Weyl0ModeORG}
    \psi_0&=\frac{1}{16}\Big[\pa{\ol{\hat{D}}_{\omega\ell m}\ol{A}^{\rm ORG}+12i\ol{\omega}MB^{\rm ORG}}e^{i\ol{\omega}t-im\phi}\hat{R}_{-\ol{\omega},\ell,-m}^{(+2)}\hat{S}_{-\ol{\omega},\ell,-m}^{(+2)}\notag\\
    &\phantom{=}\qquad+\pa{\hat{D}_{\omega\ell m}'\ol{B}^{\rm ORG}-12i\omega MA^{\rm ORG}}e^{-i\omega t+im\phi}\hat{R}_{\omega\ell m}^{(+2)}\hat{S}_{\omega\ell m}^{(+2)}\Big].
\end{align}
\end{subequations}
Setting these two expressions equal to the single mode \eqref{eq:Mode0} of $\psi_0$ determines the coefficients $A^{\rm IRG/ORG}$ and $B^{\rm IRG/ORG}$ to take the form given in Eqs.~\eqref{eq:IRGMode0Solution} and \eqref{eq:ORGMode0Solution}.
This completes the mode inversion procedure starting from a single mode \eqref{eq:Mode0} of $\psi_0$.
To find its corresponding Weyl scalar $\psi_4$, one can either plug Eq.~\eqref{eq:HertzModeIRG} into Eq.~\eqref{eq:IngoingPotential4}, or else, one can equivalently plug Eq.~\eqref{eq:HertzModeORG} into Eq.~\eqref{eq:OutgoingPotential4}.
Either way, one recovers the direct relation \eqref{eq:CorrespondingMode4}.

Likewise, plugging the two Hertz potentials \eqref{eq:HertzModeIRG} and \eqref{eq:HertzModeORG} into Eqs.~\eqref{eq:IngoingPotential4} and \eqref{eq:OutgoingPotential4}, and applying the angular and radial Teukolsky--Starobinsky identities \eqref{eq:ATSI1a} and \eqref{eq:RTSI1b} yields%
\begin{subequations}
\begin{align}
    \label{eq:Weyl4ModeIRG}
    \zeta^4\psi_4&=\frac{1}{16}\bigg[\pa{\ol{\hat{D}}_{\omega\ell m}'\ol{A}^{\rm IRG}-12i\ol{\omega}MB^{\rm IRG}}e^{i\ol{\omega}t+im\phi}\hat{R}_{-\ol{\omega},\ell,-m}^{(-2)}\hat{S}_{-\ol{\omega},\ell,-m}^{(-2)}\notag\\
    &\phantom{=}\qquad+\pa{\hat{D}_{\omega\ell m}\ol{B}^{\rm IRG}+12i\omega MA^{\rm IRG}}e^{-i\omega t+im\phi}\hat{R}_{\omega\ell m}^{(-2)}\hat{S}_{\omega\ell m}^{(-2)}\bigg],\\
    \label{eq:Weyl4ModeORG}
    \zeta^4\psi_4&=\frac{1}{64}\pa{\hat{\msc{C}}_{\omega\ell m}'\ol{B}^{\rm ORG}e^{-i\omega t+im\phi}\hat{R}_{\omega\ell m}^{(-2)}\hat{S}_{\omega\ell m}^{(-2)}+\ol{\hat{\msc{C}}'_{}}_{\!\!\!\omega\ell m}\ol{A}^{\rm ORG}e^{i\ol{\omega}t-im\phi}\hat{R}_{-\ol{\omega},\ell,-m}^{(-2)}\hat{S}_{-\ol{\omega},\ell,-m}^{(-2)}}.
\end{align}
\end{subequations}
Setting these two expressions equal to the single mode \eqref{eq:Mode4} of $\zeta^4\psi_4$ determines the coefficients $A^{\rm IRG/ORG}$ and $B^{\rm IRG/ORG}$ to take the form given in Eqs.~\eqref{eq:IRGMode4Solution} and \eqref{eq:ORGMode4Solution}.
This completes the mode inversion procedure starting from a single mode \eqref{eq:Mode4} of $\zeta^4\psi_4$.
To find its corresponding Weyl scalar $\psi_0$, one can either plug Eq.~\eqref{eq:HertzModeIRG} into Eq.~\eqref{eq:IngoingPotential0}, or else, one can equivalently plug Eq.~\eqref{eq:HertzModeORG} into Eq.~\eqref{eq:OutgoingPotential0}.
Either way, one recovers the direct relation \eqref{eq:CorrespondingMode0}.

As the short length of this section reveals, the fundamental difficulty of mode inversion lies in obtaining the Teukolsky--Starobinsky identities in their various forms.
Once they are in hand, mode inversion follows easily.

\subsection{Metric reconstruction}
\label{subsec:KerrMetricReconstruction}

Finally, we can put everything together to derive the metric reconstruction formulas presented in Secs.~\ref{subsec:MetricReconstructionIRG} and \ref{subsec:MetricReconstructionORG}.
The metric perturbations $h_{\mu\nu}$ in IRG and ORG are given by Eqs.~\eqref{eq:MetricReconstructionIRG} and \eqref{eq:MetricReconstructionORG}, respectively, with the operators ${\mc{S}_0^\dag}_{\mu\nu}$ and ${\mc{S}_4^\dag}_{\mu\nu}$ taking the forms given in Eqs.~\eqref{eq:KerrAdjointOperators}. 
We apply the operators in this form to the mode expansions \eqref{eq:HertzModes} of the IRG and ORG Hertz potentials.
Since each operator has three terms, we see that the metric components $h_{\mu\nu}$ can all be expressed in terms of three functions, which we called $H_{\omega\ell m}^{ab}(r,\theta)$ in Secs.~\ref{subsec:MetricReconstructionIRG} and \ref{subsec:MetricReconstructionORG}.
The computation of these functions is essentially the main nontrivial step in metric reconstruction.\footref{fn:Trivial}

As an example, we compute $H_{\omega\ell m}^{nm}$ in the ingoing radiation gauge.
This requires computing $h_{nm}^{\rm IRG}$ by first acting with the form \eqref{eq:AdjointOperator0} of ${\mc{S}_0^\dag}_{\mu\nu}$ on the Hertz potential $\Psi_{\rm H}$, then plugging the result into Eq.~\eqref{eq:MetricReconstructionIRG} to obtain $h_{\mu\nu}^{\rm IRG}$, before finally projecting as in Eq.~\eqref{eq:TetradProjection} to find
\begin{align}
    h_{nm}^{\rm IRG}=-\frac{\epsilon_g}{2\sqrt{2}\zeta}\pa{l\mc{L}_2+\frac{a^2\sin{2\theta}}{\Sigma}l-\frac{2r}{\Sigma}\mc{L}_2}\ol{\Psi}_{\rm H}.
\end{align}
Next, one plugs in the the modes \eqref{eq:HertzModeIRG} of the Hertz potential $\ol{\Psi}_{\rm H}$ to find
\begin{align}
    h_{nm}^{\rm IRG}&=-\frac{\epsilon_g}{\sqrt{2}\zeta} \bigg[\ol{B}^{\rm IRG}\pa{l\mc{L}_2+\frac{a^2\sin{2\theta}}{\Sigma}l-\frac{2r}{\Sigma}\mc{L}_2}e^{-i\omega t+im\phi}\ol{\hat{R}}_{-\ol{\omega},\ell,-m}^{(-2)}\ol{\hat{S}}_{-\ol{\omega},\ell,-m}^{(-2)}\notag\\
    &\phantom{=}\qquad\qquad+\ol{A}^{\rm IRG}
    \pa{l\mc{L}_2+\frac{a^2\sin{2\theta}}{\Sigma}l-\frac{2r}{\Sigma}\mc{L}_2}e^{i\ol{\omega}t-im\phi}\ol{\hat{R}}_{\omega\ell m}^{(-2)}\ol{\hat{S}}_{\omega\ell m}^{(-2)}\bigg].
\end{align}
Each of these terms takes the form of a second-order differential operator applied to one of the modes of the Hertz potential.
We can now replace all the operators $l$, $\frac{\Sigma}{\Delta}n$, $\mc{L}_n$, and $\ol{\mc{L}}_n$ by their ``mode versions'' \eqref{eq:ModeVersions}, which show how they act on modes behaving like $e^{-i\omega t+im\phi}$.
If instead, they act on complex-conjugated modes behaving like $e^{i\ol{\omega}t-im\phi}$, then one must replace $(\omega,m)\to(-\ol{\omega},-m)$ in the operators \eqref{eq:Dn} and \eqref{eq:Ln}, which amounts to sending
\begin{align}
    \msc{D}_n\to\ol{\msc{D}}_n,\qquad 
    \msc{D}_n^\dag\to\ol{\msc{D}}_n^\dag,\qquad
    \msc{L}_n\to\ol{\msc{L}}^\dag_n,\qquad 
    \msc{L}_n^\dag\to\ol{\msc{L}}_n.
\end{align}
Armed with these expressions, we then find
\begin{align}
    h^{\rm IRG}_{nm}&=-\frac{\epsilon_g}{2\sqrt{2}\zeta}\bigg[\ol{B}^{\rm IRG}e^{-i\omega t+im\phi}\pa{\msc{D}_0\msc{L}_2+\frac{a^2\sin{2\theta}}{\Sigma}\msc{D}_0-\frac{2r}{\Sigma}\msc{L}_2}\ol{\hat{R}}_{-\ol{\omega},\ell,-m}^{(-2)}\ol{\hat{S}}_{-\ol{\omega},\ell,-m}^{(-2)}\notag\\
    &\phantom{=}\qquad\qquad\ \ +\ol{A}^{\rm IRG}e^{i\ol{\omega}t-im\phi}\pa{\ol{\msc{D}}_0\ol{\msc{L}}_2^\dag+\frac{a^2\sin{2\theta}}{\Sigma}\ol{\msc{D}}_0-\frac{2r}{\Sigma}\ol{\msc{L}}_2^\dag}\ol{\hat{R}}_{\omega\ell m}^{(-2)}\ol{\hat{S}}_{\omega\ell m}^{(-2)}\bigg].
\end{align}
This is of the form \eqref{eq:TetradComponentsIRG} with 
\begin{align}
     h_{nm}^{(+)}=\ol{B}^{\rm IRG}\ol{H}_{-\ol{\omega},\ell,-m}^{nm}, \qquad
    h_{nm}^{(-)}=\ol{A}^{\rm IRG}\ol{H}_{\omega\ell m}^{nm},
\end{align}
where we defined
\begin{align}
    H_{\omega\ell m}^{nm}\equiv-\frac{\epsilon_g}{2\sqrt{2}\ol{\zeta}}\pa{\msc{D}_0\msc{L}^\dag_2+\frac{a^2\sin{2\theta}}{\Sigma}\msc{D}_0-\frac{2r}{\Sigma}\msc{L}^\dag_2}\hat{R}_{\omega\ell m}^{(-2)}\hat{S}_{\omega\ell m}^{(-2)}.
\end{align}
This is how we obtained these IRG relations as they appear in Eqs.~\eqref{eq:TetradProjectionsIRG} and \eqref{eq:HIRG}.
Carrying out the same procedure for the other tetrad projections yields the remaining IRG relations.
Explicit forms of the functions $H_{\omega\ell m}^{ab}(r,\theta)$ are presented in App.~\ref{app:NoDerivatives}.
Likewise, doing the same computations starting with the form \eqref{eq:AdjointOperator4} of ${S_4^\dag}_{\mu\nu}$ acting on the Hertz potential $\Psi_{\rm H}'$ leads to the ORG relations \eqref{eq:TetradProjectionsORG} and \eqref{eq:HORG}, with explicit forms of the $H_{\omega\ell m}^{ab}(r,\theta)$ presented in App.~\ref{app:NoDerivatives}.

We omit these computations for the sake of brevity.

\section*{Acknowledgments}

This work was supported by NSF grant AST-2307888 and NSF CAREER award PHY-2340457.
We thank Leo Stein for many useful discussions, and are grateful to Sam Gralla, Lennox Keeble, Achilleas Porfyriadis, and Frans Pretorius for their useful comments on drafts of this manuscript.

\clearpage
\appendix
\addtocontents{toc}{\protect\setcounter{tocdepth}{1}}

\section{Linearized gravity}
\label{app:LinearizedGravity}

This appendix collects standard results in linearized gravity.
In App.~\ref{subsec:LinearizedEE}, we re-derive the linearized Einstein equations for metric perturbations $h_{\mu\nu}$ around an arbitrary background $g_{\mu\nu}$.
Specializing to the case of the Kerr geometry, this allows us to recover the form \eqref{eq:LinearizedEE} of the equations.
Then, in App.~\ref{subsec:LinearizedWeyl}, we derive an explicit formula for the linearized Weyl tensor $C_{\mu\nu\rho\sigma}^{(1)}$ in terms of $h_{\mu\nu}$.
The resulting expression is useful for carrying out the consistency checks in Sec.~\ref{subsec:ConsistencyChecks}.
These derivations make use of these linearized identities that we derive in App.~\ref{app:Identities}:
\begin{subequations}
\label{eq:Linearizations}
\begin{align}
    \label{eq:LinearizedRiemann}
    R_{\mu\nu\rho\sigma}^{(1)}&=\nabla_{\nu}\nabla_{[\rho}h_{\sigma]\mu}-\nabla_\mu\nabla_{[\rho} h_{\sigma]\nu}-{h^\tau}_{[\rho} R_{\sigma]\tau\mu\nu},\\
    \label{eq:LinearizedRicci}
    R_{\mu\nu}^{(1)}&=\nabla^\rho\nabla_{(\mu}h_{\nu)\rho}-\frac{1}{2}\nabla^2h_{\mu\nu}-\frac{1}{2}\nabla_\mu \nabla_\nu h,\\
    R^{(1)}&=\nabla^\mu\nabla^\nu h_{\mu\nu}-\nabla^2h-h^{\mu\nu}R_{\mu\nu}.
\end{align}
\end{subequations}

\subsection{Linearized Einstein equations and radiation gauge}
\label{subsec:LinearizedEE}

To first order, vacuum perturbations $h_{\mu\nu}$ of an arbitrary spacetime background $g_{\mu\nu}$ obey the linearized Einstein equations $G_{\mu\nu}^{(1)}=0$, where $G_{\mu\nu}^{(1)}=R_{\mu\nu}^{(1)}-\frac{1}{2}g_{\mu\nu}R^{(1)}-\frac{1}{2}h_{\mu\nu}R$ is the linearized Einstein tensor.
Using the identities \eqref{eq:Linearizations}, this takes the explicit form
\begin{align}
    -\nabla^2h_{\mu\nu}+2\nabla^\rho\nabla_{(\mu}h_{\nu)\rho}-\nabla_\mu\nabla_\nu h-h_{\mu\nu}R+g_{\mu\nu}\pa{-\nabla^\rho\nabla^\sigma h_{\rho\sigma}+\nabla^2h+R^{\rho\sigma}h_{\rho\sigma}}=0,
\end{align}
where $h=g^{\mu\nu}h_{\mu\nu}$ denotes the trace of the metric perturbation.
If the background is Ricci-flat, as is the case with the Kerr spacetime \eqref{eq:Kerr}, then $R_{\mu\nu}=0$, and so this simplifies to
\begin{align}
    \label{eq:LinearizedEEbis}
    -\nabla^2h_{\mu\nu}+2\nabla^\rho\nabla_{(\mu}h_{\nu)\rho}-\nabla_\mu\nabla_\nu h+g_{\mu\nu}\pa{\nabla^2h-\nabla^\rho\nabla^\sigma h_{\rho\sigma}}=0,
\end{align}
which is precisely Eq.~\eqref{eq:LinearizedEE} in the Introduction.
As discussed below it, the symmetric tensor $h_{\mu\nu}=h_{\nu\mu}$ does not really have ten independent components, since four of them are fixed by the Bianchi identities and four more can be gauged away via the diffeomorphisms \eqref{eq:Diffeomorphisms}.
This leaves $h_{\mu\nu}$ with only two independent propagating degrees of freedom, as befits a massless spin-2 field.

In this paper, we solve these equations in radiation gauge: either the ingoing version \eqref{eq:IRG} or the outgoing version \eqref{eq:ORG}.
Both impose tracelessness ($h=0$) and simplify the equations to
\begin{align}
    \label{eq:TracelessEE}
    -\nabla^2h_{\mu\nu}+2\nabla^\rho\nabla_{(\mu}h_{\nu)\rho}-g_{\mu\nu}\nabla^\rho\nabla^\sigma h_{\rho\sigma}=0.
\end{align}
At first, it may seem alarming that each of these gauge choices actually imposes five constraints instead of the expected four.
However, it can be shown that for perturbations of the Kerr vacuum---or more generally, of any background of Petrov type II (which includes Petrov type D)---either set of constraints can be satisfied by an appropriate choice of generator $\xi=\xi^\mu\pd_\mu$ in the diffeomorphism \eqref{eq:Diffeomorphisms}.
In fact, the Petrov type II spacetimes form the most general class admitting a radiation gauge, with Petrov type D spacetimes admitting both IRG and ORG \cite{Price2007}.

A similar phenomenon occurs in flat spacetime, where one typically imposes the transverse traceless gauge with $\nabla^\mu h_{\mu\nu}=0$ and $g^{\mu\nu}h_{\mu\nu}=0$.
Despite appearing overconstrained, this gauge exists for any vacuum perturbation of a vacuum Einstein spacetime (see, e.g., Sec.~7.5 of \cite{Wald1984}).

\subsection{Linearized Weyl tensor}
\label{subsec:LinearizedWeyl}

As we reviewed in the Introduction, the linearized Weyl tensor $C_{\mu\nu\rho\sigma}^{(1)}$ contains two Weyl scalars \eqref{eq:WeylScalars} that encode the two degrees of freedom carried by a perturbation $h_{\mu\nu}$ of the Kerr spacetime.
Here, we derive an expression for $C_{\mu\nu\rho\sigma}^{(1)}$ in terms of the metric perturbation $h_{\mu\nu}$, which is useful for performing the consistency checks in Sec.~\ref{subsec:ConsistencyChecks}.

In general, the Weyl tensor is defined as the completely traceless part of the Riemann tensor,
\begin{align}
    C_{\mu\nu\rho\sigma}=R_{\mu\nu\rho\sigma}+\frac{1}{2}\pa{R_{\mu\sigma}g_{\nu\rho}-R_{\mu\rho}g_{\nu\sigma}+R_{\nu\rho}g_{\mu\sigma}-R_{\nu\sigma}g_{\mu\rho}}+\frac{1}{6}R\pa{g_{\mu\rho}g_{\nu\sigma}-g_{\mu\sigma}g_{\nu\rho}}.
\end{align}
Taking $g_{\mu\nu}\to g_{\mu\nu}+h_{\mu\nu}$ and extracting the pieces linear in $h_{\mu\nu}$ yields the linearized Weyl tensor
\begin{align}
    C_{\mu\nu\rho\sigma}^{(1)}&=R_{\mu\nu\rho\sigma}^{(1)}+\frac{1}{2}\pa{R_{\mu\sigma}^{(1)}g_{\nu\rho}-R_{\mu\rho}^{(1)}g_{\nu\sigma}+R_{\nu\rho}^{(1)}g_{\mu\sigma}-R_{\nu\sigma}^{(1)}g_{\mu\rho}}\notag\\
    &\phantom{=}+\frac{1}{2}\pa{R_{\mu\sigma}h_{\rho\nu}-R_{\mu\rho}h_{\nu\sigma}+R_{\nu\rho}h_{\mu\sigma}-R_{\nu\sigma}h_{\mu\rho}}+\frac{1}{6}R^{(1)}\pa{g_{\mu\rho}g_{\nu\sigma}-g_{\mu\sigma}g_{\nu\rho}}\notag\\
    &\phantom{=}+\frac{1}{6}R\pa{g_{\mu\rho}h_{\nu\sigma}-g_{\mu\sigma}h_{\nu\rho}+g_{\nu\sigma}h_{\mu\rho}-g_{\nu\rho}h_{\mu\sigma}}.
\end{align}
If the background $g_{\mu\nu}$ is Ricci-flat, so $R_{\mu\nu}=0$, and the linearized metric perturbation $h_{\mu\nu}$ solves the vacuum equations \eqref{eq:LinearizedEEbis}, then $R_{\mu\nu}^{(1)}-\frac{1}{2}g_{\mu\nu}R^{(1)}=0$ and the above expression simplifies to
\begin{align}
    C_{\mu\nu\rho\sigma}^{(1)}=R_{\mu\nu\rho\sigma}^{(1)}-\frac{1}{3}R^{(1)}\pa{g_{\mu\rho}g_{\nu\sigma}-g_{\mu\sigma}g_{\nu\rho}}.
\end{align}
Using the identities \eqref{eq:Linearizations}, this can be rewritten explicitly in terms of $h_{\mu\nu}$ as
\begin{align}
    \label{eq:LinearizedWeyl}
    C_{\mu\nu\rho\sigma}^{(1)}&=\nabla_{\nu}\nabla_{[\rho}h_{\sigma]\mu}-\nabla_\mu\nabla_{[\rho} h_{\sigma]\nu}-{h^\tau}_{[\rho} R_{\sigma]\tau\mu\nu}-\frac{1}{3}\pa{\nabla^\alpha\nabla^\beta h_{\alpha\beta}-\nabla^2h}\pa{g_{\mu\rho}g_{\nu\sigma}-g_{\mu\sigma}g_{\nu\rho}}.
\end{align}

\subsection{Linearized Riemann and Ricci tensors}
\label{app:Identities}

Here, we derive for completeness the identities \eqref{eq:Linearizations} (given also, e.g., in Sec.~3.4 of \cite{Ortin2015}).

The Christoffel symbol and Riemann tensor are defined by
\begin{align}
    {\Gamma^\mu}_{\nu\rho}=\frac{1}{2}g^{\mu\sigma}\pa{\pd_\nu g_{\rho\sigma}+\pd_\rho g_{\nu\sigma}-\pd_\sigma g_{\nu\rho}},\qquad
    {R^\mu}_{\nu\rho\sigma}=2\pd_{[\rho}{\Gamma^\mu}_{\sigma]\nu}+2{\Gamma^\mu}_{\tau[\rho}{\Gamma^\tau}_{\sigma]\nu}.
\end{align}
Sending $g_{\mu\nu}\to g_{\mu\nu}+h_{\mu\nu}$, we can extract the linearized Christoffel symbol:
\begin{subequations}
\begin{align}
    {}^{(1)}{\Gamma^\mu}_{\nu\rho}&=\frac{1}{2}g^{\mu\sigma}\pa{\pd_\nu h_{\rho\sigma}+\pd_\rho h_{\nu\sigma}-\pd_\sigma h_{\nu\rho}}-\frac{1}{2}h^{\mu\sigma}\pa{\pd_\nu g_{\rho\sigma}+\pd_\rho g_{\nu\sigma}-\pd_\sigma g_{\nu\rho}}\\
    &=\frac{1}{2}g^{\mu\sigma}\pa{\pd_\nu h_{\rho\sigma}+\pd_\rho h_{\nu\sigma}-\pd_\sigma h_{\nu\rho}-2{\Gamma^\tau}_{\nu\rho}h_{\sigma\tau}}\\
    \label{eq:LinearizedChristoffel}
    &=\frac{1}{2}g^{\mu\sigma}\pa{\nabla_\nu h_{\rho\sigma}+\nabla_\rho h_{\nu\sigma}-\nabla_\sigma h_{\nu\rho}}.
\end{align}
\end{subequations}
Even though the Christoffel symbol is not a tensor, its first-order variation manifestly is, since it can be expressed in terms of covariant derivatives of $h_{\mu\nu}$.
The linearized Riemann tensor is
\begin{subequations}
\begin{align}
    {}^{(1)}{R^\mu}_{\nu\rho\sigma}&=2\pa{\pd_{[\rho}{}^{(1)}{\Gamma^\mu}_{\sigma]\nu}+{}^{(1)}{\Gamma^\mu}_{\tau[\rho}{\Gamma^\tau}_{\sigma]\nu}+{\Gamma^\mu}_{\tau[\rho}{}^{(1)}{\Gamma^\tau}_{\sigma]\nu}}\\
    &=\pd_{\rho}{}^{(1)}{\Gamma^\mu}_{\nu\sigma}+{\Gamma^\mu}_{\rho\tau}{}^{(1)}{\Gamma^\tau}_{\sigma\nu}-{\Gamma^\tau}_{\nu\rho}{}^{(1)}{\Gamma^\mu}_{\tau\sigma}-{\Gamma^\tau}_{\rho\sigma}{}^{(1)}{\Gamma^\mu}_{\nu\tau}\notag\\
    &\phantom{=}-\pd_\sigma{}^{(1)}{\Gamma^\mu}_{\nu\rho}-{\Gamma^\mu}_{\sigma\tau}{}^{(1)}{\Gamma^\tau}_{\rho\nu}+{\Gamma^\tau}_{\nu\sigma}{}^{(1)}{\Gamma^\mu}_{\tau\rho}+{\Gamma^\tau}_{\rho\sigma}{}^{(1)}{\Gamma^\mu}_{\nu\tau}\\
    &=2\nabla_{[\rho}{}^{(1)}{\Gamma^\mu}_{\sigma]\nu}.
\end{align}
\end{subequations}
Inserting Eq.~\eqref{eq:LinearizedChristoffel}, this becomes
\begin{subequations}
\begin{align}
    {}^{(1)}{R^\mu}_{\nu\rho\sigma}
    &=\nabla_{[\rho|}\nabla_\nu{h_{|\sigma]}}^\mu-\nabla_{[\rho}\nabla^\mu h_{\sigma]\nu}+\nabla_{[\rho}\nabla_{\sigma]}{h_\nu}^\mu\\
    &=\nabla_{[\rho|}\nabla_\nu{h_{|\sigma]}}^\mu-\nabla_{[\rho}\nabla^\mu h_{\sigma]\nu}-\frac{1}{2}h^{\mu\tau}R_{\tau\nu\rho\sigma}+\frac{1}{2}{h_\nu}^\tau{R^\mu}_{\tau\rho\sigma}\\
    &= \nabla_{\nu}\nabla_{[\rho}{h_{\sigma]}}^\mu-\nabla^\mu\nabla_{[\rho} h_{\sigma]\nu}-h^{\mu\tau} R_{\tau\nu\rho\sigma}-\frac{1}{2}{h^\tau}_\rho{R^\mu}_{\nu\sigma\tau}+\frac{1}{2}{h^\tau}_\sigma{R^\mu}_{\nu\rho\tau},
\end{align}
\end{subequations}
where in the last step, we used the commutator identity $2\nabla_{[\mu}\nabla_{\nu]}h_{\rho\sigma}=-h_{\rho\tau}{R^\tau}_{\sigma\mu\nu}-h_{\sigma\tau}{R^\tau}_{\rho\mu\nu}$ to reorder covariant derivatives.
Lastly, lowering the upper index yields
\begin{align}
    R_{\mu\nu\rho\sigma}^{(1)}=g_{\mu\tau} {}^{(1)}{R^\tau}_{\nu\rho\sigma}+h_{\mu\tau}{R^\tau}_{\nu\rho\sigma}
    =\nabla_{\nu}\nabla_{[\rho}h_{\sigma]\mu}-\nabla_\mu\nabla_{[\rho} h_{\sigma]\nu}-{h^\tau}_{[\rho} R_{\sigma]\tau\mu\nu},
\end{align}
which completes the derivation of the identity \eqref{eq:LinearizedRiemann}.
Contracting with the full metric yields
\begin{align}
    R^{(1)}_{\mu\nu}=g^{\rho\sigma}R_{\rho\mu\sigma\nu}^{(1)}-h^{\rho\sigma}R_{\rho\mu\sigma\nu}
    =\nabla^\rho\nabla_{(\mu}h_{\nu)\rho}-\frac{1}{2}\nabla^2h_{\mu\nu}-\frac{1}{2}\nabla_\nu\nabla_\mu h,
\end{align}
which completes the derivation of the identity \eqref{eq:LinearizedRicci}.
Finally, contracting once more yields
\begin{align}
    R^{(1)}=g^{\mu\nu}R_{\mu\nu}^{(1)}-h^{\mu\nu} R_{\mu\nu}
    =\nabla^\mu\nabla^\nu h_{\mu\nu}-\nabla^2h-h^{\mu\nu}R_{\mu\nu}.
\end{align}
This completes the proof of the identities \eqref{eq:Linearizations}.

\section{Metric reconstruction in ingoing and outgoing coordinates}
\label{app:IngoingOutgoing}

In this appendix, we take the reconstructed metrics in ingoing and outgoing radiation gauge (given in Secs.~\ref{subsec:MetricReconstructionIRG} and \ref{subsec:MetricReconstructionORG}, respectively) and transform their Boyer-Lindquist components $h_{\mu\nu}^{\rm IRG/ORG}$ to ingoing coordinates (in App.~\ref{subsec:IngoingMetricReconstruction}) and to outgoing coordinates (in App.~\ref{subsec:OutgoingMetricReconstruction}).
We expect these forms of the metric components to be particularly useful for analyses of the perturbations at (or near) the horizon, where the Boyer-Lindquist coordinates become singular, while the ingoing and outgoing coordinates remain regular.
Another advantage they provide is that many components of $h_{\mu\nu}^{\rm ORG}$ (or $h_{\mu\nu}^{\rm IRG}$) vanish in ingoing (or resp., outgoing) coordinates.

\subsection{Metric reconstruction in ingoing coordinates}
\label{subsec:IngoingMetricReconstruction}

Ingoing Kerr coordinates $(v,r,\theta,\psi)$ are related to Boyer-Lindquist coordinates $(t,r,\theta,\phi)$ via
\begin{align}
    \label{eq:IngoingCoordinates}
    v=t+r_*,\qquad
    \psi=\phi+r_\sharp,
\end{align}
where the tortoise coordinate $r_*$ and the radius $r_\sharp$ are defined in Eqs.~\eqref{eq:TortoiseCoordinate} and \eqref{eq:SharpCoordinate}.

Under this transformation, the Kerr line element \eqref{eq:Kerr} becomes
\begin{subequations}
\begin{gather}
    \frac{ds^2}{\epsilon_g}=-\pa{1-\frac{2Mr}{\Sigma}}\ed v^2+2\ed v\ed r+\Sigma\ed\theta^2-2a\sin^2{\theta}\pa{\frac{2Mr}{\Sigma}\ed v+\ed r}\ed\psi+g_{\psi\psi}\ed\psi^2,\\
    \label{eq:AzimuthalMetric}
    g_{\psi\psi}=\frac{\sin^2{\theta}}{\Sigma}\br{\pa{r^2+a^2}^2-a^2\Delta\sin^2{\theta}}.
\end{gather}
\end{subequations}
Since the $(r,\theta)$ coordinates are untouched, modes transform simply as follows:
\begin{align}
    \label{eq:IngoingModes}
    e^{-i\omega t+im\phi}R(r)S(\theta)\to e^{i\omega r_*-imr_\sharp}e^{-i\omega v+im\psi}R(r)S(\theta).
\end{align}

In ingoing coordinates \eqref{eq:IngoingCoordinates}, the IRG metric components \eqref{eq:MetricComponentsIRG} become
\begin{subequations}
\begin{align}
    h_{vv}^{\rm IRG}&=-a^2\sin^2{\theta}\,\mathcal{M}_+-2a\sin{\theta}\,\mathcal{N}_-+h_{nn},\\ 
    h_{rr}^{\rm IRG}&=\frac{4\Sigma^2}{\Delta^2}h_{nn},\\
    h_{\theta\theta}^{\rm IRG}&=\Sigma^2\mathcal{M}_+,\\
    h_{\psi\psi}^{\rm IRG}&=-\sin^2{\theta}\br{\pa{r^2+a^2}^2\mathcal{M}_++2a\pa{r^2+a^2}\sin{\theta}\,\mathcal{N}_--a^2\sin^2{\theta}\,h_{nn}},\\
    h_{vr}^{\rm IRG}&=\frac{2\Sigma}{\Delta}\br{a\sin{\theta}\,\mathcal{N}_--h_{nn}},\\
    h_{v\theta}^{\rm IRG}&=\Sigma\br{\mathcal{N}_+-a\sin{\theta}\,\mathcal{M}_-},\\
    h_{v\psi}^{\rm IRG}&=a\sin^2{\theta}\br{\pa{r^2+a^2}\mathcal{M}_++\pa{\frac{\Sigma}{a\sin{\theta}}+2a\sin{\theta}}\,\mathcal{N}_--h_{nn}},\\
    h_{r\theta}^{\rm IRG}&=-\frac{2\Sigma^2}{\Delta}\mathcal{N}_+,\\ 
    h_{r\psi}^{\rm IRG}&=\frac{2\Sigma\sin{\theta}}{\Delta}\br{-\pa{r^2+a^2}\mathcal{N}_-+a\sin{\theta}h_{nn}},\\
    h_{\theta\psi}^{\rm IRG}&=\Sigma\sin{\theta}\br{\pa{r^2+a^2}\mc{M}_--a\sin{\theta}\,\mc{N}_+},
\end{align}
\end{subequations}
while the ORG metric components \eqref{eq:MetricComponentsORG} become (the $h_{r\mu}$ all vanish)
\begin{subequations}
\begin{align}
    h_{vv}^{\rm ORG}&=-a^2\sin^2{\theta}\,\mc{M}_+-\frac{a\Delta\sin{\theta}}{\Sigma}\mc{L}_-+\frac{\Delta^2}{4\Sigma^2}h_{ll},\\
    h_{r\mu}^{\rm ORG}&=0,\\
    h_{\theta\theta}^{\rm ORG}&=\Sigma^2\mc{M}_+,\\
    h_{\psi\psi}^{\rm ORG}&=-\sin^2{\theta}\br{\pa{r^2+a^2}^2\mc{M}_++\frac{a\Delta\pa{r^2+a^2}\sin{\theta}}{\Sigma}\mc{L}_--\frac{a^2\Delta^2\sin^2{\theta}}{4\Sigma^2}h_{ll}},\\
    h_{v\theta}^{\rm ORG}&=\frac{\Delta}{2}\mc{L}_+-a\Sigma\sin{\theta}\,\mc{M}_-,\\
    h_{v\psi}^{\rm ORG}&=a\sin^2{\theta}\br{\pa{r^2+a^2}\mc{M}_++\frac{\Delta}{2\Sigma}\pa{\frac{\Sigma}{a\sin{\theta}}+2a\sin{\theta}}\mc{L}_--\frac{\Delta^2}{4\Sigma^2}h_{ll}},\\
    h_{\theta\psi}^{\rm ORG}&=\Sigma\sin{\theta}\br{\pa{r^2+a^2}\mc{M}_--\frac{a\Delta\sin{\theta}}{2\Sigma}\mc{L}_+},
\end{align}
\end{subequations}
with $\mathcal{L}_\pm$, $\mathcal{M}_\pm$, and $\mathcal{N}_\pm$ as defined in Eqs.~\eqref{eq:RealProjections} and \eqref{eq:RealProjectionsBis}.
Now, we only need to specify the tetrad components \eqref{eq:TetradProjection} in these new coordinates.
Since the projections $h_{ab}=a^\mu b^\nu h_{\mu\nu}$ are spacetime scalars, they transform simply by substitution of \eqref{eq:IngoingCoordinates}.
By \eqref{eq:IngoingModes}, the result is just
\begin{align}
    h_{ab}=e^{i\omega r_*-imr_\sharp}e^{-i\omega v+im\psi}h_{ab}^{(+)}(r,\theta)+e^{-i\ol{\omega}r_*+imr_\sharp}e^{i\ol{\omega}v-im\psi}h_{ab}^{(-)}(r,\theta),
\end{align}
where the $h_{ab}^{(\pm)}(r,\theta)$ are the same as in Boyer-Lindquist coordinates---and are given in Eqs.~\eqref{eq:TetradProjectionsIRG} and \eqref{eq:TetradProjectionsORG}---since the coordinates $r$ and $\theta$ are unchanged under the transformation \eqref{eq:IngoingCoordinates}.

\subsection{Metric reconstruction in outgoing coordinates}
\label{subsec:OutgoingMetricReconstruction}

Outgoing Kerr coordinates $(u,r,\theta,\psi)$ are related to Boyer-Lindquist coordinates $(t,r,\theta,\phi)$ via
\begin{align}
    \label{eq:OutgoingCoordinates}
    u=t-r_*,\qquad
    \psi=\phi-r_\sharp,
\end{align}
where the tortoise coordinate $r_*$ and the radius $r_\sharp$ are defined in Eqs.~\eqref{eq:TortoiseCoordinate} and \eqref{eq:SharpCoordinate}.

Under this transformation, the Kerr line element \eqref{eq:Kerr} becomes, with $g_{\psi\psi}$ in Eq.~\eqref{eq:AzimuthalMetric}, 
\begin{align}
    \frac{ds^2}{\epsilon_g}=-\pa{1-\frac{2Mr}{\Sigma}}\ed u^2-2\ed u\ed r+\Sigma\ed\theta^2-2a\sin^2{\theta}\pa{\frac{2Mr}{\Sigma}\ed u-\ed r}\ed\psi+g_{\psi\psi}\ed\psi^2,
\end{align}
Since the $(r,\theta)$ coordinates are untouched, modes transform simply as follows:
\begin{align}
    \label{eq:OutgoingModes}
    e^{-i\omega t+im\phi}R(r)S(\theta)\to e^{-i\omega r_*+imr_\sharp}e^{-i\omega u+im\psi}R(r)S(\theta).
\end{align}
In outgoing coordinates \eqref{eq:OutgoingCoordinates}, the IRG metric components \eqref{eq:MetricComponentsIRG} become (the $h_{r\mu}$ all vanish)
\begin{subequations}
\begin{align}
    h_{uu}^{\rm IRG}&=-a^2\sin^2{\theta}\,\mc{M}_+-2a\sin{\theta}\,\mc{N}_-+h_{nn},\\
    h_{r\mu}^{\rm IRG}&=0,\\
    h_{\theta\theta}^{\rm IRG}&=\Sigma^2\mc{M}_+,\\
    h_{\psi\psi}^{\rm IRG}&=-\sin^2{\theta}\br{\pa{r^2+a^2}^2\mc{M}_++2a\pa{r^2+a^2}\sin{\theta}\,\mc{N}_--a^2\sin^2{\theta}\,h_{nn}},\\
    h_{u\theta}^{\rm IRG}&=\Sigma\br{\mc{N}_+-a\sin{\theta}\,\mc{M}_-},\\
    h_{u\psi}^{\rm IRG}&=a\sin^2{\theta}\br{\pa{r^2+a^2}\mc{M}_++\pa{\frac{\Sigma}{a\sin{\theta}}+2a\sin{\theta}}\mc{N}_--h_{nn}},\\
    h_{\theta\psi}^{\rm IRG}&=\Sigma\sin{\theta}\br{\pa{r^2+a^2}\mc{M}_--a\sin{\theta}\,\mc{N}_+},
\end{align}
\end{subequations}
while the ORG components \eqref{eq:MetricComponentsORG} become
\begin{subequations}
\begin{align}
    h_{uu}^{\rm ORG}&=-a^2\sin^2{\theta}\,\mc{M}_+-\frac{a\Delta\sin{\theta}}{\Sigma}\mc{L}_-+\frac{\Delta^2}{4\Sigma^2}h_{ll},\\
    h_{rr}^{\rm ORG}&=h_{ll},\\
    h_{\theta\theta}^{\rm ORG}&=\Sigma^2\mc{M}_+,\\
    h_{\psi\psi}^{\rm ORG}&=-\sin^2{\theta}\br{\pa{r^2+a^2}^2\mc{M}_++\frac{a\Delta\pa{r^2+a^2}\sin{\theta}}{\Sigma}\mc{L}_--\frac{a^2\Delta^2\sin^2{\theta}}{4\Sigma^2}h_{ll}},\\
    h_{ur}^{\rm ORG}&= -a\sin{\theta}\,\mathcal{L}_- + \frac{\Delta}{2 \Sigma} h_{ll},\\
    h_{u\theta}^{\rm ORG}&=\frac{\Delta}{2}\mc{L}_+-a\Sigma\sin{\theta}\,\mc{M}_-,\\
    h_{u\psi}^{\rm ORG}&=a\sin^2{\theta}\br{\pa{r^2+a^2}\mc{M}_++\frac{\Delta}{2\Sigma}\pa{\frac{\Sigma}{a\sin{\theta}}+2a\sin{\theta}}\mc{L}_--\frac{\Delta^2}{4\Sigma^2}h_{ll}},\\
    h_{r\theta}^{\rm ORG}&=\Sigma\,\mathcal{L}_+,\\
    h_{r\psi}^{\rm ORG}&=\sin{\theta} \br{
        \pa{r^2+a^2}\,\mathcal{L}_-
        - \frac{a\sin{\theta} \Delta}{2 \Sigma} h_{ll}
    },\\
    h_{\theta\psi}^{\rm ORG}&=\Sigma\sin{\theta}\br{\pa{r^2+a^2}\mc{M}_--\frac{a\Delta\sin{\theta}}{2\Sigma}\mc{L}_+},
\end{align}
\end{subequations}
with $\mathcal{L}_\pm$, $\mathcal{M}_\pm$, and $\mathcal{N}_\pm$ as defined in Eqs.~\eqref{eq:RealProjections} and \eqref{eq:RealProjectionsBis}.
Now, we only need to specify the tetrad components \eqref{eq:TetradProjection} in these new coordinates.
Since the projections $h_{ab}=a^\mu b^\nu h_{\mu\nu}$ are spacetime scalars, they transform simply by substitution of \eqref{eq:OutgoingCoordinates}.
By \eqref{eq:OutgoingModes}, the result is just
\begin{align}
    h_{ab}=e^{-i\omega r_*+imr_\sharp}e^{-i\omega u+im\psi}h_{ab}^{(+)}(r,\theta)+e^{i\ol{\omega}r_*-imr_\sharp}e^{i\ol{\omega}u-im\psi}h_{ab}^{(-)}(r,\theta).
\end{align}
where the $h_{ab}^{(\pm)}(r,\theta)$ are the same as in Boyer-Lindquist coordinates---and are given in Eqs.~\eqref{eq:TetradProjectionsIRG} and \eqref{eq:TetradProjectionsORG}---since the coordinates $r$ and $\theta$ are unchanged under the transformation \eqref{eq:OutgoingCoordinates}.

\section{Eliminating derivatives of the mode functions}
\label{app:FirstDerivatives}

The angular and radial Teukolsky--Starobinsky identities \eqref{eq:ATSI1} and \eqref{eq:RTSI1} are fourth-order relations.
Using the equations of motion \eqref{eq:RadialODE} and \eqref{eq:AngularODE} to remove the second (and higher) derivatives leads to new first-order identities between radial and angular modes of opposite spin.

These new relations take the form
\begin{subequations}
\label{eq:FirstDerivatives}
\begin{align}
    \hat{\msc{C}}_{\omega\ell m}\Delta^3\hat{R}_{\omega\ell m}^{(+2)}&=\mathbb{A}_{\omega\ell m}\msc{D}_0\hat{R}_{\omega\ell m}^{(-2)}+\mathbb{B}_{\omega\ell m}\hat{R}_{\omega\ell m}^{(-2)},\\
    \hat{\msc{C}}_{\omega\ell m}'\hat{R}_{\omega\ell m}^{(-2)}&= -\mathbb{A}_{\omega\ell m}\Delta\msc{D}_0^\dag\hat{R}_{\omega\ell m}^{(+2)}+\mathbb{C}_{\omega\ell m}\hat{R}_{\omega\ell m}^{(+2)},\\
    \hat{D}'_{\omega\ell m}\hat{S}_{\omega\ell m}^{(+2)}&=\mathbb{X}_{\omega\ell m}\msc{L}_2^\dag\hat{S}_{\omega\ell m}^{(-2)}+\mathbb{Y}_{\omega\ell m}\hat{S}_{\omega\ell m}^{(-2)},\\
    \hat{D}_{\omega\ell m}\hat{S}_{\omega\ell m}^{(-2)}&=-\mathbb{X}_{\omega\ell m}\msc{L}_2\hat{S}_{\omega\ell m}^{(+2)}+\mathbb{Y}_{-\omega,\ell,-m}\hat{S}_{\omega\ell m}^{(+2)},
\end{align}
\end{subequations}
where we (yet again) introduced some new coefficients\footref{fn:TS}
\begin{subequations}
\begin{align}
    \mathbb{A}_{\omega\ell m}&=8iK\br{K^2+\pa{r-M}^2}-\br{4iK\pa{\lambda_{\omega\ell m}^{(+2)}+6}+8i\omega r\pa{r-M}}\Delta+8i\omega\Delta^2,\\
    \mathbb{B}_{\omega\ell m}&=\br{\pa{\lambda_{\omega\ell m}^{(+2)}+6-2i\omega r}\pa{\lambda_{\omega\ell m}^{(+2)}+4+6i\omega r}- 12i\omega\pa{iK+r-M}}\Delta\notag\\
    &\phantom{=}\quad+4iK\pa{iK+r-M}\pa{\lambda_{\omega\ell m}^{(+2)}+4+6i\omega r},\\
    %\mathbb{C}^\text{old}_{\omega\ell m}&=-8iK^3+K\cu{4i\Delta\lambda_{\omega\ell m}^{(+2)}+8i\br{3\Delta-\pa{r-M}^2}}-8i\omega\Delta\br{\Delta-r\pa{r-M}},\\
    \mathbb{C}_{\omega\ell m}&=\br{\pa{\lambda_{\omega\ell m}^{(+2)}+2i\omega r+4}\pa{\lambda_{\omega\ell m}^{(+2)}-6i\omega r+6}-4i\omega \pa{9r-5M}}\Delta^2\notag\\
    &\phantom{=}\quad+32i\omega r\Delta\pa{r-M}^2-32iK^3\pa{r-M}-8\Delta K^2(2-3i\omega r)\notag\\
    &\phantom{=}\quad+4iK\cu{4\pa{r-M}\br{5\Delta-2\pa{r-M}^2}-3i\omega\Delta\br{\Delta-2r\pa{r-M}}}\notag\\
    &\phantom{=}\quad+4i\lambda_{\omega\ell m}^{(+2)}K\br{iK+3\pa{r-M}}\Delta,\\
    \mathbb{X}_{\omega\ell m}&=4\br{Q\pa{\lambda_{\omega\ell m}^{(+2)}+2\cot^2{\theta}-2Q^2+6} + 2a\omega\csc{\theta}},\\
    \mathbb{Y}_{\omega\ell m}&=\pa{\lambda_{\omega\ell m}^{(+2)}+6a\omega\cos{\theta}+4}\pa{\lambda_{\omega\ell m}^{(+2)}-2a\omega\cos{\theta}-4Q^2-4Q\cot{\theta}+6}\notag\\
    &\phantom{=}\quad-12a\omega\pa{Q\sin{\theta}+\cos{\theta}}.
    %\\
    %\mathbb{Z}_{\omega\ell m}&=-4\br{Q\pa{\lambda_{\omega\ell m}^{(+2)}+2\cot^2{\theta}-2Q^2+6}+2a\omega\csc{\theta}},\\
    %\mathbb{W}_{\omega\ell m}&=\pa{\lambda_{\omega\ell m}^{(+2)}-6a\omega\cos{\theta}+4}\pa{\lambda_{\omega\ell m}^{(+2)}+2a\omega\cos{\theta}-4Q^2+4Q\cot{\theta}+6}\notag\\&\phantom{=}\quad-12a\omega\pa{Q\sin{\theta}-\cos{\theta}}.
\end{align}
\end{subequations}

\section{Metric components with no derivatives of the mode functions}
\label{app:NoDerivatives}

In this appendix, we give the explicit forms of the functions $H_{\omega\ell m}^{ab}$ introduced in Secs.~\ref{subsec:MetricReconstructionIRG} and \ref{subsec:MetricReconstructionORG}, from which all the metric components $h_{\mu\nu}$ in both IRG and ORG are built.
These functions all take the form of a second-order differential operator acting on mode functions.
We give them in this form in App.~\ref{subsec:MetricSecondDerivatives}.
Then, in App.~\ref{subsec:MetricFirstDerivatives}, we use the radial and angular ODEs \eqref{eq:RadialODE} and \eqref{eq:AngularODE} to remove all (non-mixed) second derivatives.
Finally, in App.~\ref{subsec:MetricNoDerivatives}, we use the identities \eqref{eq:FirstDerivatives} to remove all remaining derivatives from the metric components $h_{\mu\nu}$.
In certain situations, this elimination of derivatives may help with the numerical evaluation of these components.

\subsection{Explicit metric components including second derivatives}
\label{subsec:MetricSecondDerivatives}

Here, we give explicit forms of the functions $H_{\omega\ell m}^{ab}$.
In the ingoing radiation gauge, the three functions $H_{\omega\ell m}^{nn}$, $H_{\omega\ell m}^{nm}$, and $H_{\omega\ell m}^{mm}$ are
\begin{subequations}
\begin{align}
    H_{\omega\ell m}^{nn}&\equiv-\frac{\epsilon_g}{4\ol{\zeta}^2}\pa{\msc{L}_1^\dag-\frac{2ia\sin{\theta}}{\zeta}}\msc{L}_2^\dag\hat{R}_{\omega\ell m}^{(-2)}\hat{S}_{\omega\ell m}^{(-2)}\\
    &=-\frac{\epsilon_g}{4\ol{\zeta}^2}\pa{\pd_\theta^2+A^{nn}\pd_\theta+B^{nn}}\hat{R}_{\omega\ell m}^{(-2)}\hat{S}_{\omega\ell m}^{(-2)},\\
    H_{\omega\ell m}^{nm}&\equiv-\frac{\epsilon_g}{2\sqrt{2}\ol{\zeta}}\pa{\msc{D}_0\msc{L}_2^\dag+\frac{a^2\sin{2\theta}}{\Sigma}\msc{D}_0-\frac{2r}{\Sigma}\msc{L}_2}\hat{R}_{\omega\ell m}^{(-2)}\hat{S}_{\omega\ell m}^{(-2)}\\
    &=-\frac{\epsilon_g}{2\sqrt{2}\zeta}\pa{\pd_r\pd_\theta+A^{nm}\pd_r+B^{nm}\pd_\theta+C^{nm}}\hat{R}_{\omega\ell m}^{(-2)}\hat{S}_{\omega\ell m}^{(-2)},\\
    H_{\omega\ell m}^{mm}&\equiv-\frac{\epsilon_g}{2}\pa{\msc{D}_0-\frac{2}{\ol{\zeta}}}\msc{D}_0\hat{R}_{\omega\ell m}^{(-2)}\hat{S}_{\omega\ell m}^{(-2)}\\
    &=-\frac{\epsilon_g}{2}\pa{\pd_r^2+A^{mm}\pd_r+B^{mm}}\hat{R}_{\omega\ell m}^{(-2)}\hat{S}_{\omega\ell m}^{(-2)},
\end{align}
\end{subequations}
where the coefficients of the derivatives are (here and hereafter, $\Delta'\equiv\pd_r\Delta$)
\begin{subequations}
\begin{align}
    A^{nn}&=-2Q+3\cot{\theta}-\frac{2ia\sin{\theta}}{\zeta},\\
    B^{nn}&=\pa{Q-2\cot{\theta}}\pa{Q-\cot{\theta}+\frac{2ia\sin{\theta}}{\zeta}}+Q\cot{\theta}+2a\omega\cos{\theta}-2\csc^2{\theta},\\
    A^{nm}&=-Q+2\cot{\theta}+\frac{a^2\sin{2\theta}}{\Sigma},\\
    B^{nm}&=-\frac{iK}{\Delta}-\frac{2r}{\Sigma},\\
    C^{nm}&=\pa{\frac{iK}{\Delta}+\frac{2r}{\Sigma}}\pa{Q-2\cot{\theta}}-\frac{iKa^2\sin{2\theta}}{\Sigma\Delta},\\
    A^{mm}&=-2\pa{\frac{iK}{\Delta}+\frac{1}{\zeta}},\\
    B^{mm}&=\frac{iK}{\Delta}\pa{\frac{iK+\Delta'}{\Delta}+\frac{2}{\zeta}}-\frac{2i\omega r}{\Delta}.
\end{align}
\end{subequations}
In the outgoing radiation gauge, the three functions $H_{\omega\ell m}^{ll}$, $H_{\omega\ell m}^{lm}$, and $H_{\omega\ell m}^{mm}$ are
\begin{subequations}
\begin{align}
    H_{\omega\ell m}^{ll}&\equiv-\frac{\epsilon_g}{4}\zeta^2\pa{\msc{L}_1-\frac{2ia\sin{\theta}}{\zeta}}\msc{L}_2\hat{R}_{\omega\ell m}^{(+2)}\hat{S}_{\omega\ell m}^{(+2)}\\
    &=-\frac{\epsilon_g}{4}\zeta^2\pa{\pd_\theta^2+A^{ll}\pd_\theta+B^{ll}}\hat{R}_{\omega\ell m}^{(+2)}\hat{S}_{\omega\ell m}^{(+2)},\\
    H_{\omega\ell m}^{lm}&\equiv\frac{\epsilon_g}{4\sqrt{2}}\frac{\zeta^2}{\ol{\zeta}\Delta}\pa{\msc{D}_0^\dag\msc{L}_2+\frac{a^2\sin{2\theta}}{\Sigma}\msc{D}_0^\dag-\frac{2r}{\Sigma}\msc{L}_2}\Delta^2\hat{R}_{\omega\ell m}^{(+2)}\hat{S}_{\omega\ell m}^{(+2)}\\
    &=\frac{\epsilon_g}{4\sqrt{2}}\frac{\zeta^2}{\ol{\zeta}}\Delta\pa{\pd_\theta\pd_r+A^{lm}\pd_r+B^{lm}\pd_\theta+C^{lm}}\hat{R}_{\omega\ell m}^{(+2)}\hat{S}_{\omega\ell m}^{(+2)},\\
    H_{\omega\ell m}^{mm}&\equiv-\frac{\epsilon_g}{8}\frac{\zeta^2}{\ol{\zeta}^2}\pa{\msc{D}_0^\dag-\frac{2}{\zeta}}\msc{D}_0^\dag\Delta^2\hat{R}_{\omega\ell m}^{(+2)}\hat{S}_{\omega\ell m}^{(+2)}\\
    &=-\frac{\epsilon_g}{8}\frac{\zeta^2}{\ol{\zeta}}\Delta^2\pa{\pd_r^2+A^{mm}\pd_r+B^{mm}}\hat{R}_{\omega\ell m}^{(+2)}\hat{S}_{\omega\ell m}^{(+2)}.
\end{align}
\end{subequations}
Here, the coefficients of the derivatives are
\begin{subequations}
\begin{align}
    A^{ll}&=2Q+3\cot{\theta}-\frac{2ia\sin{\theta}}{\zeta},\\
    B^{ll}&=\pa{Q+2\cot{\theta}}\pa{Q+\cot{\theta}-\frac{2ia\sin{\theta}}{\zeta}}-Q\cot{\theta}-2a\omega\cos{\theta}-2\csc^2{\theta},\\
    A^{lm}&=Q+2\cot{\theta}+\frac{a^2\sin{2\theta}}{\Sigma},\\
    B^{lm}&=\frac{iK+2\Delta'}{\Delta}-\frac{2r}{\Sigma},\\
    C^{lm}&=\pa{\frac{iK+2\Delta'}{\Delta}-\frac{2r}{\Sigma}}\pa{Q+2\cot{\theta}}+\pa{\frac{iK+2\Delta'}{\Delta}}\frac{a^2\sin{2\theta}}{\Sigma},\\
    A^{mm}&=\frac{2iK+4\Delta'}{\Delta}-\frac{2}{\zeta},\\
    B^{mm}&=\frac{iK}{\Delta}\pa{\frac{iK+3\Delta'}{\Delta}-\frac{2}{\zeta}}+\frac{2\Delta'}{\Delta}\pa{\frac{\Delta'}{\Delta}-\frac{2}{\zeta}}+\frac{2i\omega r+4}{\Delta}.
\end{align}
\end{subequations}

\subsection{Explicit metric components using only first derivatives}
\label{subsec:MetricFirstDerivatives}

We now use the radial and angular ODEs \eqref{eq:RadialODE} and \eqref{eq:AngularODE} to remove all (non-mixed) second derivatives.
(The mixed $\pd_r\pd_\theta$ derivatives appearing in the ingoing radiation gauge $H_{\omega\ell m}^{nm}$  and in the outgoing radiation gauge $H_{\omega\ell m}^{lm}$ cannot be removed using these equations of motion.)
Doing so changes the coefficients multiplying the remaining derivatives.
In the ingoing radiation gauge,%
\begin{subequations}
\begin{align}
    H_{\omega\ell m}^{nn}&=-\frac{\epsilon_g}{4\ol{\zeta}^2}\pa{D^{nn}\pd_\theta+E^{nn}}\hat{R}_{\omega\ell m}^{(-2)}\hat{S}_{\omega\ell m}^{(-2)},\\
    H_{\omega\ell m}^{mm} 
    &= -\frac{\epsilon_g}{2} \pa{D^{mm}\pd_r+E^{mm}}\hat{R}^{(-2)}_{\omega\ell m}\hat{S}_{\omega\ell m}^{(-2)},
\end{align}
\end{subequations}
where the new coefficients are
\begin{subequations}
\begin{align}
    D^{nn}&=-2Q+2\cot{\theta}-\frac{2ia\sin{\theta}}{\zeta},\qquad
    D^{mm}=-\frac{2iK-\Delta'}{\Delta}-\frac{2}{\ol{\zeta}},\\
    E^{nn}&=2\pa{Q-2\cot{\theta}}\pa{Q-\cot{\theta}+\frac{ia\sin{\theta}}{\zeta}}-6a\omega\cos{\theta}-\lambda_{\omega\ell m}^{(-2)},\\
    E^{mm}&=\frac{iK}{\Delta}\pa{\frac{2iK-\Delta'}{\Delta}+\frac{2}{\ol{\zeta}}}+\frac{6i\omega r+\lambda_{\omega\ell m}^{(-2)}}{\Delta}.
\end{align}
\end{subequations}
In the outgoing radiation gauge,
\begin{subequations}
\begin{align}
    H_{\omega\ell m}^{ll}&=-\frac{\epsilon_g}{4}\zeta^2\pa{D^{ll}\pd_\theta+E^{ll}}\hat{R}_{\omega\ell m}^{(+2)}\hat{S}_{\omega\ell m}^{(+2)},\\
    H_{\omega\ell m}^{mm}&=-\frac{\epsilon_g}{8}\frac{\zeta^2}{\ol{\zeta}^2}\Delta^2\pa{D^{mm}\pd_r+E^{mm}}\hat{R}_{\omega\ell m}^{(+2)}\hat{S}_{\omega\ell m}^{(+2)},
\end{align}
\end{subequations}
where the new coefficients are 
\begin{subequations}
\begin{align}
    D^{ll}&=2Q+2\cot{\theta}-\frac{2ia\sin{\theta}}{\zeta},\qquad
    D^{mm}=\frac{2iK+\Delta'}{\Delta}-\frac{2}{\zeta},\\
    E^{ll}&=2\pa{Q+2\cot{\theta}}\pa{Q+\cot{\theta}-\frac{ia\sin{\theta}}{\zeta}}+6a\omega\cos{\theta}-\lambda_{\omega\ell m}^{(-2)},\\
    E^{mm}&=\frac{iK}{\Delta}\pa{\frac{2iK+5\Delta'}{\Delta}-\frac{2}{\zeta}}+\frac{2\Delta'}{\Delta}\pa{\frac{\Delta'}{\Delta}-\frac{2}{\zeta}}+\frac{\lambda_{\omega\ell m}^{(-2)}-6i\omega r}{\Delta}.
\end{align}
\end{subequations}

\subsection{Explicit metric components with no derivatives at all}
\label{subsec:MetricNoDerivatives}

Finally, we can entirely remove all the derivatives from the functions $H_{\omega\ell m}^{ab}$ using the identities \eqref{eq:FirstDerivatives}.
In the ingoing radiation gauge, we find
\begin{subequations}
\begin{align}
    H_{\omega\ell m}^{nn}&=-\frac{\epsilon_g}{4\ol{\zeta}^2}\pa{F^{nn}\hat{S}_{\omega\ell m}^{(+2)}+G^{nn}\hat{S}_{\omega\ell m}^{(-2)}}\hat{R}_{\omega\ell m}^{(-2)},\\
    H_{\omega\ell m}^{nm}&=-\frac{\epsilon_g}{2\sqrt{2}\ol{\zeta}}\bigg[\pa{F^{nm}\hat{S}_{\omega\ell m}^{(+2)}+G^{nm}\hat{S}_{\omega\ell m}^{(-2)}}\hat{R}_{\omega\ell m}^{(+2)}
    +\pa{I^{nm}\hat{S}_{\omega\ell m}^{(+2)}+J^{nm}\hat{S}_{\omega\ell m}^{(-2)}}\hat{R}_{\omega\ell m}^{(-2)}\bigg],\\
    H_{\omega\ell m}^{mm}&=-\frac{\epsilon_g}{2}\pa{F^{mm}\hat{R}_{\omega\ell m}^{(+2)}+G^{mm}\hat{R}_{\omega\ell m}^{(-2)}}\hat{S}_{\omega\ell m}^{(-2)},
\end{align}
\end{subequations}
where the new coefficients are
\begin{subequations}
\begin{align}
    F^{nn}&=-\frac{2\hat{D}_{\omega\ell m}'}{\mathbb{X}_{\omega\ell m}}\pa{Q-\cot{\theta}+\frac{ia\sin{\theta}}{\zeta}},\\
    G^{nn}&=\frac{2\mathbb{Y}_{\omega\ell m}}{\mathbb{X}_{\omega\ell m}}\pa{Q-\cot{\theta}+\frac{ia\sin{\theta}}{\zeta}}-6a\omega\cos{\theta}-\lambda_{\omega\ell m}^{(-2)},\\
    F^{nm}&=\frac{\Delta^3\hat{\msc{C}}_{\omega\ell m}\hat{D}'_{\omega\ell m}}{\mathbb{A}_{\omega\ell m}\mathbb{X}_{\omega\ell m}},\\
    G^{nm}&=-\frac{\Delta^3\hat{\msc{C}}_{\omega\ell m}}{\mathbb{A}_{\omega\ell m}}\pa{\frac{\mathbb{Y}_{\omega\ell m}}{\mathbb{X}_{\omega\ell m}}-\frac{a^2\sin{2\theta}}{\Sigma}},\\
    I^{nm}&=-\frac{\hat{D}'_{\omega\ell m}}{\mathbb{X}_{\omega\ell m}}\pa{\frac{\mathbb{C}_{\omega\ell m}}{\mathbb{A}_{\omega\ell m}}+\frac{2r}{\Sigma}},\\
    J^{nm}&=\frac{\mathbb{Y}_{\omega\ell m}}{\mathbb{X}_{\omega\ell m}}\pa{\frac{\mathbb{B}_{\omega\ell m}}{\mathbb{A}_{\omega\ell m}}+\frac{2r}{\Sigma}}-\frac{\mathbb{B}_{\omega\ell m}}{\mathbb{A}_{\omega\ell m}}\frac{a^2\sin{2\theta}}{\Sigma},\\
    F^{mm}&=-\frac{\Delta^3\msc{\hat{C}}_{\omega\ell m}}{\mathbb{A}_{\omega\ell m}}\pa{\frac{2iK-\Delta'}{\Delta}+\frac{2}{\zeta}},\\
    G^{mm}&=\frac{\mathbb{B}_{\omega\ell m}}{\mathbb{A}_{\omega\ell m}}\pa{\frac{2iK-\Delta'}{\Delta}+\frac{2}{\zeta}}+\frac{6i\omega r+\lambda_{\omega\ell m}^{(-2)}}{\Delta}.
\end{align}
\end{subequations}
In the outgoing radiation gauge, we find
\begin{subequations}
\begin{align}
    H_{\omega\ell m}^{ll}&=-\frac{\epsilon_g}{4}\zeta^2\pa{F^{ll}\hat{S}_{\omega\ell m}^{(+2)}+G^{ll}\hat{S}_{\omega\ell m}^{(-2)}}\hat{R}_{\omega\ell m}^{(+2)},\\
    H_{\omega\ell m}^{lm}&=\frac{\epsilon_g}{4\sqrt{2}}\frac{\zeta^2}{\ol{\zeta}}\Delta\br{\pa{F^{lm}\hat{S}_{\omega\ell m}^{(+2)}+G^{lm}\hat{S}_{\omega\ell m}^{(-2)}}\hat{R}_{\omega\ell m}^{(+2)}+\pa{I^{lm}\hat{S}_{\omega\ell m}^{(+2)}+J^{lm}\hat{S}_{\omega\ell m}^{(-2)}}\hat{R}_{\omega\ell m}^{(-2)}},\\
    H_{\omega\ell m}^{mm}&=-\frac{\epsilon_g}{8}\frac{\zeta^2}{\ol{\zeta}^2}\Delta\pa{F^{mm}\hat{R}_{\omega\ell m}^{(+2)}+G^{mm}\hat{R}_{\omega\ell m}^{(-2)}}\hat{S}_{\omega\ell m}^{(+2)}.
\end{align}
\end{subequations}
Here, the new coefficients are
\begin{subequations}
\begin{align}
    F^{ll}&=\frac{2\mathbb{Y}_{-\omega,\ell,-m}}{\mathbb{X}_{\omega\ell m}}\pa{Q+\cot{\theta}-\frac{ia\sin{\theta}}{\zeta}}+6a\omega\cos{\theta}-\lambda_{\omega\ell m}^{(-2)},\\
    G^{ll}&=-\frac{2\hat{D}_{\omega\ell m}}{\mathbb{X}_{\omega\ell m}}\pa{Q+\cot{\theta}-\frac{ia\sin{\theta}}{\zeta}},\\
    F^{lm}&=\pa{\frac{\mathbb{C}_{\omega\ell m}}{\mathbb{A}_{\omega\ell m}}+2\Delta'}\pa{\frac{\mathbb{Y}_{-\omega,\ell,-m}}{\mathbb{X}_{\omega\ell m}}+\frac{a^2\sin{2\theta}}{\Sigma}}-\frac{\mathbb{Y}_{-\omega,\ell,-m}}{\mathbb{X}_{\omega\ell m}}\frac{2r\Delta}{\Sigma},\\
    G^{lm}&=-\frac{\hat{D}_{\omega\ell m}}{\mathbb{X}_{\omega\ell m}}\pa{2\Delta'-\frac{2r\Delta}{\Sigma}+\frac{\mathbb{C}_{\omega\ell m}}{\mathbb{A}_{\omega\ell m}}},\\
    I^{lm}&=-\frac{\hat{\msc{C}}_{\omega\ell m}'}{\mathbb{A}_{\omega\ell m}}\pa{\frac{a^2\sin{2\theta}}{\Sigma}+\frac{\mathbb{Y}_{-\omega,\ell,-m}}{\mathbb{X}_{\omega\ell m}}},\\
    J^{lm}&=\frac{\hat{\msc{C}}_{\omega\ell m}'\hat{D}_{\omega\ell m}}{\mathbb{A}_{\omega\ell m}\mathbb{X}_{\omega\ell m}},\\
    F^{mm}&=\pa{2\Delta'+\frac{\mathbb{C}_{\omega\ell m}}{\mathbb{A}_{\omega\ell m}}}\pa{\frac{2iK+\Delta'}{\Delta}-\frac{2}{\zeta}}-6i\omega r+\lambda_{\omega\ell m}^{(-2)},\\
    G^{mm}&=-\frac{\hat{\msc{C}}_{\omega\ell m}'}{\mathbb{A}_{\omega\ell m}}\pa{\frac{2iK+\Delta'}{\Delta}-\frac{2}{\zeta}}.
\end{align}
\end{subequations}

\section{Confluent Heun functions}
\label{app:ConfluentHeun}

In this appendix, we discuss important properties of the special function $\HeunC(z)$ used to express both the angular modes \eqref{eq:AngularModes} and the radial modes \eqref{eq:RadialModes}.
Throughout, we closely follow Becker's study of the full Heun equation \cite{Becker1997} but adapt it to the confluent case.
First, we present the confluent Heun equation and its local Frobenius solutions in App.~\ref{app:CanonicalHeun}. 
In App.~\ref{app:Wronskian}, we find the functional form of the Wronskian between any two local Frobenius solutions around the two regular singular points $z=0$ and $z=1$. 
Then, in App.~\ref{app:ConfluentHeunFunctions}, we establish the condition that a solution of the confluent Heun equation must satisfy to be a confluent Heun \textit{function}.
In App.~\ref{app:HeunOrthogonality}, we establish orthogonality relations between these confluent Heun functions. 
Lastly, in App.~\ref{app:HeunNormalization}, we derive a formula for the normalization integral of a confluent Heun function.

\subsection{Canonical form of the confluent Heun equation}
\label{app:CanonicalHeun}

The canonical form of the confluent Heun equation with parameters $(q,\alpha,\gamma,\delta,\epsilon)$ is
\begin{align}
    \label{eq:ConfluentHeun}
    \frac{d^2f}{dz^2}+\pa{\frac{\gamma}{z}+\frac{\delta}{z-1}+\epsilon}\frac{df}{dz}+\frac{\alpha z-q}{z(z-1)}f(z)=0.
\end{align}
$q$ is known as the \textit{accessory parameter}.
The full Heun equation has four regular singular points.
This confluent version arises when two of them coalesce, leaving two regular singular points at $z=0$ and $z=1$ and an irregular singular point at $z=\infty$.
As a second-order ODE, it admits two independent solutions.
The lowest powers of $z$ in the series expansions of the two solutions about $z=0$ are 0 and $1-\gamma$.
As for the two series expansions about $z=1$, their lowest powers of $1-z$ are 0 and $1-\delta$.
We let $\HeunC(q,\alpha,\gamma,\delta,\epsilon;z)$ denote the unique solution whose power series expansion around $z=0$ has the leading term $z^0$ with coefficient 1.
This function is hard-coded in \textsc{Mathematica} as \texttt{HeunC}$[q,\alpha,\gamma,\delta,\epsilon,z]$ and is a ``special function'' of mathematical physics.

In terms of this special function, the local Frobenius solutions around $z=0$ are\footnote{\label{fn:Log}When $1-\gamma$ is integer, one of the solutions is replaced by a more complicated expression involving a different power series plus a piece proportional to the other solution multiplied by $\ln(z)$.
If $1-\gamma>0$, then the leading term in the series is $z^0$, and the solution multiplying the logarithm has leading term $z^{1-\gamma}$.
The opposite holds for $1-\gamma<0$.
Similar statements can be made for the solutions around $z=1$ when $1-\delta$ is integer.}
\begin{subequations}
\label{eq:Frobenius0}
\begin{align}
    \label{eq:Frobenius0a}
    &\HeunC(q,\alpha,\gamma,\delta,\epsilon;z)=1-\frac{q}{\gamma}z+\frac{1}{2(\gamma+1)}\pa{\alpha+\frac{q}{\gamma}(q-\gamma-\delta+\epsilon)}z^2+\mc{O}\pa{z^3},\\
    \label{eq:Frobenius0b}
    &z^{1-\gamma}\HeunC\pa{q+(\epsilon-\delta)(1-\gamma),\alpha+\epsilon(1-\gamma),2-\gamma,\delta,\epsilon;z}.
\end{align}
\end{subequations}
Meanwhile, the local Frobenius solutions around $z=1$ are
\begin{subequations}
\label{eq:Frobenius1}
\begin{align}
    \label{eq:Frobenius1a}
    &\HeunC(q-\alpha,-\alpha,\delta,\gamma,-\epsilon;1-z),\\
    \label{eq:Frobenius1b}
    &(1-z)^{1-\delta}\HeunC\pa{q-\alpha-(\epsilon+\gamma)(1-\delta),-\alpha-\epsilon(1-\delta),2-\delta,\gamma,-\epsilon;1-z}.
\end{align}
\end{subequations}

\subsection{Functional form of the Wronskian}
\label{app:Wronskian}

Fixing $\alpha$, $\gamma$, $\delta$, and $\epsilon$, but allowing the accessory parameter $q$ to vary, let $f_0(q,z)$ denote either one of the local Frobenius solutions around $z=0$ given in Eq.~\eqref{eq:Frobenius0}.
Likewise, let $f_1(q,z)$ denote either one of the local Frobenius solutions around $z=1$ given in Eq.~\eqref{eq:Frobenius1}.
Their Wronskian is
\begin{align} 
\label{eq:Wronskian}
    W(q,z)\equiv f_0(q,z)\frac{\pd f_1}{\pd z}(q,z)-f_1(q,z)\frac{\pd f_0}{\pd z}(q,z).
\end{align}
(This definition also holds for any two solutions more generally.)
Following Becker \cite{Becker1997}, we can determine the functional form of this Wronskian.
We start by rewriting the Heun equation as
\begin{align}
    \label{eq:HeunOperator}
    \br{\msc{H}-qw(z)}f=0,
\end{align}
where the specific weight function $w(z)$ associated with the confluent Heun equation \eqref{eq:ConfluentHeun} is
\begin{align}
    w(z)\equiv z^{\gamma-1}(z-1)^{\delta-1}e^{\epsilon z},
\end{align}
while the Heun operator $\msc{H}$ (which Becker denotes by $\msc{L}$) is defined by
\begin{align}
    \msc{H}f\equiv\frac{d}{dz}\pa{p(z)\frac{df}{dz}}+\alpha zw(z)f,\qquad
    p(z)\equiv z^\gamma(z-1)^\delta e^{\epsilon z}.
\end{align}
Since $f_0$ and $f_1$ are solutions of Eq.~\eqref{eq:ConfluentHeun} with the same $q$, it follows from Eq.~\eqref{eq:HeunOperator} that
\begin{align}
    \label{eq:HeunIdentity}
    f_0\br{\msc{H}-qw(z)}f_1- f_1\br{\msc{H}-qw(z)}f_0=0,
\end{align}
which we may rewrite as
\begin{align}
    \frac{d}{dz}\pa{p(z)W(q,z)}=0.
\end{align}
This implies the existence of some function $D(q)$ such that
\begin{align}
    \label{eq:WronskianDecomposition}
    W(q,z)=\frac{D(q)}{p(z)}.
\end{align}

\subsection{Confluent Heun functions}
\label{app:ConfluentHeunFunctions}

Any Frobenius solution around $z=0$ is a linear combination of the two Frobenius solutions around $z=1$, and vice versa.
A solution that is simultaneously Frobenius around both points is a confluent Heun \textit{function}.
Such functions are classified by their exponents at $z=0$ and $1$:
\begin{table}[h]
\centering
\begin{tabular}{| c | c |}
    \hline
        Class I & $(0,0)$ \\
        Class II & $(1-\gamma,0)$ \\
        Class III & $(0,1-\delta)$ \\
        Class IV & $(1-\gamma,1-\delta)$ \\
    \hline
\end{tabular}
\caption{Classes of confluent Heun functions}
\label{tbl:Classes}
\end{table}

Given a choice of parameters $\alpha$, $\gamma$, $\delta$, and $\epsilon$, there is a discrete but infinite set of values $\cu{q_n}$ of the accessory parameter for which a Frobenius solution becomes a confluent Heun function.
There is no closed form for these values, but they can be numerically determined as follows.
When $q = q_n$, the local Frobenius $f_0(q_n,z)$ and $f_1(q_n,z)$ are linearly dependent, so we can write
\begin{align}
    f_0(q_n,z)=A(q_n)f_1(q_n,z),
\end{align}
and as a result, the Wronskian \eqref{eq:Wronskian} vanishes:
\begin{align}
    \label{eq:Quantization}
    W(q_n,z)=0.
\end{align}
The $n^\text{th}$ confluent Heun function associated with the confluent Heun equation \eqref{eq:ConfluentHeun} is 
\begin{align}
    \label{eq:ConfluentHeunFunctions}
    H_n(z)\equiv f_0(q_n,z).
\end{align}

\subsection{Orthogonality of the confluent Heun functions}
\label{app:HeunOrthogonality}

The confluent Heun functions \eqref{eq:ConfluentHeunFunctions} satisfy orthogonality relations, which we can now derive.
By analogy with Eq.~\eqref{eq:HeunIdentity}, these functions obey the identity
\begin{align}
    H_n(z)\br{\msc{H}-q_mw(z)}H_m(z)-H_m(z)\br{\msc{H}-q_nw(z)}H_n(z)=0,
\end{align}
where each term separately vanishes by Eq.~\eqref{eq:HeunOperator}.
We may rewrite this as 
\begin{align} 
    \frac{d}{dz}\br{p(z)\pa{H_m(z)\frac{dH_n}{dz}-H_n(z)\frac{dH_m}{dz}}}=(q_n-q_m)w(z)H_n(z)H_m(z).
\end{align}
Integrating both sides from $z=0$ to $z=1$ yields
\begin{align} 
    (q_n-q_m)\int_0^1w(z)H_n(z)H_m(z)\ed z=p(z)\left.\pa{H_m(z)\frac{dH_n}{dz}-H_n(z)\frac{dH_m}{dz}}\right|_0^1.
\end{align}
By examining the behavior of the confluent Heun function near the regular singular points $z=0$ and $z=1$, we can see that both the $z=0$ and $z=1$ pieces of the right-hand side vanish, provided that the following class-dependent \textit{existence conditions} are met:
\begin{table}[h]
\centering
\begin{tabular}{| c | c c |}
    \hline
        Class I & $\re[\gamma]>0$ & $\re[\delta]>0$ \\
        Class II & $\re[\gamma]<2$ & $\re[\delta]>0$ \\
        Class III & $\re[\gamma]>0$ & $\re[\delta]<2$ \\
        Class IV & $\re[\gamma]<2$ & $\re[\delta]<2$ \\
    \hline
\end{tabular}
\caption{Existence conditions for orthogonal confluent Heun functions}
\label{tbl:ExistenceConditions}
\end{table}

If the existence conditions are met, then we obtain the following orthogonality relation:
\begin{align} 
    (q_n-q_m)\int_0^1w(z)H_n(z)H_m(z)\ed z=0.
\end{align}

\subsection{Normalization of the confluent Heun functions}
\label{app:HeunNormalization}

We can also derive a formula for the normalization integral
\begin{align} 
    I_n\equiv\int_0^1w(z)\br{H_n(z)}^2\ed z.
\end{align}
Let $q$ denote an arbitrary accessory parameter, and let $q_n$ satisfy the quantization condition \eqref{eq:Quantization}.
Then $f_0(q_n,z)$ is a confluent Heun \textit{function}, while $f_0(q,z)$ is not, even though they both solve the confluent Heun equation \eqref{eq:ConfluentHeun} with respect to their individual accessory parameters.
Applying Eq.~\eqref{eq:HeunOperator} to each of them implies that
\begin{align} 
    f_0(q,z)\br{\msc{H}-q_nw(z)}f_0(q_n,z)-f_0(q_n,z)\br{\msc{H}-qw(z)}f_0(q,z)=0,
\end{align}
which may be rewritten as
\begin{align} 
    \frac{d}{dz}\br{p(z)\pa{f_0(q_n,z)\frac{\pd f_0}{\pd z}(q,z)-f_0(q,z)\frac{\pd f_0}{\pd z}(q_n,z)}}=(q-q_n)w(z)f_0(q,z)f_0(q_n,z).
\end{align}
Integrating both sides with respect to $z$ then yields
\begin{align} 
    (q-q_n)\int_0^zw(\zeta)f_0(q,\zeta)f_0(q_n,\zeta)\ed\zeta=p(z)\pa{f_0(q_n,z)\frac{\pd f_0}{\pd z}(q,z)-f_0(q,z)\frac{\pd f_0}{\pd z}(q_n,z)},
\end{align}
where we have used the fact that the right-hand side vanishes at $z=0$ provided that the existence conditions in Table~\ref{tbl:ExistenceConditions} are satisfied.
It is clear that both sides of this equation vanish at $q = q_n$, so we can use L'H\^{o}pital's rule to determine the value of the integral at $q = q_n$:
\begin{subequations}
\begin{align}
    \int_0^zw(\zeta)\br{f_0(q_n,\zeta)}^2\ed\zeta&=\lim_{q\to q_n}\br{\frac{p(z)}{q-q_n}\pa{f_0(q_n,z)\frac{\pd f_0}{\pd z}(q,z)-f_0(q,z)\frac{\pd f_0}{\pd z}(q_n,z)}}\\
    &=\lim_{q\to q_n}\br{p(z)\pa{f_0(q_n,z)\frac{\pd^2f_0}{\pd q\pd z}(q,z)-\frac{\pd f_0}{\pd q}(q,z)\frac{\pd f_0}{\pd z}(q_n,z)}}\\
    &=p(z)\pa{f_0(q_n,z)\frac{\pd^2f_0}{\pd q\pd z}(q_n,z)-\frac{\pd f_0}{\pd q}(q_n,z)\frac{\pd f_0}{\pd z}(q_n,z)}.
\end{align}
\end{subequations}
To determine a formula for the normalization integral, we must take $z\to1$ in this expression.
This naturally requires an expression for $f_0(q,z)$ in the neighborhood of $z=1$.
As mentioned previously, $f_0(q,z)$ must be some linear combination of the Frobenius solutions around $z=1$:
\begin{align}
    \label{eq:FrobeniusSolutions}
    f_0(q,z)=A(q)f_1(q,z)+B(q)\wt{f}_1(q,z),
\end{align}
where $f_1(q,z)$ denotes the Frobenius solution \eqref{eq:Frobenius1a} with exponent 0, and $\wt{f}_1(q,z)$ denotes the Frobenius solution \eqref{eq:Frobenius1b} with exponent $1-\delta$.
For the time being, we now restrict our attention to confluent Heun functions of class I or II, for which $B(q_n)=0$.
Then we find that
\begin{align}
    I_n&=\lim_{z\to1}\left.\cu{p(z)\br{Af_1\pa{\frac{dA}{dq}\frac{\pd f_1}{\pd z}+A\frac{\pd^2 f_1}{\pd q\pd z}+\frac{dB}{dq}\frac{\pd\wt{f}_1}{\pd z}}-A\frac{\pd f_1}{\pd z}\pa{\frac{dA}{dq}f_1+A\frac{\pd f_1}{\pd q}+\frac{dB}{dq}\wt{f}_1}}}\right|_{q=q_n}\notag\\
    &=\lim_{z\to1}\left.\cu{p(z)\br{Af_1\pa{A\frac{\pd^2 f_1}{\pd q\pd z}+\frac{dB}{dq}\frac{\pd\wt{f}_1}{\pd z}}-A\frac{\pd f_1}{\pd z}\pa{A\frac{\pd f_1}{\pd q}+\frac{dB}{dq}\wt{f}_1}}}\right|_{q=q_n}.
\end{align}
Since $f_1\stackrel{z\to1}{\sim}(z-1)^0$ and $\wt{f}_1\stackrel{z\to1}{\sim}(z-1)^{1-\delta}$, we find that every term but the second vanishes in the $z\to1$ limit, provided that the existence condition on $\delta$ is met (that is, $\re[\delta]>0$).
Thus,
\begin{align}
    \label{eq:IntermediateNormalization}
    I_n=\lim_{z\to1}\left.\pa{p(z)A\frac{dB}{dq}f_1\frac{\pd\wt{f}_1}{\pd z}}\right|_{q=q_n}.
\end{align}
The class I or II condition $B(q_n)=0$ together with Eq.~\eqref{eq:FrobeniusSolutions} imply that
\begin{align}
    \label{eq:HeunProportionality}
    A(q_n)=\frac{f_0(q_n,z)}{f_1(q_n,z)},
\end{align}
so we focus on the $\frac{dB}{dq}$ piece next.
Differentiating Eq.~\eqref{eq:FrobeniusSolutions} with respect to $z$ yields
\begin{align} 
    \frac{\pd f_0}{\pd z}(q,z)=A(q)\frac{\pd f_1}{\pd z}(q,z)+B(q)\frac{\pd\wt{f}_1}{\pd z}(q,z).
\end{align}
We can now invoke Eq.~\eqref{eq:FrobeniusSolutions} to eliminate $A(q)$ from this equation and then solve for $B(q)$:
\begin{align} 
    B(q)=\frac{f_0\frac{\pd f_1}{\pd z}-f_1\frac{\pd f_0}{\pd z}}{\wt{f}_1\frac{\pd f_1}{\pd z}-f_1\frac{\pd\wt{f}_1}{\pd z}}
    =\frac{W(q,z)}{\wt{f}_1\frac{\pd f_1}{\pd z}-f_1\frac{\pd\wt{f}_1}{\pd z}}.
\end{align}
Recalling that $W(q_n,z)=0$, if we differentiate with respect to $q$ and then set $q=q_n$, we find
\begin{subequations}
\begin{align} 
    \frac{dB}{dq}(q_n)&=\left.\frac{\pa{\wt{f}_1\frac{\pd f_1}{\pd z}-f_1\frac{\pd\wt{f}_1}{\pd z}}\frac{\pd W}{\pd q}-W(q,z)\frac{\pd}{\pd q}\pa{\wt{f}_1\frac{\pd f_1}{\pd z}-f_1\frac{\pd\wt{f}_1}{\pd z}}}{\pa{\wt{f}_1\frac{\pd f_1}{\pd z}-f_1\frac{\pd\wt{f}_1}{\pd z}}^2}\right|_{q=q_n}\\
    &=\left.\frac{\pd W}{\pd q}\pa{\wt{f}_1\frac{\pd f_1}{\pd z}-f_1\frac{\pd\wt{f}_1}{\pd z}}^{-1}\right|_{q=q_n}.
\end{align}
\end{subequations}
We can now replace both $A$ and $\frac{dB}{dq}$ in Eq.~\eqref{eq:IntermediateNormalization} to obtain
\begin{align} 
    I_n=\left.\lim_{z\to1}\br{p(z)f_0\frac{\pd\wt{f}_1}{\pd z}\frac{\pd W}{\pd q}\frac{1}{\wt{f}_1\frac{\pd f_1}{\pd z}-f_1\frac{\pd\wt{f}_1}{\pd z}}}\right|_{q=q_n}.
\end{align}
By Eq.~\eqref{eq:WronskianDecomposition}, $p(z)\frac{\pd W}{\pd q}=D'(q)$ is independent of $z$.
Using the known behavior of $f_1$ and $\wt{f}_1$ in the $z \to 1$ limit, we find that the $\frac{\pd\wt{f}_1}{\pd z}$ term in the denominator dominates, and so we obtain
\begin{align} 
    \label{eq:HeunNormalization}
    I_n=-p(z)\frac{\pd W}{\pd q}(q_n,z)\frac{f_0(q_n,z)}{f_1(q_n,z)},
\end{align}
where we were able to drop the limit because both $p(z)\frac{\pd W}{\pd q}$ and $\frac{f_0(q_n,z)}{f_1(q_n,z)}$ are independent of $z$.
Indeed, recalling Eqs.~\eqref{eq:WronskianDecomposition} and \eqref{eq:HeunProportionality}, this can also be expressed as
\begin{align} 
    I_n=-D'(q_n)A(q_n),
\end{align}
which drives home that each piece is a constant (that is, independent of $z$).

This argument can be extended to confluent Heun functions of class III or IV, essentially exchanging $A$ and $B$ in the above derivation.
Becker \cite{Becker1997} worked this procedure for the full Heun equation, resulting in an expression identical to Eq.~\eqref{eq:HeunNormalization}, but with $f_1$ now representing the local Frobenius solution with exponent $1-\delta$ (provided the existence conditions are satisfied).

\section{Angular modes as confluent Heun functions}
\label{app:AngularHeun}

In this appendix, we recast the angular modes \eqref{eq:AngularModes} as confluent Heun functions, which allows us to apply some of the results derived in App.~\ref{app:ConfluentHeun}.
First, we map the angular ODE \eqref{eq:AngularODE} to the confluent Heun equation in App.~\ref{app:HeunAngularODE}.
Then, in App.~\ref{app:AngularBoundaryConditions}, we impose physical boundary conditions on its solutions to derive the spectrum of angular modes \eqref{eq:AngularModes} and their separation constants.
Finally, we determine their normalization in App.~\ref{app:AngularNormalization} using the formulas of App.~\ref{app:HeunNormalization}.

\subsection{Mapping the angular ODE to the confluent Heun equation}
\label{app:HeunAngularODE}

In this section, we closely follow Borissov and Fiziev \cite{Borissov2009} and map the angular ODE \eqref{eq:AngularODE} to the confluent Heun equation.
First, we change to a variable $u=\cos{\theta}$, such that Eq.~\eqref{eq:AngularODE} becomes
\begin{align}
    \label{eq:AngularVariableChange}
    \cu{\frac{d}{du}\br{\pa{1-u^2}\frac{d}{du}}+a^2\omega^2u^2-\frac{(m+su)^2}{1-u^2}-2sa\omega u+s+A}S_{\omega\ell m}^{(s)}(u)=0. 
\end{align}
This equation has two regular singular points at $u=\pm1$ and an irregular singular point at $u=\infty$.
This means that via a change of variable and a field redefinition, it can be mapped to the canonical form \eqref{eq:ConfluentHeun} of the confluent Heun equation, for some parameters $\gamma$, $\delta$, $\epsilon$, $\alpha$, and $q$.
To put it into this canonical form, we first perform the redefinition\footnote{The $+1$ in the exponent has no effect on what follows and simply scales the function by a constant.
We add it here to match Eq.~\eqref{eq:AngularModes} and prevent factors of $e^{a\omega}$ from appearing in many equations.}
\begin{align}
    \label{eq:AngularCosineMode}
    S_{\omega\ell m}^{(s)}(u)=(1-u)^{\mu_1}(1+u)^{\mu_2}e^{\mu_3(u+1)}H(u),
\end{align}
where the $\mu_i$ are yet to be fixed.
This transformation turns Eq.~\eqref{eq:AngularVariableChange} into
\begin{align}
    \label{eq:IntermediateHeun}
    &H''(u)+\pa{\frac{2\mu_1+1}{u-1}+\frac{2\mu_2+1}{u+1}+2\mu_3}H'(u)\notag\\
    &+\pa{\frac{\mu_1^2-\frac{1}{4}(m+s)^2}{(u-1)^2}+\frac{\mu_2^2-\frac{1}{4}(m-s)^2}{(u+1)^2}+\mu_3^2-a^2\omega^2+\frac{p+\beta u}{(u-1)(u+1)}}H(u)=0, 
\end{align}
where
\begin{subequations}
\begin{gather}
    \beta=2\br{sa\omega+\mu_3(\mu_1+\mu_2+1)},\\
    p=-\lambda_{\omega\ell m}^{(s)}-2am\omega+2\mu_3(\mu_1-\mu_2)+2\mu_1 \mu_2+\mu_1+\mu_2+\frac{m^2-s^2}{2}-s.
\end{gather}
\end{subequations}
To match onto the canonical form \eqref{eq:ConfluentHeun} of the confluent Heun equation, we must cancel out the first three terms on the second line of Eq.~\eqref{eq:IntermediateHeun} by choosing
\begin{align}
    \label{eq:MuConditions}
    \mu_1=\pm\frac{s+m}{2},\qquad
    \mu_2=\pm\frac{s-m}{2},\qquad
    \mu_3=\pm a\omega.
\end{align}
At this point, we have not imposed any boundary conditions on the solution $S_{\omega\ell m}^{(s)}$, so any choice of $\mu_i$ obeying the conditions \eqref{eq:MuConditions} is suitable; we will handle the boundary conditions below.
We note that the conditions \eqref{eq:MuConditions} can be used to rewrite $p$ in the simpler form
\begin{align}
    \label{eq:HeunP}
    p=-\lambda_{\omega\ell m}^{(s)}-2am\omega+2\mu_3(\mu_1-\mu_2)+ (\mu_1+\mu_2)^2+\mu_1+\mu_2-s(s+1).
\end{align}
To finally recover precisely the canonical form \eqref{eq:ConfluentHeun} of the confluent Heun equation, we still have do a M\"obius transformation $u\to z$ to map the regular singular points $u=\pm1$ to their canonical positions $z=0$ and $z=1$.
There are two natural choices:
\begin{align}
    u\to z_+=\frac{1+u}{2},\qquad
    u\to z_-=\frac{1-u}{2}.
\end{align}
The map to $z_+$ sends the north pole $\theta=0$ to $z_+=1$ and the south pole $\theta=\pi$ to $z_+=0$, while the map to $z_-$ does the reverse.
Both choices transform Eq.~\eqref{eq:IntermediateHeun} into Eq.~\eqref{eq:ConfluentHeun},
\begin{align}
    \label{eq:zAngularHeun}
    \frac{d^2H}{dz_\pm^2}+\pa{\frac{\gamma_\pm}{z_\pm}+\frac{\delta_\pm}{z_\pm-1}+\epsilon_\pm}\frac{dH}{dz_\pm}+\frac{\alpha_\pm z_\pm-q_\pm}{z_\pm(z_\pm-1)}H(z_\pm)=0,
\end{align}
but with different parameters:
\begin{subequations}
\label{eq:AngularHeunParameters}
\begin{gather}
    \label{eq:HeunGammaDelta}
    \gamma_+=\delta_-=2\mu_2+1,\qquad 
    \delta_+=\gamma_-=2\mu_1+1,\qquad
    \epsilon_\pm=\pm 4\mu_3,\\
    \label{eq:HeunAlphaAndQ}
    \alpha_\pm=\pm 2\beta,\qquad
    q_\pm=-p\pm\beta.
\end{gather}
\end{subequations}

\subsection{Boundary conditions and quantization of the mode spectrum}
\label{app:AngularBoundaryConditions}

The variables $z_+$ and $z_-$ are related by $u\to-u$, or $\theta\to\pi-\theta$.
In other words, they are related by a reflection through the equatorial plane $\theta=\frac{\pi}{2}$.
Since the Kerr metric \eqref{eq:Kerr} enjoys an equatorial reflection symmetry, it follows that the choice of either $z_+$ or $z_-$ is arbitrary, so from now on, we consider only the mapping to $z_+$ and drop the $+$ subscript.

The angular modes $S_{\omega\ell m}^{(s)}(z)$ must be regular at both the north pole and the south pole on the sphere.
At the south pole $z=0$, $H(z)$ must be a linear combination of the local Frobenius solutions \eqref{eq:Frobenius0a} and \eqref{eq:Frobenius0b}, which we now denote by $f_0(z)$ and $\wt{f}_0(z)$, respectively.
Plugging in $H(z)=Af_0(z)+B\wt{f}_0(z)$ into Eq.~\eqref{eq:AngularCosineMode} and expanding around $z=0$, we find
\begin{align}
    \label{eq:BoundaryConditions0}
    S_{\omega\ell m}^{(s)}(z)\stackrel{z\to0}{\sim}z^{\mu_2}\pa{Az^0+Bz^{-2\mu_2}}.
\end{align}
If $\mu_2>0$, then regularity requires $B=0$.
Conversely, if $\mu_2<0$, then regularity requires $A=0$.
If $\mu_2=0$, then one of the solutions is logarithmically divergent and must be discarded.
Either way, regularity at $z=0$ forces $H(z)$ to precisely equal one of the local Frobenius solutions \eqref{eq:Frobenius0}.

Likewise, at the north pole $z=1$, $H(z)$ must be a linear combination of the local Frobenius solutions \eqref{eq:Frobenius1a} and \eqref{eq:Frobenius1b}, which we now denote by $f_1(z)$ and $\wt{f}_1(z)$, respectively.
Plugging in $H(z)=Cf_1(z)+D\wt{f}_1(z)$ into Eq.~\eqref{eq:AngularCosineMode} and expanding around $z=1$, we find
\begin{align}
    \label{eq:BoundaryConditions1}
    S_{\omega\ell m}^{(s)}(z)\stackrel{z\to1}{\sim}(1-z)^{\mu_1}\br{C(1-z)^0+D(1-z)^{-2\mu_1}}.
\end{align}
If $\mu_1>0$, then regularity requires $D=0$.
Conversely, if $\mu_1<0$, then regularity requires $C=0$.
If $\mu_1=0$, then one of the solutions is logarithmically divergent and must be discarded.
Either way, regularity at $z=1$ forces $H(z)$ to precisely equal one of the local Frobenius solutions \eqref{eq:Frobenius1}.

Putting these two statements together, we see that the regularity conditions on $S_{\omega\ell m}^{(s)}(z)$ force $H(z)$ to be a confluent Heun \textit{function}, as described in App.~\ref{app:ConfluentHeunFunctions}.

At this stage, we can choose $H(z)$ to lie in any one of the four classes of confluent Heun functions listed in Table~\ref{tbl:Classes}.
The simplest choice is to take $H(z)$ to be of class I, so that it goes as $z^0$ and $(1-z)^0$ at the two singular points.
In that case, $B=D=0$ in Eqs.~\eqref{eq:BoundaryConditions0} and \eqref{eq:BoundaryConditions1}.
As a result, we must choose $\mu_1\ge0$ and $\mu_2\ge0$, which fixes their signs in Eq.~\eqref{eq:MuConditions}:\footnote{Choosing $H(z)$ to be of class II, III, or IV leads to the other three possible sign combinations for $\mu_1$ and $\mu_2$.}
\begin{align} 
    \label{eq:MuChoice}
    \mu_1=\frac{|s+m|}{2},\qquad
    \mu_2=\frac{|s-m|}{2}.
\end{align}
With this choice, the Heun parameters $\gamma$ and $\delta$ given in Eq.~\eqref{eq:HeunGammaDelta} obey the existence conditions given in Table.~\ref{tbl:ExistenceConditions}.
Therefore, the confluent Heun functions $H(z)$ appearing in the modes $S_{\omega\ell m}^{(s)}(z)$ are orthogonal and obey the normalization derived in App.~\ref{app:HeunNormalization} and given in App.~\ref{app:AngularNormalization} below.

The choice of sign for $\mu_3$ in Eq.~\eqref{eq:MuConditions} changes the behavior of $H(z)$ at the irregular singular point $z=\infty$, which does not lie on the sphere.
It is therefore arbitrary, and we choose $\mu_3=+a\omega$.

As discussed in App.~\ref{app:ConfluentHeunFunctions}, a given solution of the confluent Heun equation \eqref{eq:ConfluentHeun} is a confluent Heun \textit{function} if and only if its accessory parameter $q$ belongs to the infinite but discrete set of values $\cu{q_n}$ such that the Wronskian \eqref{eq:Wronskian} vanishes.
Since the accessory parameter $q$ in Eq.~\eqref{eq:HeunAlphaAndQ} is related to the separation constant $\lambda_{\omega\ell m}^{(s)}$ via the parameter $p$ given in Eq.~\eqref{eq:HeunP}, it follows that the quantization condition $q\in\cu{q_n}$ imposes a quantization condition on the separation constant $\lambda_{\omega\ell m}^{(s)}$, whose discrete values we label using $\ell$ rather than $n$.

As shown in App.~\ref{app:ConfluentHeunFunctions}, the quantization condition on $q$, or equivalently $\lambda_{\omega\ell m}^{(s)}$, is the vanishing of the Wronskian
\begin{align}
    \label{eq:ExplicitWronskian}
    f_0(q_n,z)\frac{\pd f_1}{\pd z}(q_n,z)-f_1(q_n,z)\frac{\pd f_0}{\pd z}(q_n,z)=0,
\end{align}
where for a confluent Heun function of class I,
\begin{subequations}
\label{eq:ClassI}
\begin{align}
    f_0(q,z)&=\HeunC(q,\alpha,\gamma,\delta,\epsilon;z),\\
    f_1(q,z)&=\HeunC(q-\alpha,-\alpha,\delta,\gamma,-\epsilon;1-z).
\end{align}
\end{subequations}
In summary, the boundary conditions (and our choices) fix the angular modes $S_{\omega\ell m}^{(s)}(z)$ to take the form \eqref{eq:AngularCosineMode} with $\mu_1$ and $\mu_2$ given in Eq.~\eqref{eq:MuChoice}, $\mu_3=a\omega$, and $H(z)=\HeunC(q,\alpha,\gamma,\delta,\epsilon;z)$ a confluent Heun function with parameters given with a $+$ in Eq.~\eqref{eq:AngularHeunParameters}.
Moreover, the accessory parameter $q$, and hence the separation constant $\lambda_{\omega\ell m}^{(s)}$, are quantized via the condition \eqref{eq:ExplicitWronskian}.

\subsection{Normalization of the angular modes}
\label{app:AngularNormalization}

Since the confluent Heun functions $H(z)$ in the modes $S_{\omega\ell m}^{(s)}(z)$ obey the existence conditions given in Table.~\ref{tbl:ExistenceConditions}, we can compute the normalization constants $I_{\omega\ell m}^{(s)}$, defined in Eq.~\eqref{eq:Orthogonality} as
\begin{align} 
    I_{\omega\ell m}^{(s)}\equiv\int_0^\pi\br{\hat{S}_{\omega\ell m}^{(s)}(\theta)}^2\sin{\theta}\ed\theta
    =C\int_0^1w(z)[H(z)]^2\ed z,
\end{align}
where $H(z)$ is given in Eq.~\eqref{eq:AngularHeun}, while
\begin{align}
    C=2^{|s+m|+|s-m|+1}
    =2^{2\max\pa{|s|,|m|}+1}.
\end{align}
The results of App.~\ref{app:HeunNormalization} enable us to express this normalization as
\begin{subequations}
\begin{align}
    I_{\omega\ell m}^{(s)}&=-Cp(z)\frac{dW}{dq}(q_n,z)\frac{f_0(q_n,z)}{f_1(q_n,z)}\\
    &=-Cp(z)\left.\frac{d}{dq}\pa{f_0\frac{\pd f_1}{\pd z}-f_1\frac{\pd f_0}{\pd z}}\right|_{q=q_n}\frac{f_0(q_n,z)}{f_1(q_n,z)},
\end{align}
\end{subequations}
where $f_0(q,z)$ and $f_1(q,z)$ are given in Eq.~\eqref{eq:ClassI}, and a $q$ without an $n$ subscript is \textit{not} assumed to be a zero of the Wronskian (otherwise the derivative with respect to $q$ would be meaningless).

\section{Radial modes as confluent Heun functions}
\label{app:RadialHeun}

In this appendix, we recast the radial modes \eqref{eq:RadialModes} as solutions of the confluent Heun equation \eqref{eq:ConfluentHeun}.
We note, however, these these solutions are generically \textit{not} confluent Heun \textit{functions} as defined in App.~\ref{app:ConfluentHeunFunctions}.
After mapping the radial ODE \eqref{eq:RadialODE} to the confluent Heun equation in App.~\ref{app:HeunRadialODE}, we derive the radial Teukolsky--Starobinsky constants \eqref{eq:ConstantsC} in App.~\ref{app:RTSDerivation} using an elegant method developed by Ori \cite{Ori2003}.
His method allows us to compute not only the constants associated with the ``in'' and ``out'' modes \eqref{eq:RadialModes}, but also those of the ``up'' and ``down'' modes.

\subsection{Mapping the radial ODE to the confluent Heun equation}
\label{app:HeunRadialODE}

In this section, we closely follow Borissov and Fiziev \cite{Borissov2009} and map the radial ODE \eqref{eq:RadialODE} to the confluent Heun equation.
This equation has two regular singular points at $r=r_\pm$ and an irregular singular point at $r=\infty$.
Once again, this means that via a change of variable and a field redefinition, it can be mapped to the canonical form \eqref{eq:ConfluentHeun} of the confluent Heun equation, for some parameters $\gamma$, $\delta$, $\epsilon$, $\alpha$, and $q$.
To put it into this canonical form, we first transform
\begin{align}
    R_{\omega\ell m}^{(s)}(r)=\pa{r-r_+}^{\xi_1}\pa{r-r_-}^{\xi_2}e^{\xi_3r}H(r),
\end{align}
where the $\xi_i$ are yet to be fixed.
This transformation turns Eq.~\eqref{eq:RadialODE} into
\begin{align}
    \label{eq:RadialHeun}
    &H''(r)+\pa{\frac{2\xi_1+s+1}{r-r_+}+\frac{2\xi_2+s+1}{r-r_-}+2\xi_3}H'(r)\notag\\
    &+\left\{\xi_3^2+\omega^2+\frac{1}{\pa{r-r_+}^2}\br{\xi_1-ic_+\pa{\omega-m\Omega_+}}\br{\xi_1+s+ic_+\pa{\omega-m\Omega_+}}\right.\notag\\
    &\qquad+\frac{1}{\pa{r-r_-}^2}\br{\xi_2+ic_-\pa{\omega-m\Omega_-}}\br{\xi_2+s-ic_-\pa{\omega-m\Omega_-}}\notag\\
    &\qquad\left.+\frac{1}{\pa{r-r_+}\pa{r-r_-}}\pa{2\beta r+p}\right\}H(r)
    =0,
\end{align}
where
\begin{subequations}
\begin{gather}
    c_\pm=\frac{2M r_\pm}{r_+-r_-},\qquad
    \Omega_\pm=\frac{a}{2Mr_\pm},\qquad
    \beta=2\omega^2M+is\omega+\xi_3\pa{\xi_1+\xi_2+s+1},\\
    p=-\lambda_{\omega\ell m}^{(s)}-2am\omega+4M^2\omega^2-c_+^2(\omega-m\Omega_+)^2-c_-^2(\omega-m\Omega_-)^2\notag\\
    \phantom{=}\quad\,+(s+1)(\xi_1+\xi_2)+2\xi_1\xi_2-2\xi_3\pa{\xi_1r_-+\xi_2r_+}-2\xi_3M(s+1).
\end{gather}
\end{subequations}
To match onto the canonical form \eqref{eq:ConfluentHeun} of the confluent Heun equation, we must cancel out the second and third lines of Eq.~\eqref{eq:RadialHeun} by choosing
\begin{subequations}
\label{eq:XiConstraints}
\begin{gather}
    \xi_1=ic_+\pa{\omega-m\Omega_+}
    \qquad\text{or}\qquad 
    \xi_1=-s-ic_+\pa{\omega-m\Omega_+},\\
    \xi_2=-ic_-\pa{\omega-m\Omega_-}
    \qquad\text{or}\qquad 
    \xi_2=-s+ic_-\pa{\omega-m\Omega_-},\\
    \xi_3=\pm i\omega.
\end{gather}
\end{subequations}
These conditions can be used to rewrite $p$ in the simpler form
\begin{align}
    p&=-\lambda_{\omega\ell m}^{(s)}-2is\omega M-2am\omega+4\omega^2M^2+\pa{\xi_1+\xi_2}^2+(2s+1)(\xi_1+\xi_2)\notag\\&
    \phantom{=}\ -2\xi_3\pa{\xi_1r_-+\xi_2r_+}-2\xi_3M(s+1).
\end{align}
To finally recover precisely the canonical form \eqref{eq:ConfluentHeun} of the confluent Heun equation, we still have do a M\"obius transformation $r\to z$ to map the regular singular points $r=r_\pm$ to their canonical positions $z=0$ and $z=1$.
Once again, there are two natural choices:
\begin{align}
    r\to z_+=-\frac{r-r_+}{r_+-r_-},\qquad
    r\to z_-=\frac{r-r-}{r_+-r_-}.
\end{align}
The map to $z_+$ sends the outer horizon $r=r_+$ to $z_+=0$ and the inner horizon $r=r_-$ to $z_+=1$, while the map to $z_-$ does the reverse.
Both choices transform Eq.~\eqref{eq:RadialHeun} into Eq.~\eqref{eq:ConfluentHeun},
\begin{align}
    \label{eq:RadialHeunBis}
    \frac{d^2H}{dz_\pm^2}+\pa{\frac{\gamma_\pm}{z_\pm}+\frac{\delta_\pm}{z_\pm-1}+\epsilon_\pm}\frac{dH}{dz_\pm}+\frac{\alpha_\pm z_\pm-q_\pm}{z_\pm(z_\pm-1)}H(z_\pm)=0,
\end{align}
but with different parameters:
\begin{subequations}
\begin{gather}
    \gamma_+=\delta_-
    =2\xi_1+s+1,\qquad 
    \delta_+=\gamma_-
    =2\xi_2+s+1,\qquad
    \epsilon_\pm=\mp 2(r_+-r_-)\xi_3,\\
    \alpha_\pm=\mp 2(r_+-r_-)\beta,\qquad
    q_\pm=-2r_\pm\beta-p.
\end{gather}    
\end{subequations}
\filbreak
\noindent In the absence of additional boundary conditions, Eq.~\eqref{eq:RadialHeunBis} has two mode solutions, which take the simple forms \eqref{eq:Frobenius0} when expanded around $z_\pm=0$.
If we map $r\to z_+$, then we obtain the ``in'' and ``out'' radial modes \eqref{eq:RadialModes}, which have a definite behavior at the outer horizon $z_+=0$.

\subsection{Derivation of the radial Teukolsky--Starobinsky constants}
\label{app:RTSDerivation}

Finally, we derive the constants appearing in the radial Teukolsky--Starobinsky identities \eqref{eq:RTSI1},%
\begin{subequations}
\begin{align}
    \msc{D}_0^4\hat{R}_{\omega\ell m}^{(-2)\,{\rm in/out}}&=\hat{\msc{C}}_{\omega\ell m}^{\rm in/out}\hat{R}_{\omega\ell m}^{(+2)\,{\rm in/out}},\\
    \Delta^2\pa{\msc{D}_0^\dag}^4\Delta^2\hat{R}_{\omega\ell m}^{(+2)\,{\rm in/out}}&=\hat{\msc{C}}_{\omega\ell m}^{{\rm in/out}\,\prime}
    \hat{R}_{\omega\ell m}^{(-2)\,{\rm in/out}}.
\end{align}
\end{subequations}
As discussed in Sec.~\ref{subsec:TeukolskyStarobinsky}, their product $\mc{C}_{\omega\ell m}=\hat{\msc{C}}_{\omega\ell m}\hat{\msc{C}}_{\omega\ell m}^\prime$ can be determined from the second form \eqref{eq:RTSI2} of the radial Teukolsky--Starobinsky identities and is independent of the choice of modes.
Here, we compute $\hat{\msc{C}}_{\omega\ell m}$ and $\hat{\msc{C}}_{\omega\ell m}'$ for the particular modes we have chosen in Eq.~\eqref{eq:RadialModes}. 

Ori \cite{Ori2003} developed an elegant method for carrying out this computation.
The idea is to exploit the known behavior \eqref{eq:HorizonBehavior} of the ``in'' and ``out'' modes near the outer horizon:\footnote{Caveat lector: What we call the ``in'' and ``out'' modes, Ori calls ``down'' and ``up'' modes, and vice versa.}
\begin{subequations}
\label{eq:HorizonBehaviorBis}
\begin{align}
    \hat{R}_{\omega\ell m}^{(s)\,{\rm in}}(r)&\stackrel{r\to r_+}{\approx}c^{\rm in}\Delta^{-s}e^{-ikr_*},
    &&c^{\rm in}=\br{e^{ikr_+}\frac{\pa{r_+-r_-}^{-ic_-k}}{\pa{r_++r_-}^{2ikM}}},\\
    \hat{R}_{\omega\ell m}^{(s)\,{\rm out}}(r)&\stackrel{r\to r_+}{\approx}c^{\rm out}e^{ik r_*},
    &&c^{\rm out}=\br{e^{-ikr_+}\frac{\pa{r_+-r_-}^{ic_-k}}{\pa{r_++r_-}^{-2ikM}}},
\end{align}
\end{subequations}
where $r_*$ is the tortoise coordinate \eqref{eq:TortoiseCoordinate}.
Near the horizon, the operators \eqref{eq:Dn} simplify to 
\begin{align}
    \msc{D}_0\approx\frac{2Mr_+}{\Delta}\pa{\pd_{r_*}-ik},\qquad
    \msc{D}_0^\dag\approx\frac{2Mr_+}{\Delta}\pa{\pd_{r_*}+ik}.
\end{align}
It follows that at leading order near the outer horizon, we have in terms of $w=4Mkr_+$:\footnote{This corrects Ori's version of these equations, appearing between Eqs.~(33) and (34), which are missing terms.}
\begin{subequations}
\begin{align}
    \msc{D}_0\pa{f(r)e^{ikr_*}}&\approx f'(r)e^{ikr_*},\\
    \msc{D}_0\pa{f(r)e^{-ikr_*}}&\approx\pa{f'(r)-\frac{iw}{\Delta}f(r)}e^{-ikr_*},\\
    \msc{D}_0^\dag\pa{f(r)e^{ikr_*}}&\approx\pa{f'(r)+\frac{iw}{\Delta}f(r)}e^{ikr_*},\\
    \msc{D}_0^\dag\pa{f(r)e^{-ikr_*}}&\approx f'(r)e^{-ikr_*}.
\end{align}
\end{subequations}
When $\msc{D}_0$ acts on $\hat{R}_{\omega\ell m}^{(-2)\,{\rm out}}$ or when $\msc{D}_0^\dag$ acts on $\Delta^2\hat{R}_{\omega\ell m}^{(+2)\,{\rm in}}$, the leading-order terms vanish by Eq.~\eqref{eq:HorizonBehaviorBis}.
As such, we focus on the following two expressions:
\begin{subequations}
\begin{align}
    \msc{D}_0^4\hat{R}_{\omega\ell m}^{(-2)\,{\rm in}}&\approx c^{\rm in}\msc{D}_0^4\pa{\Delta^2e^{-ikr_*}}
    \approx c^{\rm in}\frac{\Gamma}{\Delta^2}e^{-ikr_*},\\
    \Delta^2\pa{\msc{D}_0^\dag}^4\Delta^2\hat{R}_{\omega\ell m}^{(+2)\,{\rm out}}&\approx c^{\rm out}\Delta^2\pa{\msc{D}_0^\dag}^4\pa{\Delta^2e^{ikr_*}}
    \approx c^{\rm out}\wt{\Gamma}e^{ikr_*},
\end{align}
\end{subequations}
where $\Gamma$ and $\wt{\Gamma}$ are defined in Eq.~\eqref{eq:GammaSigmaW}.
We can now read off $\hat{\msc{C}}_{\omega\ell m}^{\rm in}$ and $\hat{\msc{C}}_{\omega\ell m}^{{\rm out}\,\prime}$.
Since the product of $\hat{\msc{C}}_{\omega\ell m}^{\rm in/out}$ and $\hat{\msc{C}}_{\omega\ell m}^{{\rm in/out}\,\prime}$ is fixed to be the known quantity $\mc{C}_{\omega\ell m}$ given in Eq.~\eqref{eq:RadialConstant}, this then also determines $\hat{\msc{C}}_{\omega\ell m}^{\rm out}$ and $\hat{\msc{C}}_{\omega\ell m}^{{\rm in}\,\prime}$, reproducing all the constants given in Eq.~\eqref{eq:ConstantsC}.

Ori \cite{Ori2003} showed that this method also works for the ``up'' and ``down'' modes, which we have ignored as we do not know how to represent them in terms of the function $\HeunC(z)$.\footref{fn:UpDown}
However, such a representation is not needed here, as we only need their behavior at infinity,
\begin{subequations}
\label{eq:InfinityBehavior}
\begin{align}
    \hat{R}_{\omega\ell m}^{(s)\,{\rm up}}(r)&\stackrel{r\to \infty}{\approx}c^{\rm up}\frac{e^{+i\omega r_*}}{r^{2s+1}},\\
    \hat{R}_{\omega\ell m}^{(s)\,{\rm down}}(r) &\stackrel{r\to \infty}{\approx}c^{\rm down}\frac{e^{-i\omega r_*}}{r}.
\end{align}
\end{subequations}
At large radius $r\to\infty$, the operators \eqref{eq:Dn} simplify to
\begin{align}
    \msc{D}_0\approx\pd_{r}-i\omega,\qquad
    \msc{D}_0^\dag\approx \pd_{r}-i\omega.
\end{align}
It follows that at leading order in $1/r$,
\begin{subequations}
\begin{gather}
    \msc{D}_0\pa{\frac{e^{i\omega r_*}}{r^n}}\approx\msc{D}_0^\dag\pa{\frac{e^{-i\omega r_*}}{r^n}}
    \approx0,\\
    \msc{D}_0\pa{\frac{e^{-i\omega r_*}}{r^n}}\approx -2i\omega\frac{e^{-i\omega r_*}}{r^n},\qquad
    \msc{D}_0^\dag\pa{\frac{e^{i\omega r_*}}{r^n}}\approx 2i\omega\frac{e^{i\omega r_*}}{r^n}.
\end{gather}
\end{subequations}
Hence, $\msc{D}_0\hat{R}^{(\pm2)\,\rm up}_{\omega\ell m}$ and $\msc{D}_0^\dag \Delta^2 \hat{R}^{(\pm2)\,\rm down}_{\omega\ell m}$ vanish at leading order.
As such, we focus on the following two expressions:
\begin{subequations}
\begin{align}
    \msc{D}_0^4\hat{R}_{\omega\ell m}^{(-2)\,\rm down}&\approx c^{\rm down}\msc{D}_0^4\pa{\frac{e^{-i\omega r_*}}{r}}
    \approx16\omega^4c^{\rm down}\frac{e^{-i\omega r_*}}{r},\\
    \Delta^2\pa{\msc{D}_0^\dag}^4\Delta^2\hat{R}_{\omega\ell m}^{(+2)\,\rm up}&\approx c^{\rm up}\Delta^2\pa{\msc{D}_0^\dag}^4\pa{\frac{e^{+i\omega r_*}}{r^3}}
    \approx16\omega^4c^{\rm up}\frac{e^{i\omega r_*}}{r}.
\end{align}
\end{subequations}
We now read off the analogues of the constants \eqref{eq:ConstantsC}:
\begin{align}
    \hat{\msc{C}}_{\omega\ell m}^{\rm up}= 16\omega^4,\qquad
    \hat{\msc{C}}_{\omega\ell m}^{{\rm up}\,\prime}
    = \frac{\mc{C}_{\omega\ell m}}{16\omega^4}, \qquad
    \hat{\msc{C}}_{\omega\ell m}^{\rm down}=\frac{\mc{C}_{\omega\ell m}}{16\omega^4},\qquad
    \hat{\msc{C}}_{\omega\ell m}^{{\rm down}\,\prime}=16\omega^4.
\end{align}
The same method applied to the angular modes \eqref{eq:AngularModes} leads to their associated constants \eqref{eq:ConstantsD}, but the expansions are more complicated.
We omit these computations for the sake of brevity.

\bibliographystyle{utphys}
\bibliography{BGL1.bib}

\end{document}